\documentclass[longauth,traditabstract]{aa}  
\usepackage{footmisc}
\usepackage{amsmath}

\usepackage{setspace}
\usepackage{graphicx}
\usepackage{placeins} 
\usepackage{txfonts}
\usepackage{multirow}
\usepackage{caption}
\usepackage{subcaption}
\usepackage{float}
\usepackage{natbib}
\bibpunct{(}{)}{;}{a}{}{,}

\usepackage[colorlinks=false]{hyperref}
\usepackage{url}
\newcommand{\hess}{H.E.S.S.}
\newcommand{\rxj}{RX~J1713.7$-$3946}
\newcommand{\gthree}{G347.3$-$0.5}
\newcommand{\xmm}{\emph{XMM-Newton}}
\newcommand{\fermi}{{\emph{Fermi}-LAT}}
\newcommand{\rosat}{\emph{ROSAT}}

\begin{document} 

\title{\hess\ observations of \rxj\ with improved angular and spectral
  resolution: Evidence for gamma-ray emission extending beyond the
  X-ray emitting shell}

\titlerunning{Observations of \rxj}
\authorrunning{\hess\ Collaboration}

\date{Received 26 Sep 2016; accepted 18 Dec 2016}

\abstract{Supernova remnants exhibit shock fronts (shells) that can
  accelerate charged particles up to very high energies. In the past
  decade, measurements of a handful of shell-type supernova remnants
  in very high-energy gamma rays have provided unique insights into
  the acceleration process. Among those objects, \rxj~(also known as
  \gthree) has the largest surface brightness, allowing us in the past
  to perform the most comprehensive study of morphology and spatially
  resolved spectra of any such very high-energy gamma-ray source.
  Here we present extensive new \hess\ measurements of \rxj, almost
  doubling the observation time compared to our previous
  publication. Combined with new improved analysis tools, the previous
  sensitivity is more than doubled. The \hess\ angular resolution of
  $0.048^\circ$ ($0.036^\circ$ above 2~TeV) is unprecedented in
  gamma-ray astronomy and probes physical scales of 0.8 (0.6) parsec
  at the remnant's location.

  The new \hess\ image of \rxj\ allows us to reveal clear
  morphological differences between X-rays and gamma rays. In
  particular, for the outer edge of the brightest shell region, we
  find the first ever indication for particles in the process of
  leaving the acceleration shock region. By studying the broadband
  energy spectrum, we furthermore extract properties of the parent
  particle populations, providing new input to the discussion of the
  leptonic or hadronic nature of the gamma-ray emission mechanism.}


\makeatletter
\renewcommand*{\fnsymbol}[1]{\ifcase#1\or*\or$\dagger$\or$\ddagger$\or**\or$\dagger\dagger$\or$\ddagger\ddagger$\fi}
\makeatother

\author{\tiny H.E.S.S. Collaboration
\and H.~Abdalla \inst{1}
\and A.~Abramowski \inst{2}
\and F.~Aharonian \inst{3,4,5}
\and F.~Ait Benkhali \inst{3}
\and A.G.~Akhperjanian\protect\footnotemark[2] \inst{6,5} 
\and T.~Andersson \inst{10}
\and E.O.~Ang\"uner \inst{7}
\and M.~Arrieta \inst{15}
\and P.~Aubert \inst{24}
% ...
}

\institute{
Centre for Space Research, North-West University, Potchefstroom 2520, South Africa \and 
Universit\"at Hamburg, Institut f\"ur Experimentalphysik, Luruper Chaussee 149, D 22761 Hamburg, Germany \and 
Max-Planck-Institut f\"ur Kernphysik, P.O. Box 103980, D 69029 Heidelberg, Germany \and 
Dublin Institute for Advanced Studies, 31 Fitzwilliam Place, Dublin 2, Ireland \and 
National Academy of Sciences of the Republic of Armenia,  Marshall Baghramian Avenue, 24, 0019 Yerevan, Republic of Armenia  \and
Yerevan Physics Institute, 2 Alikhanian Brothers St., 375036 Yerevan, Armenia \and
Institut f\"ur Physik, Humboldt-Universit\"at zu Berlin, Newtonstr. 15, D 12489 Berlin, Germany \and
University of Namibia, Department of Physics, Private Bag 13301, Windhoek, Namibia \and
GRAPPA, Anton Pannekoek Institute for Astronomy, University of Amsterdam,  Science Park 904, 1098 XH Amsterdam, The Netherlands \and
Department of Physics and Electrical Engineering, Linnaeus University,  351 95 V\"axj\"o, Sweden \and
Institut f\"ur Theoretische Physik, Lehrstuhl IV: Weltraum und Astrophysik, Ruhr-Universit\"at Bochum, D 44780 Bochum, Germany \and
GRAPPA, Anton Pannekoek Institute for Astronomy and Institute of High-Energy Physics, University of Amsterdam,  Science Park 904, 1098 XH Amsterdam, The Netherlands \and
Institut f\"ur Astro- und Teilchenphysik, Leopold-Franzens-Universit\"at Innsbruck, A-6020 Innsbruck, Austria \and
School of Physical Sciences, University of Adelaide, Adelaide 5005, Australia \and
LUTH, Observatoire de Paris, PSL Research University, CNRS, Universit\'e Paris Diderot, 5 Place Jules Janssen, 92190 Meudon, France \and
Sorbonne Universit\'es, UPMC Universit\'e Paris 06, Universit\'e Paris Diderot, Sorbonne Paris Cit\'e, CNRS, Laboratoire de Physique Nucl\'eaire et de Hautes Energies (LPNHE), 4 place Jussieu, F-75252, Paris Cedex 5, France \and
Laboratoire Univers et Particules de Montpellier, Universit\'e Montpellier, CNRS/IN2P3,  CC 72, Place Eug\`ene Bataillon, F-34095 Montpellier Cedex 5, France \and
DSM/Irfu, CEA Saclay, F-91191 Gif-Sur-Yvette Cedex, France \and
Astronomical Observatory, The University of Warsaw, Al. Ujazdowskie 4, 00-478 Warsaw, Poland \and
Aix Marseille Universit\'e, CNRS/IN2P3, CPPM UMR 7346,  13288 Marseille, France \and
Instytut Fizyki J\c{a}drowej PAN, ul. Radzikowskiego 152, 31-342 Krak{\'o}w, Poland \and
Funded by EU FP7 Marie Curie, grant agreement No. PIEF-GA-2012-332350,  \and
School of Physics, University of the Witwatersrand, 1 Jan Smuts Avenue, Braamfontein, Johannesburg, 2050 South Africa \and
Laboratoire d'Annecy-le-Vieux de Physique des Particules, Universit\'{e} Savoie Mont-Blanc, CNRS/IN2P3, F-74941 Annecy-le-Vieux, France
% ...
}

\offprints{H.E.S.S.~collaboration,
\protect\\\email{\href{mailto:contact.hess@hess-experiment.eu}{contact.hess@hess-experiment.eu}};
  \protect\\${}^*$ Corresponding authors
  \protect\\${}^\dagger$ Deceased
}

\maketitle

\makeatletter
\renewcommand*{\fnsymbol}[1]{\ifcase#1\@arabic{#1}\fi}
\makeatother

\section{Introduction}
\label{sec:intro}
Highly energetic particles with energies up to $10^{20}$
electron volts (eV, 1~eV~=~$1.6\times 10^{-19}$~J) hit the 
atmosphere of the Earth from outer space. These cosmic rays (CRs) are an important
part of the energy budget of the interstellar medium. In our Galaxy
the CR energy density is as large as the energy density of thermal gas
or magnetic fields, yet the exact connection and interaction between
these different components is poorly
understood~\citep{2015ARA&A..53..199G}.

Among the measured properties of CRs is the energy spectrum
measured at Earth, which extends over many orders of magnitude. At
least up to a few times $10^{15}$~eV, these particles are likely of
Galactic origin -- there must be objects in the Milky Way that
accelerate charged particles to these energies.  The composition of
Galactic CRs is also known~\citep{Agashe:2014kda}: at GeV to TeV
energies, they are dominantly protons. Alpha particles and heavier
ions make up only a small fraction of CRs. Electrons, positrons, gamma
rays and neutrinos contribute less than 1\%.

Establishing the Galactic sources of charged CRs is one of the main
science drivers of gamma-ray astronomy. The standard paradigm is that
young supernova remnants (SNRs), expanding shock waves following
supernova explosions, are these accelerators of high-energy Galactic
CRs~\citep[for a review, see for example][]{2013A&ARv..21...70B}. Such
events can sustain the energy flux needed to power Galactic CRs. In
addition, there exists a theoretical model of an acceleration process
at these shock fronts, known as diffusive shock acceleration
(DSA)~\citep{1977DoSSR.234.1306K,1977ICRC...11..132A,1978MNRAS.182..147B,1978ApJ...221L..29B},
which provides a good explanation of the multiwavelength data of
young SNRs. In the past decade, a number of young SNRs have been
established by gamma-ray observations as accelerators reaching
particle energies up to at least a few hundred TeV. It is difficult to
achieve unequivocal
proof, however, that these accelerated particles are protons, which emit
gamma rays via the inelastic production of neutral pions, and not
electrons, which could emit very high-energy (VHE; Energies
$E > 100$\,GeV) gamma rays via inverse Compton (IC) scattering of
ambient lower energy photons. For old SNRs,
for which the highest energy particles are believed to have already
escaped the accelerator volume, the presence of protons has been
established in at least five cases. For W28, the correlation of TeV
gamma rays with nearby molecular clouds suggests the presence of
protons~\citep{2008A&A...481..401A}. At lower GeV energies, four SNRs
(IC\,443, W44, W49B, and W51C) have recently been proven to be proton
accelerators by the detection of the characteristic pion bump in the
\emph{Fermi} Large Area Telescope (\fermi)
data~\citep{FermiPion,2016ApJ...816..100J,W49BForth}. At higher
energies and for young SNRs, this unequivocal proof remains to be
delivered. It can ultimately be found at gamma-ray energies exceeding
about 50~TeV, where electrons suffer from the so-called Klein-Nishina
suppression~(see for example \citet{2013APh....43...71A} for a
review), or via the detection of neutrinos from charged pion decays
produced in collisions of accelerated CRs with ambient gas at or near
the accelerator.

\rxj~(also known as \gthree) is the best-studied young gamma-ray
SNR~\citep{Hess1713a,Hess1713b,Hess1713c,FermiRXJ}. It was discovered
in the \rosat\ all-sky survey~\citep{Pfeffermann} and has an estimated
distance of 1\,kpc~\citep{2003PASJ...55L..61F}. It is a prominent and
well-studied example of a class of X-ray bright and radio
dim~\citep{Lazendic} shell-type SNRs\footnote{Another example with
  very similar properties is RX~J0852.0$-$4622 (Vela Junior); see
  \citet{VelaJrForth}.}. The X-ray emission of \rxj\ is completely
dominated by a non-thermal
component~\citep{Koyama,Slane,CassamXMM,UchiyamaChandra,2008ApJ...685..988T},
and in fact, the first evidence for thermal X-ray line emission was
reported only recently~\citep{2015ApJ...814...29K}. Despite the past
deep \hess\ exposure and detailed spectral and morphological studies,
the origin of the gamma-ray emission (leptonic, hadronic, or a mix of
both) is not clearly established. All scenarios have been shown to
reproduce the spectral data under certain assumptions~(as discussed by
\citet{2014MNRAS.445L..70G} and references therein). In addition to
such broadband modelling of the emission spectra of \rxj, correlation
studies of the interstellar gas with X-ray and gamma-ray emission are
argued to show evidence for hadronic gamma-ray
emission~\citep{2012ApJ...746...82F}.

We present here new, deeper, \hess\ observations, analysed with our most
advanced reconstruction techniques yielding additional performance
improvements. After a detailed presentation of the new \hess\ data
analysis results and multiwavelength studies, we update the
discussion about the origin of the gamma-ray emission.

\begin{figure*}
\centering
  \begin{subfigure}{0.49\textwidth}
    \includegraphics[width=\textwidth]{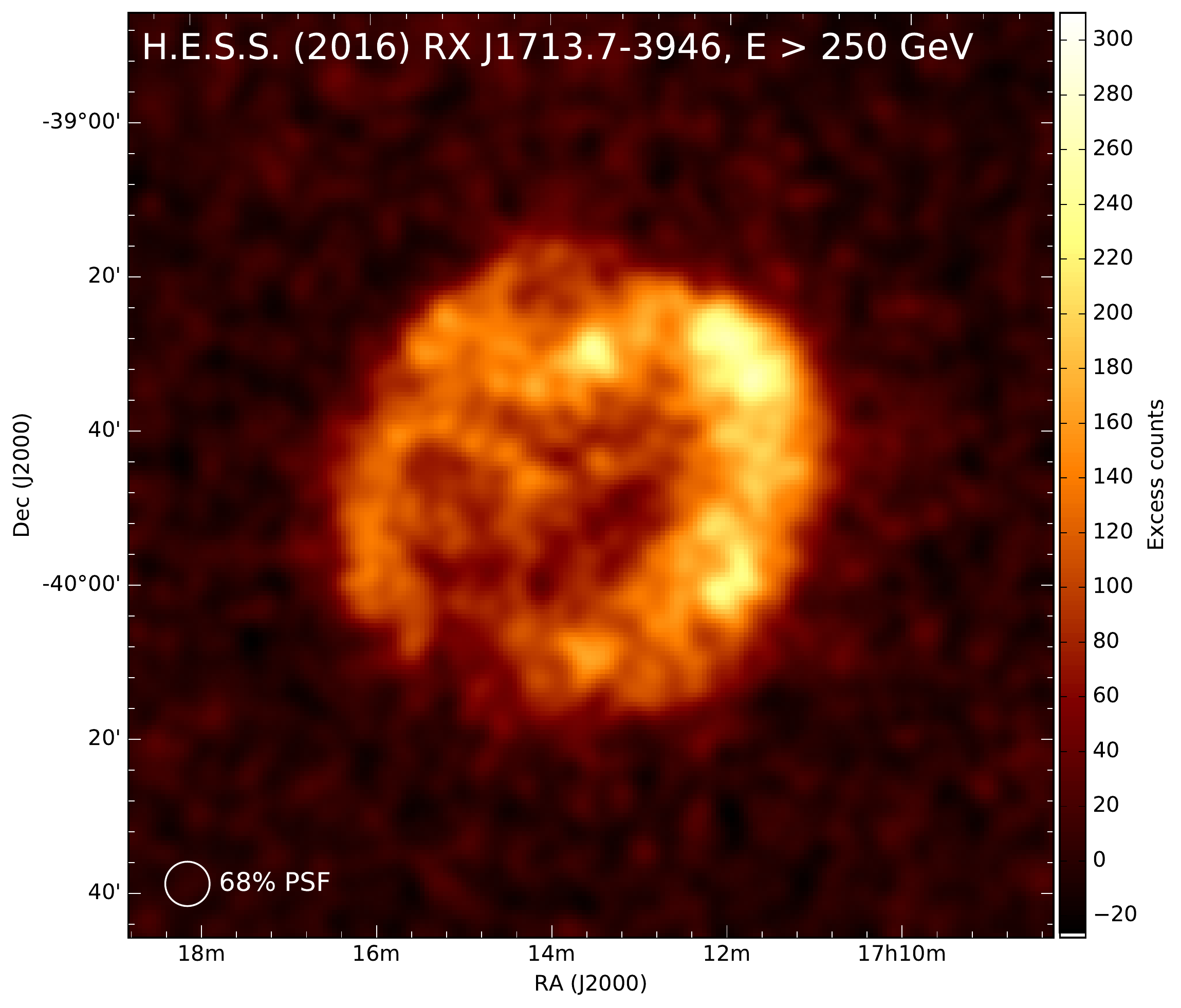}
  \end{subfigure}
  \begin{subfigure}{0.49\textwidth}
    \includegraphics[width=\textwidth]{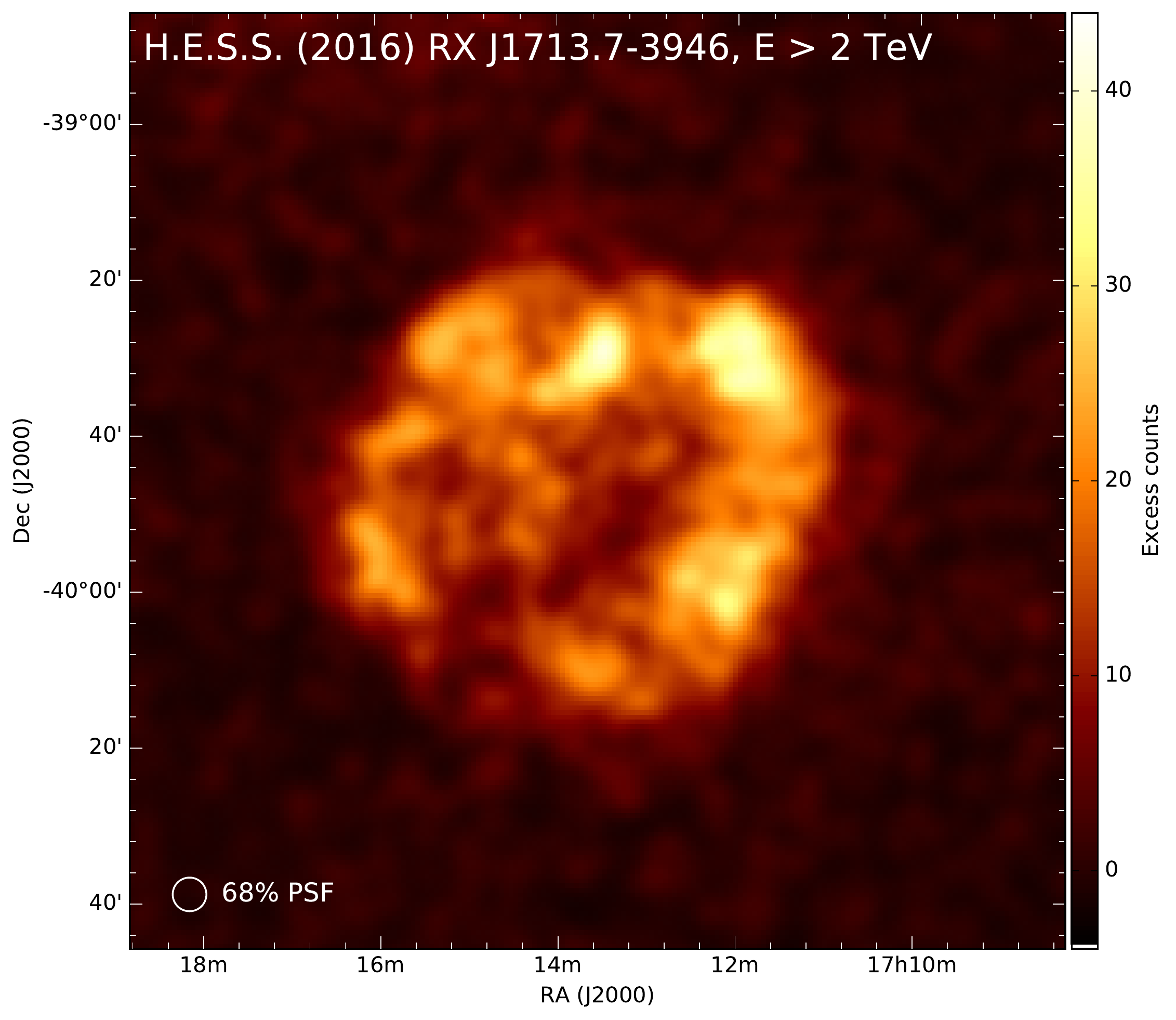}
  \end{subfigure}
  \caption{\hess\ gamma-ray excess count images of \rxj, corrected
    for the reconstruction acceptance. On the left, the image is made
    from all events above the analysis energy threshold of 250~GeV. On
    the right, an additional energy requirement of $E>2$\,TeV is
    applied to improve the angular resolution. Both images are
    smoothed with a two-dimensional Gaussian of width $0.03^{\circ}$,
    i.e.\ smaller than the 68\% containment radius of the PSF of
    the two images ($0.048^{\circ}$ and $0.036^{\circ}$,
    respectively). The PSFs are indicated by the white circles in the
    bottom left corner of the images. The linear colour scale is in
    units of excess counts per area, integrated in a circle of radius
    $0.03^{\circ}$, and adapted to the width of the Gaussian function used
    for the image smoothing.}
  \label{fig:hess-maps}
\end{figure*}

\section{\hess\ observations and analysis}
\label{sec:obsana}
The High Energy Stereoscopic System (\hess) is an array of imaging
atmospheric Cherenkov telescopes located at 1800\,m altitude in the
Khomas highlands of Namibia. The \hess\ array is designed to detect
and image the brief optical Cherenkov flash emitted from air showers,
induced by the interaction of VHE gamma rays with the 
atmosphere of the Earth. In the first phase of \hess, during which the data used
here were recorded, the array consisted of four 13\,m telescopes
placed on a square of 120\,m side length. Gamma-ray events were
recorded when at least two telescopes in the array were triggered in
coincidence~\citep{FunkTriggerPaper}, allowing for a stereoscopic
reconstruction of gamma-ray events~\citep[for further details
see][]{HessCrab}.

Recently \hess\ has entered its second phase with the addition of a
fifth, large 28\,m telescope at the centre of the array. The addition
of this telescope, which is able to trigger both independently and in
concert with the rest of the array, increases the energy coverage of
the array to lower energies. The work presented in the following
sections does not use data recorded with the large telescope. 

\begin{table}[htb]
  \caption[]{Overview of the \hess\ observation campaigns. The livetime
    given in hours corresponds to the data fulfilling quality
    requirements.}
\begin{center}
\begin{tabular}{lccc}
\hline\hline\noalign{\smallskip}
\multicolumn{1}{l}{year} &
\multicolumn{1}{l}{mean offset$^{(1)}$} &
\multicolumn{1}{l}{mean zenith angle} &
\multicolumn{1}{l}{livetime} \\
\multicolumn{1}{l}{} &
\multicolumn{1}{c}{(degrees)} &
\multicolumn{1}{c}{(degrees)} &
\multicolumn{1}{c}{(hours)} \\
\noalign{\smallskip}\hline\noalign{\smallskip}
2004 & 0.74 & 30 & 42.7 \\
2005 & 0.77 & 48 & 42.1 \\
2011 & 0.73 & 42 & 65.3 \\
2012 & 0.90 & 28 & 13.4 \\
\hline\noalign{\smallskip}
\end{tabular}
\label{table:obstimes}
\end{center}
$^{(1)}$ Mean angular distance between the \hess\ observation position
and the nominal centre of the SNR taken to be at R.A.:
17$^\mathrm{h}$13$^\mathrm{m}$33.6$^\mathrm{s}$, Dec.: 
$-$39$^\mathrm{d}$45$^\mathrm{m}$36$^\mathrm{s}$.
\end{table}

The \rxj\ data used here are from two distinct observation campaigns. The first
took place during the years 2003--2005 and resulted in
three \hess\ publications~\citep{Hess1713a,Hess1713b,Hess1713c}. The
second campaign took place in 2011 and 2012 and is published here for
the first time. In this new combined analysis of all \hess\
observations the two-telescope data from 2003~\citep{Hess1713a} from
the system commissioning phase are omitted to make the data set more
homogeneous. These initial \hess\ data have very limited sensitivity
compared to the rest of the data set and can therefore be safely
ignored. Details of the different campaigns are given in
Table~\ref{table:obstimes}. Only observations passing data quality
selection criteria are used, guaranteeing optimal atmospheric
conditions and correct camera and telescope tracking behaviour. This
procedure yields a total dead-time corrected exposure time of
164\,hours for the source morphology studies. For the spectral studies
of the SNR, a smaller data set of 116\,hours is used as explained
below.

The data analysis is performed with an air-shower template
technique~\citep{modelAna}, which is called the primary analysis chain
below. This reconstruction method is based on simulated gamma-ray
image templates that are fit to the measured images to derive the
gamma-ray properties. Goodness-of-fit selection criteria are applied
to reject background events that are not likely to be from gamma rays. All
results shown here were cross-checked using an independent calibration
and data analysis chain~\citep{TMVA,ImPACT}.

\section{Morphology studies}
\label{sec:morphology}
The new \hess\ image of \rxj\ is shown in Fig.~\ref{fig:hess-maps}: on
the left, the complete data set above an energy threshold of 250\,GeV
(about 31,000 gamma-ray excess events from the SNR region) and, on the
right, only data above energies of 2\,TeV. For both images an analysis
optimised for angular resolution is used~\citep[the \emph{hires}
analysis in][]{modelAna} for the reconstruction of the gamma-ray
directions, placing tighter constraints on the quality of the
reconstructed event geometry at the expense of gamma-ray
efficiency. This increased energy requirement ($E> 2$\,TeV) leads to a
superior angular resolution of $0.036^{\circ}$ ($68\%$ containment
radius of the point-spread function; PSF) compared to $0.048^{\circ}$
for the complete data set with $E > 250$\,GeV. These PSF radii are
obtained from simulations of the \hess\ PSF for this data set, where
the PSF is broadened by 20\% to account for systematic differences
found in comparisons of simulations with data for extragalactic
point-like sources such as
PKS~2155--304~\citep{2010A&A...520A..83H}. This broadening is carried out by
smoothing the PSF with a Gaussian such that the $68\%$ containment
radius increases by 20\%. To investigate the morphology of the SNR, a
gamma-ray excess image is produced employing the ring background
model~\citep{BackgroundPaper}, excluding all known gamma-ray emitting
source regions found in the latest \hess\ Galactic Plane Survey
catalogue~\citep{HGPSForth} from the background ring.

\begin{figure*}
\centering
	\begin{subfigure}[b]{0.47\textwidth}
		\centering
		\includegraphics[width=0.95\textwidth]{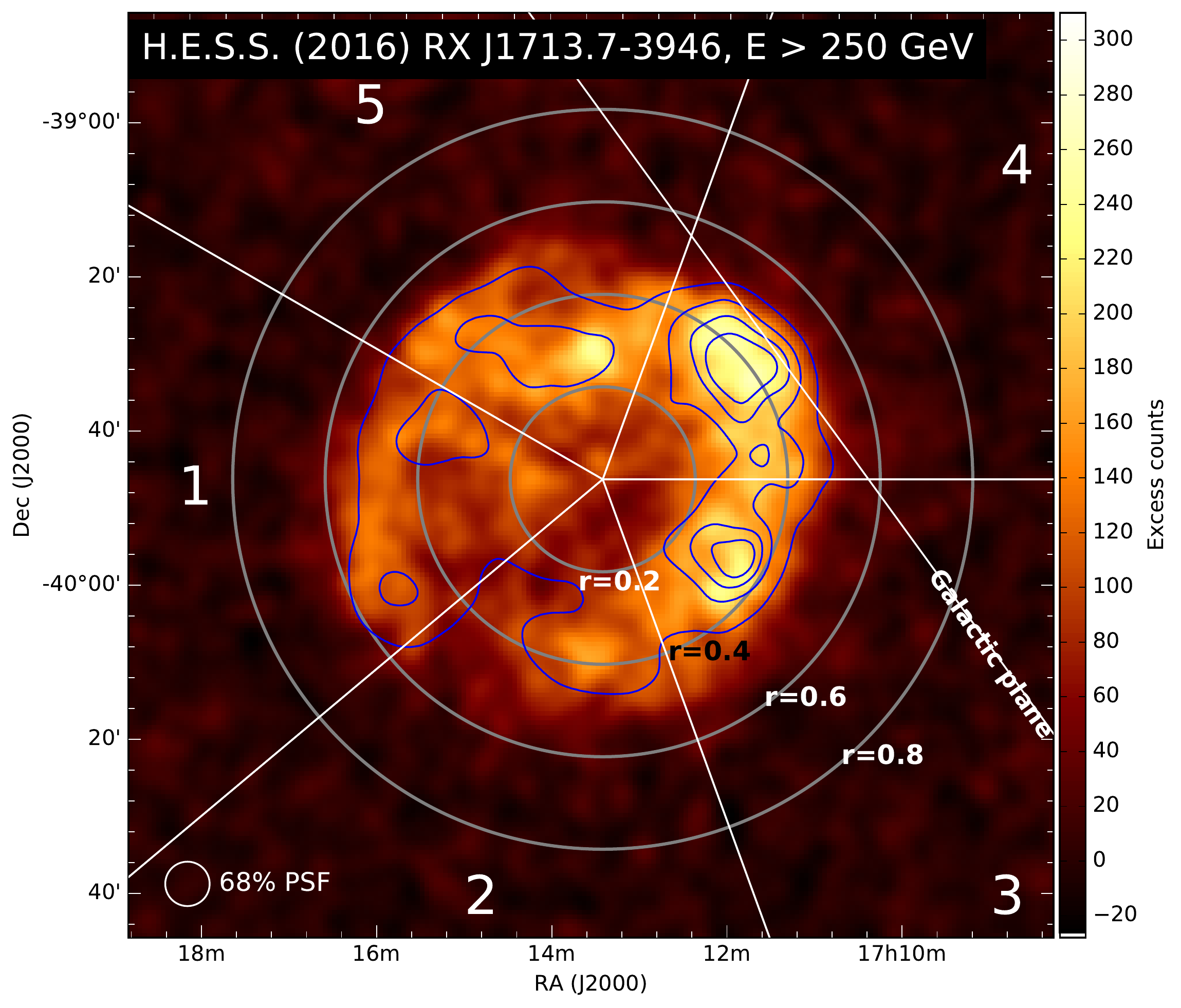}
	\end{subfigure}
	\begin{subfigure}[b]{0.47\textwidth}
		\centering
		\includegraphics[width=\textwidth]{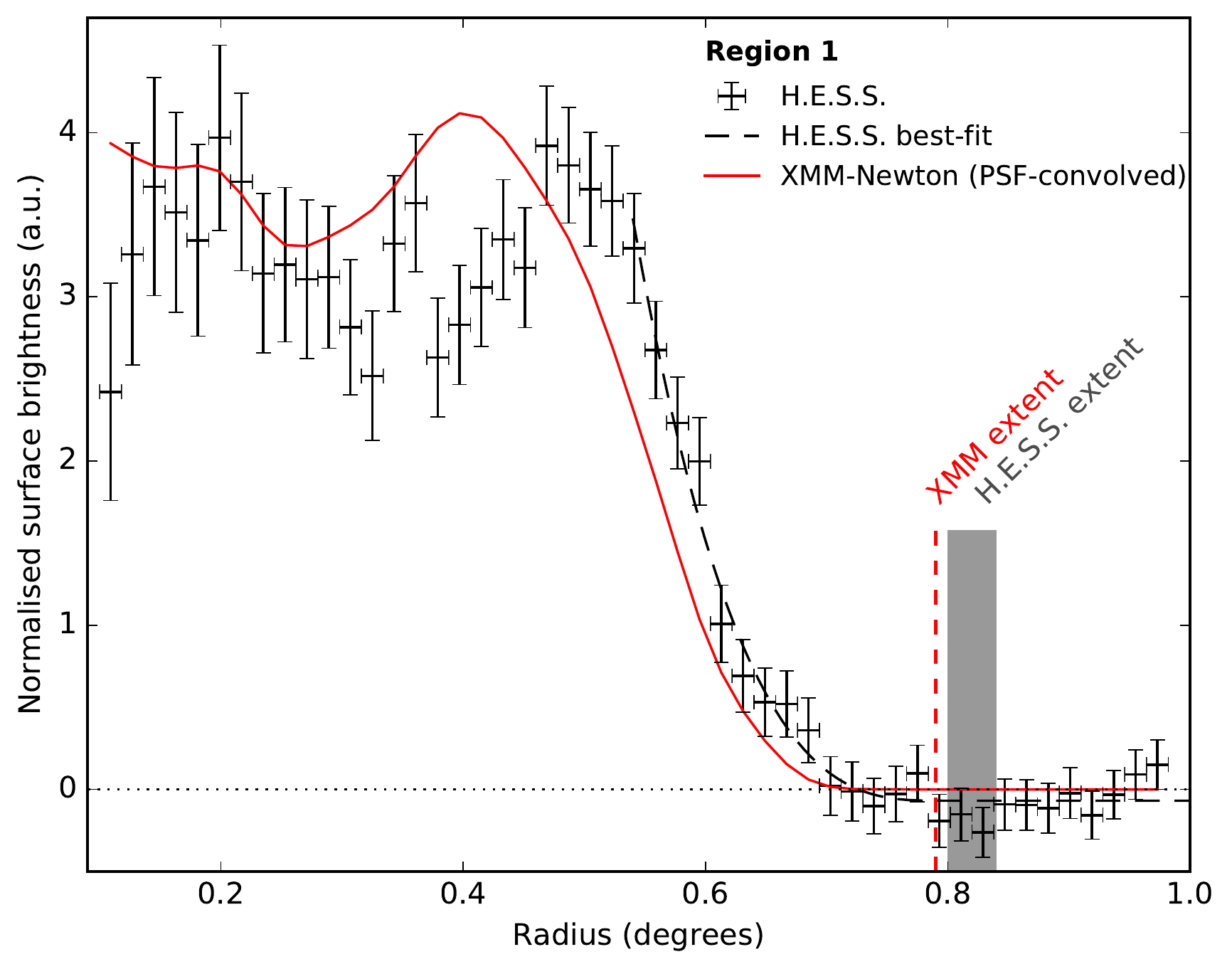}
	\end{subfigure}
	\begin{subfigure}[b]{0.47\textwidth}
		\centering
		\includegraphics[width=\textwidth]{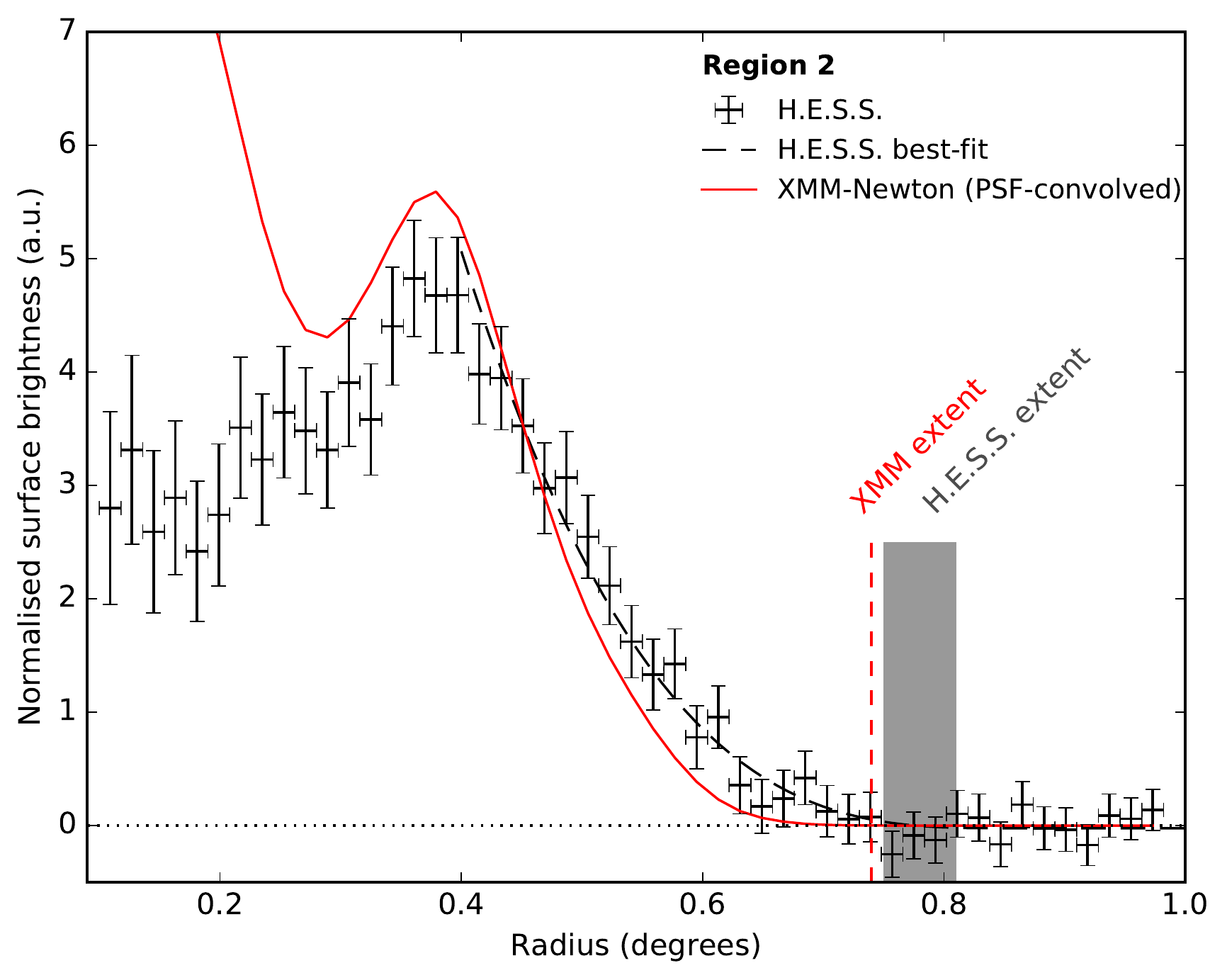}
	\end{subfigure}
	\begin{subfigure}[b]{0.47\textwidth}
		\centering
		\includegraphics[width=\textwidth]{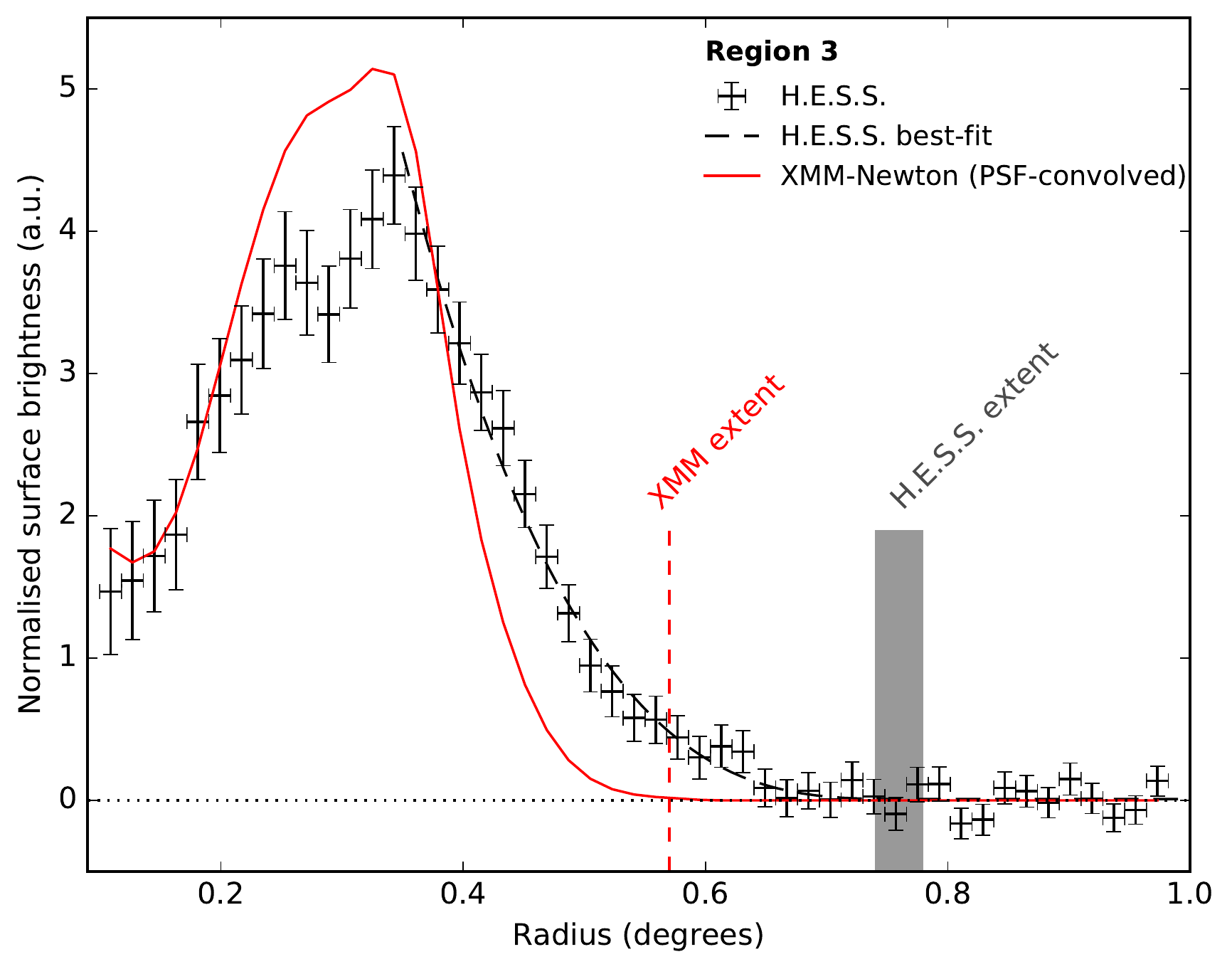}
	\end{subfigure}
	\begin{subfigure}[b]{0.47\textwidth}
		\centering
		\includegraphics[width=\textwidth]{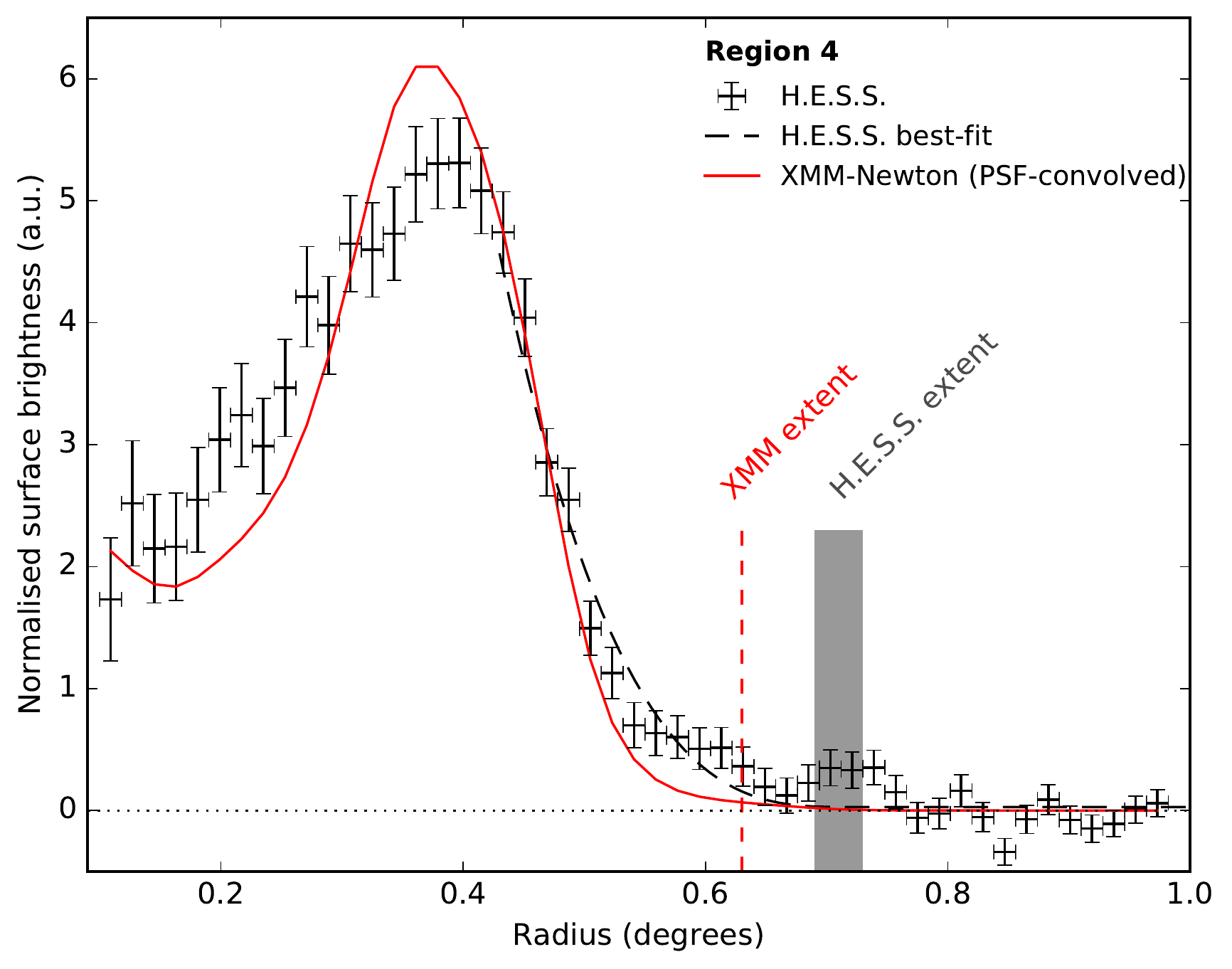}
	\end{subfigure}
	\begin{subfigure}[b]{0.47\textwidth}
		\centering
		\includegraphics[width=\textwidth]{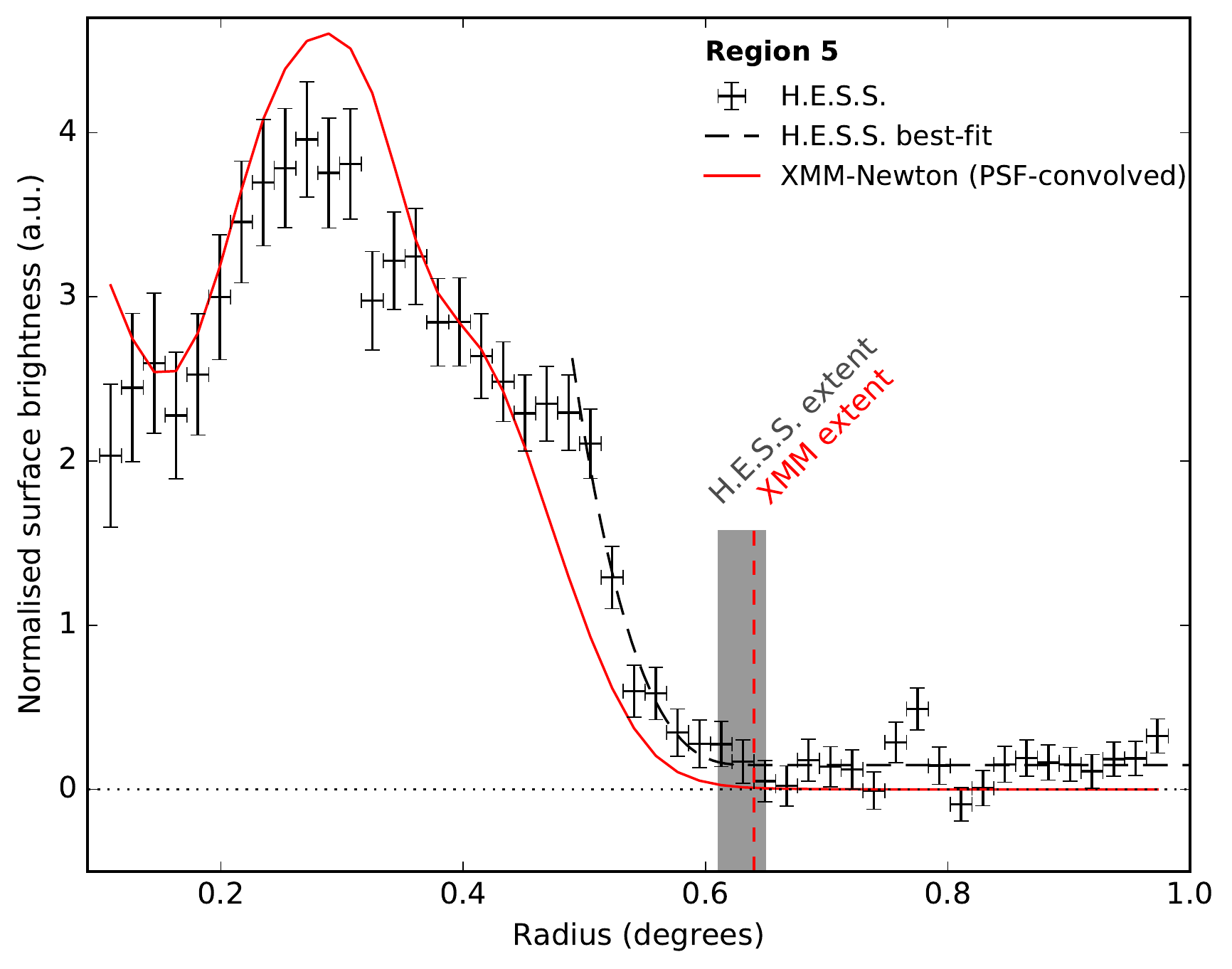}
	\end{subfigure}
	\caption{Gamma-ray excess map and radial profiles. Top left:
          the \hess\ gamma-ray count map ($E > 250$\,GeV) is shown
          with \xmm\ X-ray contours (1--10~keV, smoothed with the
          \hess\ PSF) overlaid. The five regions used to compare the
          gamma-ray and X-ray data are indicated along with concentric
          circles (dashed grey lines) with radii of $0.2^\circ$ to
          $0.8^\circ$ and centred at R.A.:
          17$^\mathrm{h}$13$^\mathrm{m}$25.2$^\mathrm{s}$, Dec.:
          $-$39$^\mathrm{d}$46$^\mathrm{m}$15.6$^\mathrm{s}$. The
          Galactic plane is also drawn. The other five panels show the
          radial profiles from these regions. The profiles are
          extracted from the \hess\ maps (black crosses) and from an
          \xmm\ map convolved with the \hess\ PSF (red line). The
          relative normalisation between the \hess\ and \xmm\ profiles
          is chosen such that for regions 1, 2, 4 the integral in
          $[0.3^\circ, 0.7^\circ]$ is the same, for regions 3, 5 in
          $[0.2^\circ, 0.7^\circ]$. The grey shaded area shows the
          combined statistical and systematic uncertainty band of the
          radial gamma-ray extension, determined as described in the
          main text. The vertical dashed red line is the radial X-ray
          extension. For the X-ray data, the statistical uncertainties
          are well below 1\% and are not shown.}
  \label{fig:radial-profiles}
\end{figure*}

The overall good correlation between the gamma-ray and X-ray image of
\rxj, which was previously found by \hess~\citep{Hess1713b}, is again
clearly visible in Fig.~\ref{fig:radial-profiles}~(top left) from the
hard X-ray contours (\xmm\ data, 1--10~keV, described further below)
overlaid on the \hess\ gamma-ray excess image. For a quantitative
comparison that also allows us to determine the radial extent of the
SNR shell both in gamma rays and X-rays, radial profiles are extracted
from five regions across the SNR as indicated in the top left plot in
Fig.~\ref{fig:radial-profiles}. To determine the optimum central
position for such profiles, a three-dimensional spherical shell model, matched to the
morphology of \rxj, is fit to the \hess\ image. This toy model
of a thick shell fits five parameters to the data as follows: the normalisation,
the $x$ and $y$ coordinates of the centre, and the inner and outer radius
of the thick shell.  The resulting centre
point is R.A.: 17$^\mathrm{h}$13$^\mathrm{m}$25.2$^\mathrm{s}$, Dec.:
$-$39$^\mathrm{d}$46$^\mathrm{m}$15.6$^\mathrm{s}$. As seen from the
figure, regions 1 and 2 cover the fainter parts of \rxj, while regions
3 and 4 contain the brightest parts of the SNR shell, closer to the
Galactic plane, including the prominent X-ray hotspots and the
densest molecular
clouds~\citep{2013PASA...30...55M,2012ApJ...746...82F}. Region 5
covers the direction along the Galactic plane to the north of \rxj.

\subsection{Production of radial profiles}
\label{subsec:radial}
The \hess\ radial profiles shown in Fig.~\ref{fig:radial-profiles} are
extracted from the gamma-ray excess image.  To produce the X-ray
radial profiles, the following procedure is applied. One mosaic image
is produced from all available archival \xmm\ data following the
method described by \citet{AceroXMM2009}. All detected point-like
sources are then removed from the map, refilling the resulting holes
using the count statistics from annular regions surrounding the
excluded regions. The cosmic-ray-induced and instrumental backgrounds
are subtracted from each observation using closed filter wheel
data sets. To subtract the diffuse Galactic astrophysical background
from the \xmm\ map, the level of the surface brightness at large
distances ($> 0.7^\circ$) from the SNR centre, well beyond the SNR
shell position, is used. Through comparison with the \rosat\ all-sky
survey map~\citep{RASS}, which covers a much larger area around \rxj,
it is confirmed that the baseline Galactic diffuse X-ray flux level is
indeed reached within the field of view of the \xmm\ coverage of
\rxj~(see Fig.~\ref{fig:xmm-vs-rosat} in the appendix). An energy
range of 1--10\,keV is chosen for the \xmm\ image to compare to the
\hess\ data to suppress Galactic diffuse emission at low energies
$<1$\,keV as much as possible while retaining good data statistics.

To compare the X-ray profiles to the \hess\ measurement, the
background is first subtracted from the \xmm\ mosaic. In a second
step, the X-ray map is then convolved with the \hess\ PSF for this
data set to account for the sizeable difference in angular resolution
of the two instruments. The resulting radial X-ray profiles are shown
in red in Fig.~\ref{fig:radial-profiles}. The relative normalisation
of the \hess\ and \xmm\ profiles, chosen such that the integral from
0.3$^\circ$ to 0.7$^\circ$ is identical for regions 1, 2, and 4 and
0.2$^\circ$ to 0.7$^\circ$ for regions 3 and 5, is arbitrary and not
relevant for the following comparison of the radial shapes, where only
relative shape differences are discussed.

\subsection{Comparisons of the gamma-ray and X-ray radial profiles}
\label{subsec:morph1}
A number of significant differences between X-rays and gamma rays
appear in the radial profiles in all five regions. In region 1, the
supposed gamma-ray shell appears as a peak around $0.5^\circ$ distance
from the SNR centre whereas it is at $0.4^\circ$ in X-rays. In region
2, pronounced differences appear below $0.3^\circ$. The central X-ray
emission in this region is brighter than the rather dim X-ray shell,
which is a
behaviour that is not present in the gamma-ray data. In region 3, the
X-ray shell peak between $0.3^\circ$ and $0.4^\circ$ is relatively
brighter and stands out above the gamma-ray peak, as was already noted
by \citet{2008ApJ...685..988T}~(see their Fig.~18 and related
discussion) and later by \citet{AceroXMM2009}. The X-ray peak is also
falling off more quickly: the decline between the peak position and a
radius of $0.7^\circ$ is significantly different. The gamma-ray data
are entirely above the X-ray data between $0.4^\circ$ and
$0.7^\circ$. Regions 4 and 5 show similar behaviour; for instance, in region 3
the X-ray peak is above the gamma-ray peak, but the X-ray data fall
off more quickly at radii beyond the peak. The gamma-ray data are then
above the X-ray data for radii $\gtrsim 0.5^\circ$.

\subsection{Determination of the SNR shell extent}
\label{subsec:morph2}
\begin{table*}[htbp]
\begin{center}
\begin{tabular}{l|cccc|ccc}
\hline\hline\noalign{\smallskip}
&\multicolumn{4}{|c|}{\hess}&\multicolumn{3}{|c}{\xmm}\\
\multicolumn{1}{l|}{Region} &
\multicolumn{1}{l}{Fit range} &
\multicolumn{1}{c}{$c$} &
\multicolumn{1}{c}{$r_0$} &
\multicolumn{1}{c|}{$\delta r_{1/e}$} &
\multicolumn{1}{l}{Fit range} &
\multicolumn{1}{c}{$r_0$} &
\multicolumn{1}{c}{$\delta r_{1/e}$} \\
\multicolumn{1}{l|}{} &
\multicolumn{1}{c}{(degrees)} &
\multicolumn{1}{c}{(10$^{-2}$ counts / pixel)} &
\multicolumn{1}{c}{(degrees)} &
\multicolumn{1}{c|}{(degrees)} &
\multicolumn{1}{c}{(degrees)} &
\multicolumn{1}{c}{(degrees)} &
\multicolumn{1}{c}{(degrees)} \\
\noalign{\smallskip}\hline\noalign{\smallskip}
1 & 0.54--1.0 & $-$3.4 $\pm$ 1.9$_\mathrm{stat}\, \pm$ 2.0$_\mathrm{sys}$ & 0.82 $\pm$ 0.02$_\mathrm{stat}$ & $-$ & 0.54--1.0 & 0.79 & $-$ \\
2 & 0.40--1.0 & $-$1.1 $\pm$ 1.9$_\mathrm{stat}\, \pm$ 2.0$_\mathrm{sys}$ & 0.78 $\pm$ 0.03$_\mathrm{stat}$ & 0.14 $\pm$ 0.02$_\mathrm{stat}$& 0.42--1.0& 0.74 & 0.10 \\
3 & 0.35--1.0 & \hspace{0.139cm} 0.6 $\pm$ 1.8$_\mathrm{stat}\, \pm$ 3.0$_\mathrm{sys}$ & 0.76 $\pm$ 0.02$_\mathrm{stat}$ & 0.11 $\pm$ 0.01$_\mathrm{stat}$ & 0.35--1.0& 0.57 & 0.06\\
4 & 0.43--1.0 & \hspace{0.139cm} 2.2 $\pm$ 1.8$_\mathrm{stat}\, \pm$ 2.0$_\mathrm{sys}$ & 0.71 $\pm$ 0.02$_\mathrm{stat}$ & 0.07 $\pm$ 0.01$_\mathrm{stat}$ & 0.43--1.0& 0.63 & 0.06\\
5 & 0.49--1.0 & \hspace{0.139cm} 9.9 $\pm$ 1.6$_\mathrm{stat}\, \pm$ 3.0$_\mathrm{sys}$ & 0.63 $\pm$ 0.02$_\mathrm{stat}$ & $-$ & 0.49--1.0& 0.64 & $-$\\
\hline\noalign{\smallskip}
\end{tabular}
\end{center}
\caption[]{Results from the radial extension measurement. The quoted
  errors indicate statistical uncertainties from the fit and
  systematic uncertainties from background normalisation.} 
\label{table:extension}
\end{table*}
Besides the general notion that the hotspots in X-rays are relatively
brighter than those in VHE gamma rays, the profiles in all regions
seem to suggest that the radial extension of the gamma-ray data exceeds
that of the X-ray data: gamma rays are measured from regions beyond
the X-ray shell. To quantify this effect, an algorithm was
developed to measure the radial extent of the SNR emission in both
data sets. For this purpose, the simple differentiable function $P(r)$,
\begin{equation}
  \label{eq:profile}
  P(r) =
  \left\{
    \begin{array}{ll} 
      A \times \left|r_0 - r\right|^n + c & \mathrm{for}~r \le r_0 \\
      c & \mathrm{for}~r > r_0
    \end{array}
  \right.\mathrm{,} 
\end{equation}
is fit to the radial profiles.

Here, $r$ is the distance from the centre, $r_0$ is the parameter
determining the radial size of the emission region, $A$ is a
normalisation factor and $c$ is a constant to account for eventual
flat residual surface brightness in the map. The exponent is fixed
empirically at a value of $n=3$ in all fits and a systematic bias due
to this choice is not found (see below). The fit ranges are chosen
such that the start always coincides with the beginning of the falling
edge of the profiles. We verified that the exact choice of the
starting value of the fit has no impact on the results as long as it
is beyond the peak or the rising part of the profile. Validation plots
demonstrating the excellent match of the best-fit model with the
gamma-ray profiles are shown in the appendix (see
Fig.~\ref{fig:radial-profile-fits:1}).

Table~\ref{table:extension} lists the best-fit results with
statistical uncertainties. In regions 1--4, the constant $c$ is
compatible with zero within errors, showing that no large-scale flat
excess emission beyond the SNR shell is seen towards these
regions. The parameter $c$ is significantly positive only for region
5, hinting at a diffuse gamma-ray excess flux along the Galactic plane
in this region.

The systematic uncertainty of the absolute normalisation of the ring
background map is at the 1--2\% level
\citep[see][]{BackgroundPaper}. To investigate its influence on the
best-fit parameters, the background normalisation is varied by
$\pm$\,2\% before subtraction from the raw event map and the fit is
repeated. The only parameter that shows a change at a comparable level
to the statistical uncertainties is the constant $c$, whereas $r_0$
remains largely unaffected. The resulting systematic uncertainty is at
a negligible level of $0.002^\circ$. This demonstrates that the free
parameter $c$ helps to mitigate the effects of systematic background
uncertainties on the value of $r_0$. The parameter offsets due to
background normalisation are quoted as systematic uncertainties in
Table~\ref{table:extension}. The choice of the exponent $n$ has a
systematic effect on the results, such that higher values of $n$ lead
to higher values of $r_0$. To avoid biassing the gamma-ray to X-ray
comparison of the $r_0$ values, the exponent is fixed at $n=3$ for
both the \hess\ and \xmm\ fits. The exact choice of $n$
influences the absolute values of $r_0$ in both wavelength regimes to
a similar degree without affecting the relative difference. The
combined statistical and systematic uncertainty for the \hess\ data is
indicated in the radial profiles in Fig.~\ref{fig:radial-profiles} as
a grey shaded band.
 
A fit with the same function (Eq.\,\ref{eq:profile})  is performed to determine the radial extent of the X-ray profiles from the
PSF-convolved \xmm\ map. Similar to the \hess\ profiles,
the model describes the data very well (see appendix,
Fig.~\ref{fig:radial-profile-fits:1}) and the fit results for $r_0$
are listed in Table~\ref{table:extension}. Statistical uncertainties
are not given in this case because the fit is performed on profiles
from an oversampled PSF-convolved map. The statistical quality of the
\xmm\ data is in any case very high such that the uncertainties (well
below 1\%) are negligible compared to those of \hess

The two dominant sources of systematic effects when measuring $r_0$ in
the \xmm\ data are uncertainties in the true angular resolution of
\hess~(relevant for the PSF smearing), and the estimation of the level
of the Galactic diffuse background in the X-ray map. The first issue
is addressed using a conservative \hess\ PSF model that is
broadened to cover systematic uncertainties. The latter issue is again
taken into account by the constant $c$ in the fit function
(Eq.\,\ref{eq:profile}) that compensates for flat homogeneous offsets
in the map.

An additional parameter, $\delta r_{1/e}$, is listed in
Table~\ref{table:extension} for regions 2, 3, and 4. This parameter is
defined as the radial distance between the peak shell emission, which is
assumed to be the position of the shock, and the position where the
peak emission has dropped to $1/e$ of its value. This parameter is
used because it is robust against systematic uncertainties of the PSF
tails and because it is often used as a measure of the diffusion length
scale of particles (see below). The difference of this parameter
$\delta r_{1/e}$ between X-rays and gamma rays is used in
Sect.~\ref{subsec:escape} below in the discussion of the radial
differences in terms of particle diffusion. For regions 1 and 5, the
shock position in X-rays in the raw un-smoothed map is not clearly
defined. We are therefore only quoting $\delta r_{1/e}$ values for
regions 2, 3, and 4.

In all regions we also tested for differences in the extent of
the gamma-ray and X-ray profiles in energy bands, splitting up the
data into $E < 1$\,TeV, $1 < E < 3$\,TeV, $E > 3$\,TeV. There is no
additional significant energy dependent difference visible in the
data: the \hess\ data behaviour is identical in the entire energy
range covered. We also verified that this energy independence
persists when using larger radii of the ring background model
($1.1^\circ$) to make sure that the standard ring radius ($0.8^\circ$)
does not cancel any energy dependence. 

In the appendix, we also present an independent approach to measure
the extension of the SNR shell using a border-detection algorithm
(see Sect.~\ref{subsec:appendix-border-finder}). The same conclusions
are reached with this algorithm: the gamma-ray emission extends beyond
the X-ray emission. We also note that the procedure described above
when performed on \hess\ maps extracted from two subsets of the data
of roughly equal exposure (years 2004/2005 and 2011/2012) shows no
significant variation of the best-fit parameters of more than
2\,$\sigma$ between the two independent data sets.

\subsection{Summary of the morphology studies: Emission beyond the
  X-ray shell}
\label{subsec:morph3}
Figure~\ref{fig:radial-profiles} demonstrates very clearly that there
are significant differences between the X-ray and gamma-ray data of
\rxj. While some of the differences within the bright western shell
region were seen before~\citep{2008ApJ...685..988T,AceroXMM2009}, we
find for the first time significant differences between X-rays and
gamma rays in the radial extent of the emission associated with the
SNR \rxj. These differences do not depend on energy; we find the same
behaviour for $E < 1$\,TeV, $1 < E < 3$\,TeV, and $E > 3$\,TeV. As
seen from Table~\ref{table:extension}, the TeV gamma-ray emission
extends significantly beyond the X-ray emission associated with the
SNR shell in region 3, and there is also similarly strong evidence in
region 4. We see the same tendency in regions 1, 2, and 5, although
the algorithm we chose to measure the radial extent does not result in
a significant radial extension difference in these regions. The
border-finder algorithm, on the other hand, yields a larger TeV
gamma-ray extension of the entire SNR (see
appendix~\ref{subsec:appendix-border-finder}).

We interpret these findings as gamma rays from VHE particles that are
in the process of leaving the main shock region. The main shock
position and extent are visible in the X-ray data, as discussed below
in Sect.~\ref{subsec:escape}, the gamma-ray emission extending further
is either due to accelerated particles escaping the acceleration shock
region or particles accelerated in the shock precursor region. This is
a major new result as it is the first such finding for any SNR shell
ever measured.

\section{Energy spectrum studies}
\label{sec:spectrum}
The gamma-ray energy spectrum of \rxj\ is investigated here in two
different ways: we first present the new \hess\ spectrum of the whole SNR
region, and second we present an update of the spatially resolved
spectral studies of the SNR region.
\subsection{Spectrum of the full remnant}
\label{subsec:full-remnant-spectrum}
The large \hess\ data set available for a spectral analysis of \rxj\
comprises observation positions that vary over the years because the
data are from both dedicated pointed campaigns of \rxj\ and the \hess\
Galactic plane survey~\citep{HGPSForth}. When applying the standard
\hess\ spectral analysis, only 23\,hours of observations are retained
because in the standard procedure the cosmic-ray background in the
gamma-ray source region (the ON region) of radius $0.6^\circ$ is
modelled with the reflected-region background
model~\citep{BackgroundPaper}. In this approach, a region of the same
size and shape as the ON region (the so-called OFF region) is
reflected about the centre of the array field of view and all events
reconstructed in the OFF region constitute the cosmic-ray background
to be subtracted from the ON region. However, for the \rxj\ data set
and sky field, close to the Galactic plane with other gamma-ray
sources nearby, many observation runs in which OFF regions overlap
partly with known gamma-ray sources are excluded from the
analysis. These 23\,hours of exposure resulting from the standard
\hess\ procedure are even below the exposure of our previous
publication~\citep{Hess1713c} because we exclude more nearby gamma-ray
source regions from the prospective OFF regions. These new
excluded gamma-ray sources were unknown at the time that we performed the
previous analysis of \rxj. The dramatic exposure loss (23\,hours
remaining from an initial 164\,hours of observation time) is therefore
calling for a new, modified approach.

\begin{figure*}
    \centering
	\includegraphics[width=0.8\textwidth]{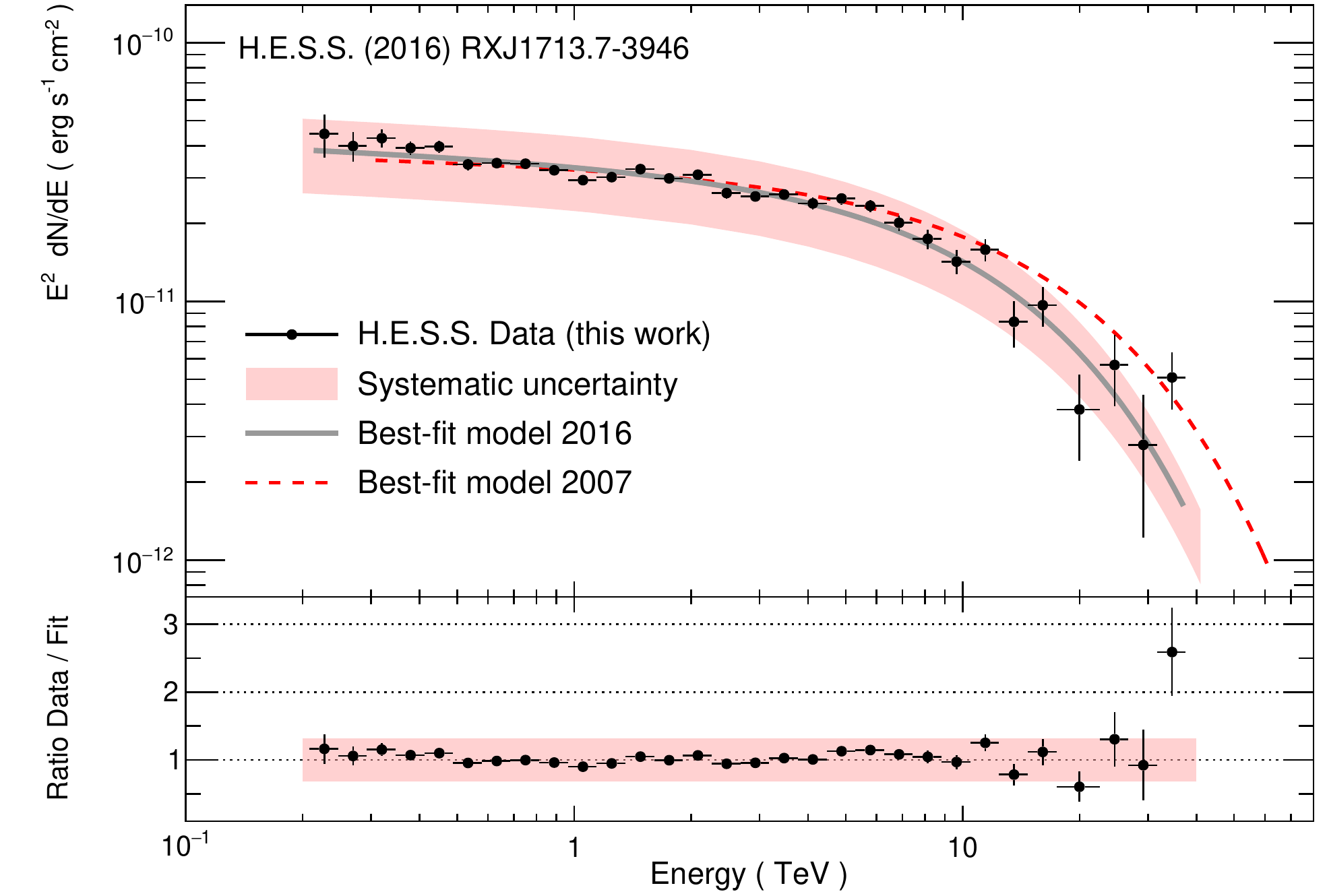}
	\caption{\hess\ energy flux spectrum. Upper panel: the black
          data points with statistical 1\,$\sigma$ error bars are the
          energy spectrum of the full SNR \rxj, using an extraction
          radius of 0.6$^\circ$ centred on R.A.:
          17$^\mathrm{h}$13$^\mathrm{m}$33.6$^\mathrm{s}$, Dec.:
          $-$39$^\mathrm{d}$45$^\mathrm{m}$36$^\mathrm{s}$. The
          binning is chosen to match the energy resolution, requiring
          a minimum significance of 2\,$\sigma$ per bin. The grey
          solid line shows the best-fit exponential cut-off power-law
          model, (2) from
          Table~\ref{tab:full-SNR-fitting-results}. The dashed red
          line shows the corresponding best-fit model from the
          previous \hess\ publication \citep{Hess1713c}. The
          experimental flux systematic uncertainty of $\pm 32\%$,
          described further in the main text, is indicated as a light
          red band. Lower panel: the residuals are shown including
          statistical and systematic uncertainties.}
	\label{fig:full-remnant-spectrum}
\end{figure*}

\begin{table*}
\caption[]{Results of the spectral fitting procedure on the full
  remnant analysis for a number of spectral models. The fits are
  performed using a finely binned SNR energy spectrum, whereas the
  spectrum shown in Fig.~\ref{fig:full-remnant-spectrum} has coarser
  binning for presentation purposes.}
\begin{center}
\begin{tabular}{lllllll}
\hline\hline\noalign{\smallskip}
\multicolumn{1}{l}{Spectral Model} &
\multicolumn{1}{l}{$\Gamma$} &
\multicolumn{1}{l}{$E_\mathrm{cut}$} &
\multicolumn{1}{l}{$F (>1$\,TeV)} &
\multicolumn{1}{l}{$F_0$, at 1\,TeV} &
\multicolumn{1}{l}{$\chi^2$ / ndf}\\[0.05cm]
\multicolumn{1}{l}{} &
\multicolumn{1}{l}{} &
\multicolumn{1}{l}{(TeV)} &
\multicolumn{1}{l}{(10$^{-11}$\,cm$^{-2}$\,s$^{-1}$)} &
\multicolumn{1}{l}{(10$^{-11}$\,cm$^{-2}$\,s$^{-1}$\,TeV$^{-1}$)} &
\multicolumn{1}{l}{} \\
\noalign{\smallskip}\hline\noalign{\smallskip}
(1) : $ F_0 E^{-\Gamma}$	& 2.32 $\pm$ 0.02	& - & 1.52 $\pm$ 0.02 & 2.02 $\pm$ 0.08 & 304 / 118 \\[0.3cm]
(2) : $ F_0 E^{-\Gamma} \exp\left(-(E / E_\mathrm{cut})^1\right)$	& 2.06 $\pm$ 0.02	& 12.9 $\pm$ 1.1	& 1.64 $\pm$ 0.02	& 2.3 $\pm$ 0.1 & 120 / 117 \\[0.3cm]
(3) : $ F_0 E^{-\Gamma} \exp\left(-(E / E_\mathrm{cut}) ^2\right)$ & 2.17 $\pm$ 0.02	& 16.5 $\pm$ 1.1	& 1.63 $\pm$ 0.02	& 2.08 $\pm$ 0.09	& 114 / 117 \\[0.3cm]
(4) : $ F_0 E^{-\Gamma} \exp\left(-(E / E_\mathrm{cut})^{1/2}\right)$	& 1.82 $\pm$ 0.04	& 2.7 $\pm$ 0.4	& 1.63 $\pm$ 0.02	& 4.0 $\pm$ 0.2	& 142 / 117 \\ 
\\
\hline\noalign{\smallskip}
\end{tabular}
\label{tab:full-SNR-fitting-results}
\end{center}
\end{table*}

For this purpose, we employed the reflected-region technique on
smaller subregions of the SNR. The circular ON region of 0.6$^\circ$
radius, centred on R.A.:
17$^\mathrm{h}$13$^\mathrm{m}$33.6$^\mathrm{s}$, Dec.:
$-$39$^\mathrm{d}$45$^\mathrm{m}$36$^\mathrm{s}$ as in
\citet{Hess1713b,Hess1713c}, is split into 18 subregions each of
similar size. The exposures of these regions vary between 97 and
130\,hours. For each of the subregions, the ON and OFF energy spectra
are extracted using the reflected-region-background model, and then
they are combined yielding the spectrum of the full SNR with improved
exposure and statistics.

During the combination procedure, we need to account for partially
overlapping OFF background regions. This is carried out by correcting the
statistical uncertainties for OFF events that are used multiple times
in the background model (60\% of the events). Moreover, the exposure
is not homogeneous across the subregions, which may lead to biasing
effects towards more exposed regions in the combined spectrum.  To
deal with this issue, the spectrum of the whole SNR is determined as
the exposure-weighted sum of all subregion spectra, rescaled to the
average exposure of the SNR before merging, and conserving the original
statistical uncertainties through error propagation.  The resulting
spectrum of the full SNR is shown in
Fig.~\ref{fig:full-remnant-spectrum}, and the energy flux points are
listed in appendix~\ref{appendix:subsection:spectra}. This spectrum
corresponds to a livetime of 116\,hours, an improvement of more than a
factor of two over our previous publications. This final energy
spectrum and in particular the split-ON-region spectral analysis has
been cross-checked with an independent analysis using the
reflected-pixel background technique, as described
in~\citet{ReflectedPixelBG}. This cross-check yields a comparable gain
in livetime and provides consistent results.

The new spectrum shown here is fit with a power-law model without and
with exponential cut-off. A best-fit model with a cut-off at 12.9\,TeV
is preferred at 13.5\,$\sigma$ over a pure power-law model as can be
seen from Table~\ref{tab:full-SNR-fitting-results}. The cut-off models
are also fit as super and sub-exponential versions (models (3) and (4)
in the table) motivated by \citet{2006PhRvD..74c4018K} who have
derived analytical expressions for the gamma-ray spectra from
inelastic proton-proton interactions, and these expressions suggest
such exponential cut-offs at varying degrees. While the
sub-exponential cut-off, model (4), is statistically disfavoured, both
the super-exponential model (3) and the simple exponential
cut-off model (2) describe the data well. The latter is used below and in
Fig.~\ref{fig:full-remnant-spectrum} as a baseline model to describe our
data.

Owing to the high statistical quality of the spectrum, the statistical
errors of the best-fit spectral parameters are smaller than the
systematic uncertainties discussed below. Taking both types of
uncertainties into account, these new results are compatible with our
previous spectrum~\citep{Hess1713c} except for a moderate discrepancy
at the highest energies. The systematic tendency of all new flux
points above about 7 TeV to lie below the previous fit seen in
Fig.~\ref{fig:full-remnant-spectrum} was traced down to the optical
efficiency correction procedure in the old
analysis~\citep{Hess1713c}. The procedure is now improved and avoids a
bias above 10\,TeV.

This experimental systematic uncertainty on the absolute reconstructed
flux is $\pm 32\%$ for the analysis and data set used here. This
uncertainty is conservatively determined as the quadratic sum of the
\hess\ flux uncertainty of $\pm 20\%$ for a point
source~\citep{HessCrab} and an extended source uncertainty that is
mainly due to the analysis method employed. This extended-source
uncertainty is found to be $25\%$, and is determined by analysing
energy spectra in 29 regions across the SNR with the primary analysis
chain~\citep{modelAna} and the cross-check
chain~\citep{TMVA,ImPACT}. These are two completely different analysis
approaches with independent calibration, reconstruction, analysis
selection requirements, and with the same reflected-region-background
model. For this purpose, both energy spectra are compared in each
region by taking the difference between the best-fit models ((2) from
Table~\ref{tab:full-SNR-fitting-results}) as measure for the
systematic uncertainty at a given energy. This is carried out for all 29
regions. The resulting differences are averaged in energy bins, and
the largest average deviation seen for any energy in the range covered
(200\,GeV to 40\,TeV) is $25\%$, which is added conservatively as an
additional $\pm 25\%$ systematic uncertainty to a total flux
uncertainty $\Delta\, \mathrm{F}_\mathrm{syst}$ of
\begin{equation}
  \label{eq:syst}
  \Delta\,\mathrm{F}_\mathrm{syst} = \sqrt{(\Delta\,
    \mathrm{F}^\mathrm{Crab}_\mathrm{syst}=20\%)^2 + (\Delta\,
    \mathrm{F}^\mathrm{Extended}_\mathrm{syst}=25\%)^2} = 32\%\, .
\end{equation}

For validation purposes, energy spectra were also extracted from two
subsets of the data (years 2004/2005 and 2011/2012 separately) to test
for a temporal evolution of the flux and spectral parameters of \rxj\
as suggested by the model in \citet{2015A&A...577A..12F}. We find that
all spectral parameters agree within their 1\,$\sigma$ uncertainty
intervals during the eight years covered by the \hess\ observations.

\subsection{Spatially resolved spectroscopy}
\label{subsec:Spatially-Resolved-Spectrum}
\begin{figure*}
  \centering
	\begin{subfigure}[b]{0.385\textwidth}
		\centering
		\includegraphics[width=\textwidth]{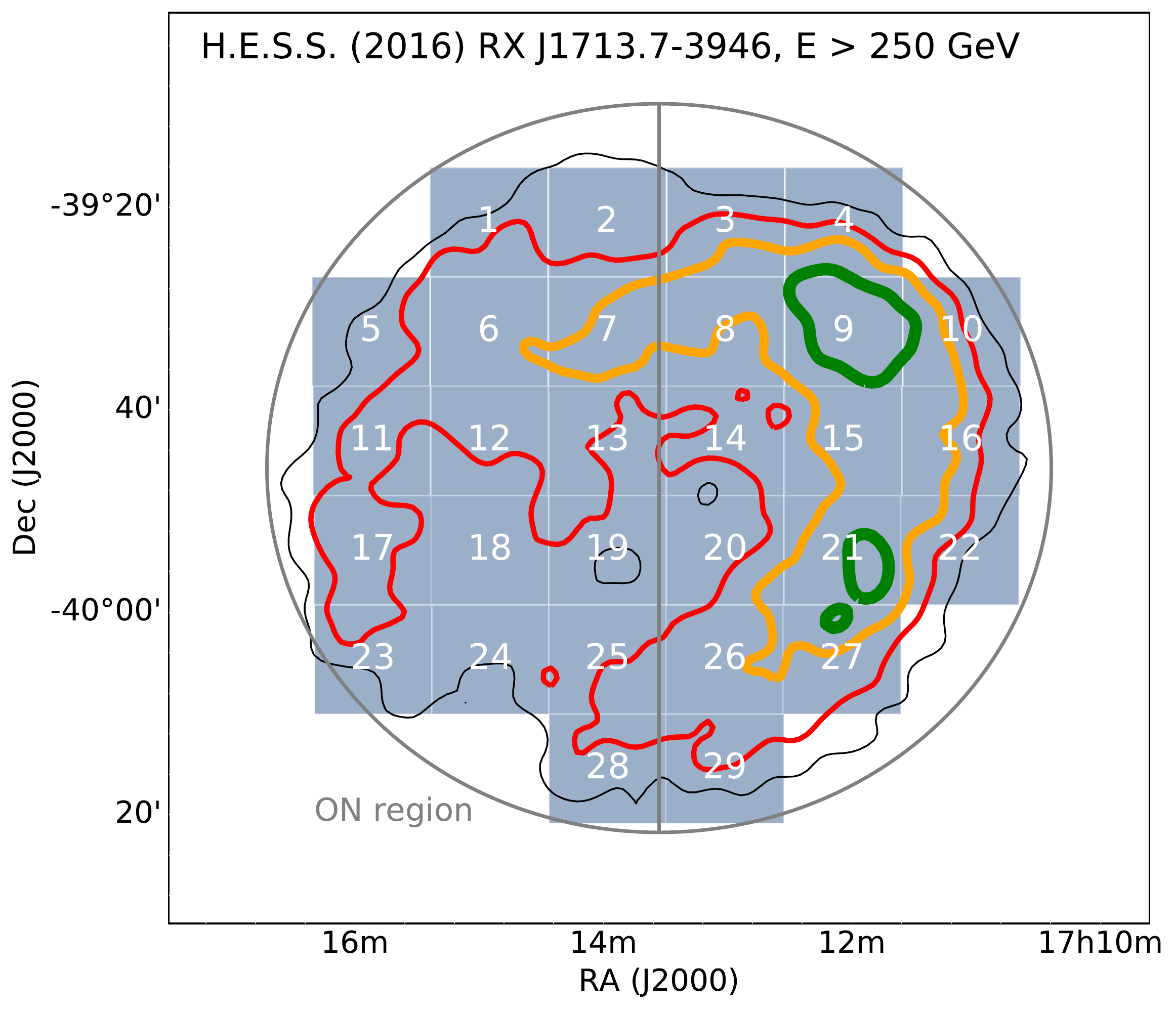}
	\end{subfigure}
	\begin{subfigure}[b]{0.58\textwidth}
		\centering
		\includegraphics[width=\textwidth]{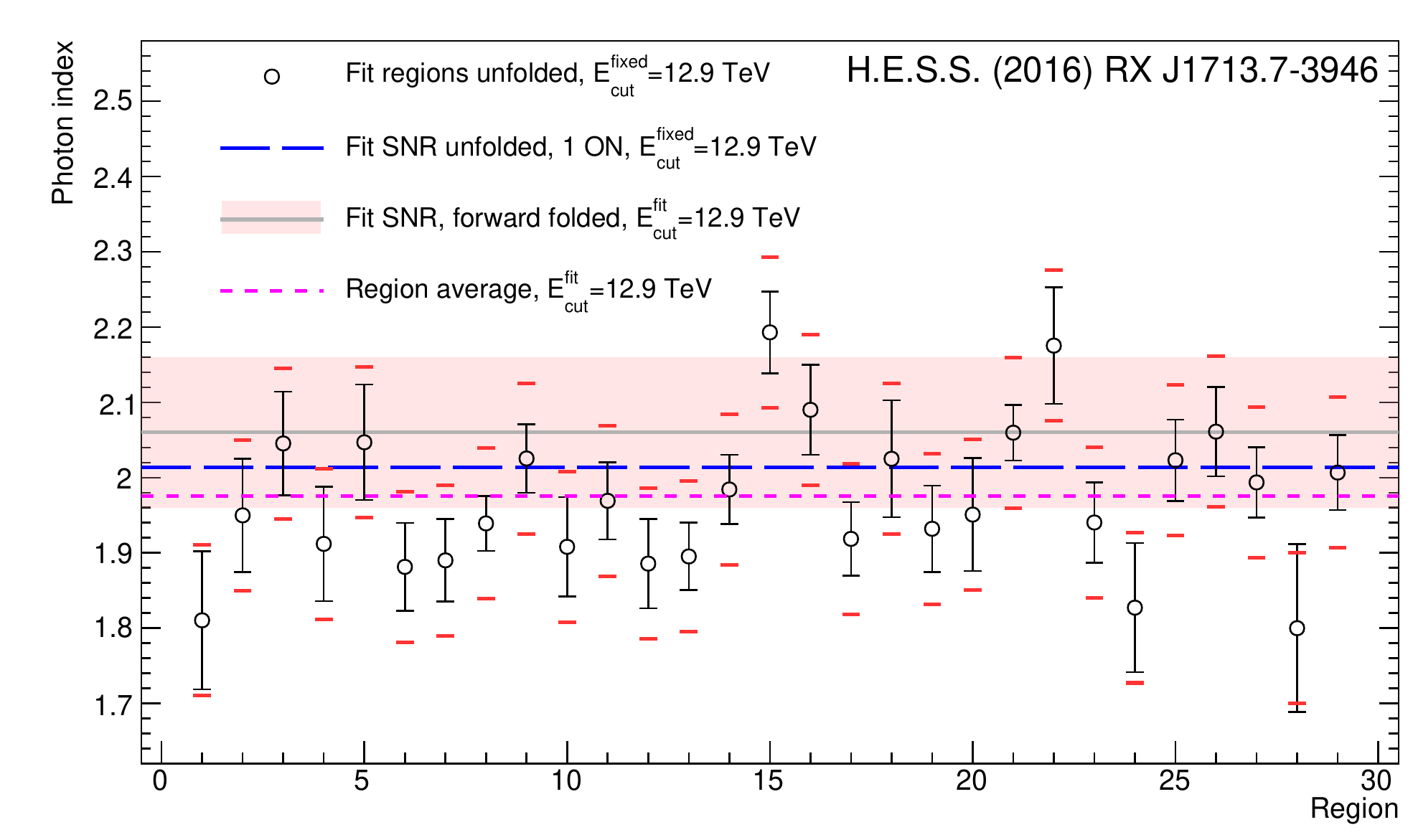}
	\end{subfigure}
        \caption{Results of the spatially resolved spectral
          analysis. On the left-hand side, the 29
          regions used for this study are shown, overlaid on the
          \hess\ gamma-ray significance contours at 3, 5, 7, and
          $9\,\sigma$ coloured in black, red, orange, and green. The
          resulting photon index distribution is shown on the
          right-hand side. The ON region used for the full SNR
          spectrum is also shown as a grey circle, the vertical grey
          line bisects the ON region into the western and eastern half
          also used for the spectral analysis. The error bars on the
          right-hand side are $1\,\sigma$ statistical fit uncertainties,
          and the additional red error intervals indicated for each point
          correspond to the systematic uncertainty on the
          reconstructed photon index of
          $\Delta \Gamma_\mathrm{sys} = 0.1$, which is also shown as
          light red band around the forward folded SNR photon index.}
\label{fig:spatialspec}
\end{figure*}

The unprecedented angular resolution and much improved sensitivity of
\hess\ has now allowed us to produce high statistics spectra of
the SNR in 29 small non-overlapping boxes with $0.18^\circ$ or 10.8
arcminute side lengths as shown in
Fig.~\ref{fig:spatialspec}~(left)\footnote{We  also verified that
  the current analysis yields consistent spectral results for the 14
  regions analysed in~\citet{Hess1713b}.}. These regions are identical
to those chosen in \citet{2008ApJ...685..988T} for the comparison of
\emph{Suzaku} X-ray and \hess\ gamma-ray data. For each box, the
spectrum is extracted by employing the reflected-region-background model,
and the resulting data are fit by a power-law model with an
exponential cut-off~((2) in
Table~\ref{tab:full-SNR-fitting-results}). The results are shown in
Table~\ref{tab:spatial-spec}. For some of the regions with low surface
brightness we cannot derive tight constrains on the exponential
cut-off energy. In these regions, a pure power-law model without
a cut-off yields a similarly good fit.

\begin{table}
  \caption[]{Spectral fitting results for the 29 \emph{Suzaku}
    regions. A power law with exponential cut-off, (2) of
    Table~\ref{tab:full-SNR-fitting-results}, was fit to the gamma-ray
    \hess\ data.}
\renewcommand{\tabcolsep}{4pt}
\begin{center}
\begin{tabular}{rcrcr}
\hline\hline\noalign{\smallskip}
\multicolumn{1}{l}{Reg.} &
\multicolumn{1}{l}{$\Gamma$} &
\multicolumn{1}{l}{$E_\mathrm{cut}$} &
\multicolumn{1}{l}{$F (>1$\,TeV)} &
\multicolumn{1}{l}{$\chi^2$ / ndf}\\[0.05cm]
\multicolumn{1}{l}{} &
\multicolumn{1}{l}{} &
\multicolumn{1}{l}{(TeV)} &
\multicolumn{1}{l}{(10$^{-13}$\,cm$^{-2}$\,s$^{-1}$)} &
\multicolumn{1}{l}{} \\
\noalign{\smallskip}\hline\noalign{\smallskip}
1 & 1.99 $\pm$ 0.16 & 20 $\pm$ 17 & 3.4 $\pm$ 0.6 & 74 / 78\\
2 & 1.95 $\pm$ 0.15 & 10.9 $\pm$ 6.3 & 4.7 $\pm$ 0.9 & 70 / 76\\
3 & 1.66 $\pm$ 0.22 & 4.2 $\pm$ 1.7 & 4.6 $\pm$ 1.3 & 58 / 77\\
4 & 1.84 $\pm$ 0.17 & 10.1 $\pm$ 5.5 & 4.1 $\pm$ 0.8 & 73 / 81\\
5 & 2.06 $\pm$ 0.13 & 25 $\pm$ 18 & 3.4 $\pm$ 0.5 & 94 / 83\\
6 & 1.72 $\pm$ 0.10 & 8.1 $\pm$ 2.1 & 8.3 $\pm$ 1.0 & 74 / 80\\
7 & 1.65 $\pm$ 0.11 & 5.8 $\pm$ 1.4 & 8.6 $\pm$ 1.2 & 97 / 79\\
8 & 1.95 $\pm$ 0.08 & 13.2 $\pm$ 3.9 & 9.6 $\pm$ 0.8 & 84 / 81\\
9 & 1.81 $\pm$ 0.08 & 7.3 $\pm$ 1.6 & 12.0 $\pm$ 1.1 & 82 / 82\\
10 & 1.90 $\pm$ 0.10 & 11.1 $\pm$ 3.9 & 6.8 $\pm$ 0.8 & 80 / 82\\
11 & 1.87 $\pm$ 0.11 & 9.8 $\pm$ 3.3 & 6.0 $\pm$ 0.7 & 119 / 80\\
12 & 1.57 $\pm$ 0.13 & 6.0 $\pm$ 1.5 & 6.6 $\pm$ 1.2 & 79 / 81\\
13 & 1.69 $\pm$ 0.12 & 7.0 $\pm$ 1.9 & 6.8 $\pm$ 1.0 & 62 / 82\\
14 & 1.97 $\pm$ 0.10 & 12.6 $\pm$ 4.6 & 6.6 $\pm$ 0.7 & 67 / 83\\
15 & 1.99 $\pm$ 0.09 & 8.4 $\pm$ 2.5 & 8.4 $\pm$ 0.9 & 89 / 77\\
16 & 2.02 $\pm$ 0.09 & 14.7 $\pm$ 5.4 & 7.6 $\pm$ 0.7 & 88 / 81\\
17 & 1.80 $\pm$ 0.11 & 9.3 $\pm$ 2.8 & 6.4 $\pm$ 0.8 & 76 / 80\\
18 & 1.34 $\pm$ 0.22 & 2.8 $\pm$ 0.8 & 5.1 $\pm$ 1.5 & 83 / 80\\
19 & 1.82 $\pm$ 0.12 & 10.3 $\pm$ 4.1 & 5.4 $\pm$ 0.8 & 73 / 78\\
20 & 1.77 $\pm$ 0.13 & 8.0 $\pm$ 2.8 & 5.3 $\pm$ 0.9 & 85 / 81\\
21 & 1.98 $\pm$ 0.09 & 9.2 $\pm$ 2.7 & 8.9 $\pm$ 0.9 & 74 / 82\\
22 & 2.14 $\pm$ 0.10 & 24 $\pm$ 15 & 5.9 $\pm$ 0.6 & 99 / 78\\
23 & 1.91 $\pm$ 0.12 & 14.0 $\pm$ 6.0 & 5.0 $\pm$ 0.6 & 81 / 80\\
24 & 1.99 $\pm$ 0.11 & 45 $\pm$ 42 & 4.0 $\pm$ 0.5 & 80 / 83\\
25 & 1.88 $\pm$ 0.15 & 6.2 $\pm$ 2.4 & 5.2 $\pm$ 0.9 & 83 / 76\\
26 & 1.76 $\pm$ 0.12 & 6.0 $\pm$ 1.6 & 7.3 $\pm$ 1.0 & 78 / 79\\
27 & 1.79 $\pm$ 0.09 & 6.2 $\pm$ 1.3 & 10.0 $\pm$ 1.0 & 84 / 81\\
28 & 1.45 $\pm$ 0.17 & 4.4 $\pm$ 1.2 & 5.5 $\pm$ 1.2 & 60 / 80\\
29 & 2.05 $\pm$ 0.09 & 17.3 $\pm$ 7.7 & 6.2 $\pm$ 0.6 & 78 / 80\\
\hline\noalign{\smallskip}
\end{tabular}
\label{tab:spatial-spec}
\end{center}
\end{table}

The following procedure is employed to test for spatial variation of
the spectral shapes measured with \hess\ Each subregion is fit with a
power law with an exponential cut-off, leaving the normalisation and
the photon index free, but fixing the cut-off at 12.9~TeV. This is the
average value obtained by fitting the energy spectrum of the whole SNR
(see (2) in Table~\ref{tab:full-SNR-fitting-results}). We verified
that the free exponential cut-off fit shown in
Table~\ref{tab:spatial-spec} does not provide a statistically
preferred best-fit model compared to the model with the cut-off fixed
in any region. This fit procedure with a fixed cut-off allows us to
directly probe all regions for a difference in the photon index
$\Gamma$; fit parameter correlations between the index and the
cut-off can then be ignored.

The resulting photon index comparison is shown in
Fig.~\ref{fig:spatialspec}~(right). Region to region, the statistical
spread around the region average is at most $3\,\sigma$. A $\chi^2$
test for a constant index results in a $p$ value of $0.02\%$, i.e.\
the index fluctuations around the region average are not compatible
with statistical fluctuations alone. To evaluate whether the indices
also vary significantly when taking the systematic uncertainties into
account, these uncertainties on the reconstructed photon index are
again determined by comparing all energy spectra, region by region,
between the primary analysis chain~\citep{modelAna} and the
cross-check chain~\citep{TMVA,ImPACT}. This approach is already
applied above for the systematic uncertainty of the full SNR energy
spectrum. For each region, both reconstructed energy spectra are fit
with a power law with the same fixed cut-off of 12.9~TeV. For all 29
regions, the resulting best-fit photon index difference distribution
has an RMS spread of 0.09. This exceeds the statistical spread
expected from the largely correlated data of the two independent
analysis chains, and we take this value as the systematic index
uncertainty. A small part of this spread may still be from statistical
uncertainties, but we conservatively take the full RMS as systematic
uncertainty.

Since this approach covers only the uncertainty related to the
calibration and analysis method, but not, for example, the atmospheric
transmission uncertainties, which are potentially as large as
0.05~\citep{2014APh....54...25H}, we conclude that
$\Delta \Gamma_\mathrm{sys} = 0.1$ is the systematic photon index
uncertainty for the \hess\ data set shown here.  With this, the
$\chi^2$ test reveals a $p$ value of 57\%, the photon index does
therefore not vary significantly across different regions of the SNR.

As confirmation of the systematic photon index uncertainty, the energy
spectrum of the whole SNR is determined in two different ways, which
demonstrates how the analysis and fitting method impacts the spectral
index reconstruction. For this, the energy spectrum of the primary
analysis chain (see Sect.~\ref{subsec:full-remnant-spectrum}),
determined by splitting the whole SNR into subregions, merging the
spectra and fitting a spectral model by forward folding it with the
instrument response functions, is compared to an alternative
approach. For this, as described above at the expense of exposure, we
use the entire SNR ON region without further splitting. We furthermore
fit the flux points that were unfolded from the instrument
response functions to derive the best-fit photon index. The resulting
two photon index measurements of the SNR are shown in
Fig.~\ref{fig:spatialspec}~(right) by the difference of
$\Delta \Gamma = 0.05$ of the grey solid and blue long-dashed lines,
which are well contained within the quoted systematic uncertainty. The
same is true for the average subregion index of 1.98 (purple short
dashed line in the right panel of Fig.~\ref{fig:spatialspec}), which is
compatible with the photon index of the entire SNR fit, 2.06, at the
$< 1\,\sigma_{\mathrm{sys}}$ level.

To conclude, there is no evidence for a spatial variation of the
photon index, either from region to region or from any region to the
entire SNR energy spectrum, within our statistical and systematic
uncertainties.

\section{Broadband modelling}
\label{sec:model}
The \hess\ results presented above provide a new opportunity to study
the origin of the VHE emission from \rxj. Even though \rxj\ is one of
the best-studied young SNRs with non-thermal emission, which clearly
indicates the acceleration of VHE particles by the shell of the
remnant, the hadronic or leptonic nature of the VHE gamma-ray emission
remains a source of disagreement in the literature
\citep{2006A&A...451..981B,2009MNRAS.392..240M,2010ApJ...708..965Z,2010ApJ...712..287E,2012ApJ...751...65F,2014MNRAS.445L..70G},
as discussed in detail for example in
\citet{2016EPJWC.12104001G}. Below, we add new information for this
discussion. We use \emph{Suzaku} X-ray, \fermi\ gamma-ray and \hess\
VHE gamma-ray data to model the broadband energy spectrum of \rxj\ and
derive the present-age parent particle spectra from the spectral
energy distribution (SED) of the SNR in both emission scenarios.  We
also model the broadband energy spectra, which are spatially resolved
into the 29 boxes described above, using the \emph{Suzaku} and \hess\
data only.

\begin{table*}
  \begin{center}
    {\def\arraystretch{1.3}
      \begin{tabular}{clccc} 
        \hline\hline\noalign{\smallskip} 
        Model & Parameter & Full & West & East \\ 
        \noalign{\smallskip}\hline\noalign{\smallskip} 
        \multirow{6}{*}{\rotatebox[origin=r]{90}{\parbox[b]{1.8cm}{\centering Hadronic}}} & $\Gamma\mathrm{_p^1}$& $1.53 \pm 0.09$ & $1.34^{+0.18}_{-0.3}$ & $1.2 \pm 0.3$  \\ 
              & $E\mathrm{_p^{break}}$ ( TeV )& $1.4^{+0.7}_{-0.4}$ & $0.78 \pm 0.19$ & $0.44 \pm 0.12$  \\ 
              & $\Gamma\mathrm{_p^2}$& $1.94 \pm 0.05$ & $2.26 \pm 0.03$ & $2.20 \pm 0.05$  \\ 
              & $E\mathrm{_p^{cutoff}}$ ( TeV )& $93 \pm 15$ & $280 \pm 70$ & $350^{+200}_{-100}$  \\ 
              & $W\mathrm{_p}$ ( $10^{49} (n_\mathrm{H} / 1\,\mathrm{cm}^{-3})^{-1}$ erg )& $5.81 \pm 0.12$ & $3.93 \pm 0.08$ & $3.15 \pm 0.09$  \\ 
              & AIC (broken PL) & 87.0 & 101.5 & 79.3 \\ 
              & AIC (simple PL) & 106.2 & 153.1 & 96.3 \\ 
              & Relative likelihood (simple vs. broken) & $7\times10^{-5}$ & 0 & $2\times10^{-4}$ \\
        \noalign{\smallskip}\hline\noalign{\smallskip} 
        \multirow{7}{*}{\rotatebox[origin=r]{90}{\parbox[b]{1.8cm}{\centering Leptonic\\ {\small (X and $\gamma$-ray)}}}} & $\Gamma\mathrm{_e^1}$& $1.78 \pm 0.12$ & $1.83 \pm 0.05$ & $1.92^{+0.06}_{-0.11}$  \\ 
              & $E\mathrm{_e^{break}}$ ( TeV )& $2.5 \pm 0.3$ & $2.15 \pm 0.11$ & $1.79^{+0.12}_{-0.2}$  \\ 
              & $\Gamma\mathrm{_e^2}$& $2.93 \pm 0.02$ & $3.10 \pm 0.02$ & $2.97 \pm 0.03$  \\ 
              & $E\mathrm{_e^{cutoff}}$ ( TeV )& $88.4 \pm 1.2$ & $84.8 \pm 1.5$ & $120 \pm 3$  \\ 
              & $W\mathrm{_e}$ ( $10^{46}$ erg )& $11.9 \pm 0.5$ & $9.0 \pm 0.4$ & $7.0 \pm 0.4$  \\ 
              & $B$ ( $\mu$G )& $14.26 \pm 0.16$ & $16.7 \pm 0.2$ & $12.0 \pm 0.2$  \\ 
              & AIC (broken PL) & 116.2 & 157.0 & 112.8 \\ 
              & AIC (simple PL) & 345.5 & 250.2 & 127.1 \\ 
              & Relative likelihood (simple vs. broken) & 0 & 0 & $8\times10^{-4}$ \\
        \noalign{\smallskip}\hline\noalign{\smallskip} 
        \multirow{6}{*}{\rotatebox[origin=r]{90}{\parbox[b]{1.8cm}{\centering Leptonic\\ {\small (only $\gamma$-ray)}}}} & $\Gamma\mathrm{_e^1}$& $1.6 \pm 0.2$ & $1.9 \pm 0.3$ & $2.1 \pm 0.3$  \\ 
              & $E\mathrm{_e^{break}}$ ( TeV )& $2.0 \pm 0.3$ & $1.7 \pm 0.3$ & $1.6 \pm 0.3$  \\ 
              & $\Gamma\mathrm{_e^2}$& $2.82 \pm 0.04$ & $3.12 \pm 0.03$ & $3.05 \pm 0.04$  \\ 
              & $E\mathrm{_e^{cutoff}}$ ( TeV )& $65 \pm 7$ & $140 \pm 30$ & $210^{+13000}_{-80}$  \\ 
              & $W\mathrm{_e}$ ( $10^{46}$ erg )& $11.8 \pm 0.5$ & $10.1 \pm 0.6$ & $7.8 \pm 0.5$  \\ 
              & AIC (broken PL) & 80.8 & 104.5 & 83.5 \\ 
              & AIC (simple PL) & 142.9 & 177.3 & 98.8 \\ 
              & Relative likelihood (simple vs. broken) & 0 & 0 & $5\times10^{-4}$ \\
        \noalign{\smallskip}\hline\noalign{\smallskip} 
      \end{tabular}
    }
  \end{center}
  \caption{Results from the hadronic and leptonic model fits. For the
    leptonic model, the fit was performed to the 
    full available broadband data from X-rays to gamma rays and
    restricted to the GeV to TeV gamma rays alone. The parameters are
    given for the full remnant, the western half, and the eastern half.
    To compare the fits of the simple and broken power-law models with
    exponential cut-off, the Akaike information
    criterion~\citep[AIC,][]{akaike1974new} is also given. The relative
    likelihood 
    (see main text), set to 0 for values below $1\times 10^{-5}$,
    illustrates clearly that in all cases the broken 
    power law is statistically favoured over the simple power law. The total energy in
    protons is given for a target density of
    $n_\mathrm{H}=1\,\mathrm{cm}^{-3}$, and both the proton and 
    electron total energy are given above 1~TeV (for graphical
    versions see Figs.~\ref{fig:sed:full} and
    \ref{fig:sed:halves}). The 1\,$\sigma$ statistical errors are
    asymmetric, unless the difference between the up- and downward
    errors is less than 25\%; in that case they are given as symmetric
    errors.
  }
  \label{tab:sed}
\end{table*}

\subsection{\emph{Suzaku} data analysis}
\label{subsec:suzaku}
\citet{2008ApJ...685..988T} have presented a detailed analysis of the
\emph{Suzaku} data that cover most of the SNR \rxj.  We performed
additional \emph{Suzaku} observations in 2010, adding four pointings with
a total exposure time of 124\,ks after the standard screening
procedure, to complete the coverage of \rxj\ with \emph{Suzaku}. After
combining the \emph{Suzaku} XIS data set from
\citet{2008ApJ...685..988T} with these additional pointings, we
perform a spatially resolved spectral analysis for the 29 regions
shown in Fig.~\ref{fig:spatialspec} to combine the synchrotron X-ray
spectrum in each region with the corresponding TeV gamma-ray spectrum
in the broadband analysis below.

\subsection{GeV data analysis}
\label{subsec:fermi}
For the \fermi\ analysis, 5.2 years of data (4 August 2008 to 25
November 2013) are used. The standard event selection, employing the
reprocessed P7REP\_SOURCE\_V15 source class, is applied to the data,
using events with zenith angles less than 100$^\circ$ and a rocking
angle of less than 52$^\circ$ to minimise the contamination from the
emission from the Earth atmosphere in the \fermi\ field of view
\citep{FermiEarthLimb}. Standard analysis tools, available from the
\fermi\ Science support centre, are used for the event selection and
analysis. In particular, a 15$^\circ$ radius for the region of
interest and a binned analysis with 0.1$^\circ$ pixel size is
used. The P7REP diffuse
model~\citep{2016ApJS..223...26A}\footnote{gll\_iem\_v05.fit and
  iso\_source\_v05.txt from
  \url{http://fermi.gsfc.nasa.gov/ssc/data/access/lat/BackgroundModels.html}}
and background source list, consistent with the official third source
catalogue~\citep[3FGL,][]{2015ApJS..218...23A} with the addition of
Source C from \citet{FermiRXJ}, are employed in the analysis. The
binned maximum-likelihood mode of \texttt{gtlike}, which is part of
the
ScienceTools\footnote{\url{http://fermi.gsfc.nasa.gov/ssc/data/analysis/software/}},
is used to determine the intensities and spectral parameters presented
in this paper. Further details of the analysis are provided in the
description in~\citet{FermiRXJ}.

For the spectral analysis, we adopt the spatial extension model based
on the \hess\ excess map. In the first step of the spectral analysis,
we perform a maximum-likelihood fit of the spectrum of \rxj\ in the
energy range between 200\,MeV and 300\,GeV using a power-law spectral
model with integral flux and spectral index as free parameters.  To
accurately account for correlations between close-by sources, we also
allow the integral fluxes and spectral indices of the nearby
background sources at less than $3^\circ$ from the centre of the
region of interest to vary in the likelihood maximisation. In
addition, the spectral parameters of identified \fermi\ pulsars,
the normalisation and the index of an energy-dependent multiplicative
correction factor of the Galactic diffuse emission model, and the
normalisation of the isotropic diffuse model are left free in the
fit. This accounts for localised variations in the spectrum of the
diffuse emission in the fit which are not considered in the global
model. For the Galactic diffuse emission, we find a normalisation
factor of $1.007 \pm 0.001$ in our region of interest.  The
normalisation factor for the remaining isotropic emission component
(extra-galactic emission plus residual backgrounds not fully
suppressed by the analysis requirements) is $1.17 \pm 0.02$.  These
factors demonstrate the reasonable consistency of the local brightness
and spectrum of the diffuse gamma-ray emission with the global diffuse
emission model. The LAT spectrum for the emission from \rxj\ is well
described by a power law with $\Gamma = 1.58 \pm 0.06$ and an
integrated flux above 1\,GeV of $I = (4.1 \pm 0.4) \times 10^{-9}$
cm$^{-2}$ s$^{-1}$. When fitting with different Galactic diffuse
models based on alternative interstellar emission
models~\citep{2016ApJS..224....8A}, the systematic uncertainty on the
photon index is approximately 0.1 and on the normalisation above
1\,GeV approximately $0.7 \times 10^{-9}$ cm$^{-2}$ s$^{-1}$
\citep[consistent with the previous study of this object,
see][]{FermiRXJ}.

To test for spatial differences in the emission, we split the \hess\
template in an eastern and a western half: the spectrum in the west
yields $\Gamma = 1.74 \pm 0.11$, $I= (3.1 \pm 0.8) \times 10^{-9}$
cm$^{-2}$ s$^{-1}$, in the east $\Gamma = 1.41 \pm 0.12$,
$I= (1.8 \pm 0.5) \times 10^{-9}$ cm$^{-2}$ s$^{-1}$. These results
show a shape agreement at the 2\,$\sigma$ level when taking the statistical
uncertainties into account. In addition there is a systematic error of
$\Delta\Gamma_\mathrm{syst}^{\mathrm{Fermi}-LAT}=0.1$ from the choice
of the diffuse model as tested for the overall remnant.

In a second step we perform a maximum-likelihood fit of the flux of
\rxj\ in 13 independent logarithmically spaced energy bands from 280
MeV to 400 GeV (using the spectral model and parameters for the
background sources obtained in the previous fit) to obtain an SED for
the SNR. We require a test statistic value of $TS \geq 4$ in each band
to draw a data point corresponding to a $2\,\sigma$ detection
significance (see Fig.~\ref{fig:sed:full} below).

\subsection{Derivation of the present-age parent particle
  distribution} 
\label{sec:pdist-derivation}
To understand the acceleration and emission processes that produce the
observed data, the present-age particle distribution can be derived by
fitting the observed energy spectra. We present below the results of
the derivation of the parent particle population for both a hadronic
and leptonic model for the whole remnant, for the two halves of the
remnant, and for the spatially resolved spectra extracted from
the 29 regions described above in
Sect.~\ref{subsec:Spatially-Resolved-Spectrum}. For the full and half
remnant spectra, the gamma-ray spectra from \fermi\ and \hess\ are
used in the fitting, whereas only the \hess\ spectra are used for the
29 spatially resolved regions because the \fermi\ data lacks
sufficient statistics when split into these regions. For leptonic
models, X-ray spectra from the \emph{Suzaku} observations are also
fitted simultaneously with the gamma-ray data.  We used the
radiative code and Markov Chain Monte Carlo (MCMC) fitting routines of
\texttt{naima}\footnote{\url{http://www.github.com/zblz/naima}}~\citep{2015arXiv150903319Z}
to derive the present-age particle distribution, using the
parametrisation of neutral pion decay by \citet{2014PhRvD..90l3014K},
and of IC up-scattering of diluted blackbody radiation by
\citet{2014ApJ...783..100K}. For the leptonic models, the magnetic
  field and exponential energy cut-off are treated as independent
  parameters. The magnetic field is constrained by the ratio of
  synchrotron to IC flux magnitude, whereas the cut-off in the parent
  electron spectrum is constrained by the cut-off in the VHE part of
  the IC gamma-ray spectrum. In the MCMC fit, the seed photon fields
considered for the IC emission are the cosmic microwave background
radiation and a far-infrared component with temperature $T = 26.5$\,K
and a density of 0.415\,eV\,cm$^{-3}$. These latter values are derived
from GALPROP by \citet{2011ApJ...727...38S} for a distance of
1\,kpc. For hadronic gamma-ray emission, the solar CR composition
is used to take the effects of heavier nuclei into account.

We note that the assumption that the emission in each of the
extraction regions arises from a single particle population under
homogeneous ambient conditions might not hold for several of the
regions selected. In particular, whereas the IC emissivity samples the
electron density, keeping in mind that the target photon field density
for IC emission does not vary on small scales, the synchrotron
emissivity samples the product of electron density and magnetic field
energy density. Considering the inhomogeneity of magnetic field in
supernova remnants, and in \rxj\ in particular
\citep{2007Natur.449..576U}, this means that the X-ray emission 
typically samples smaller volumes than IC emission.

\subsubsection{Full remnant}
\label{sec:full-remnant-pdist}

\begin{figure*}
  \centering
  \includegraphics[width=\textwidth]{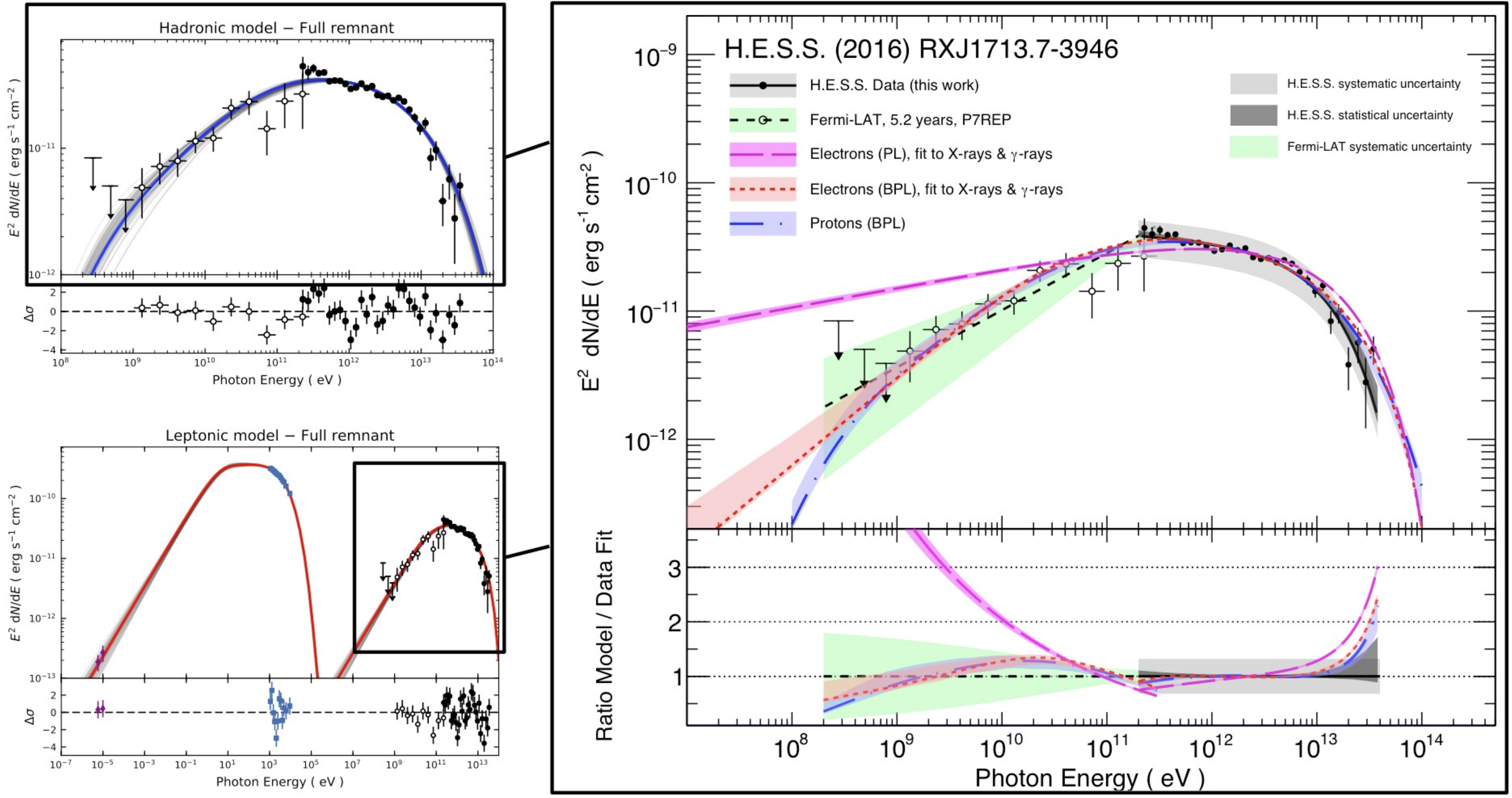}
  \caption{Comparison of hadronic and leptonic models to the
    data. \textbf{Top left:} the hadronic gamma-ray model obtained
    with our broadband fit is compared to data. \textbf{Bottom left:}
    the same plot of the leptonic gamma-ray model compared to data
    including lower-energy X-rays and radio data. The thick blue and
    red lines indicate the maximum-likelihood models, and the grey
    lines surrounding them are the models for 100 samples of the MCMC
    chain and serve to illustrate the fit uncertainties. The energy
    flux data points shown from high to low energy are the \hess\ and
    \fermi\ gamma-ray data as solid and open circles, the
    \emph{Suzaku} X-ray data and ATCA radio data~\citep{Lazendic}. The
    latter flux was determined for the northwest part of the SNR shell
    only and was scaled up by a factor of two here to represent the
    whole SNR. Owing to this ad hoc scaling, these points are not
    included in the fit, but are shown for illustration
    only. \textbf{Right-hand side:} both leptonic and hadronic models
    are compared to the \fermi\ and \hess\ data points including
    statistical and systematic uncertainties. In addition to the
    preferred best-fit models of a broken power law with a cut-off
    (BPL), a power law without cut-off is also shown for the leptonic
    model to demonstrate that this model cannot describe the \fermi\
    gamma-ray data.}
  \label{fig:sed:full}
\end{figure*}

\begin{figure*}
  \centering
  \begin{subfigure}[c]{0.48\textwidth}
    \centering
    \includegraphics[width=\textwidth]{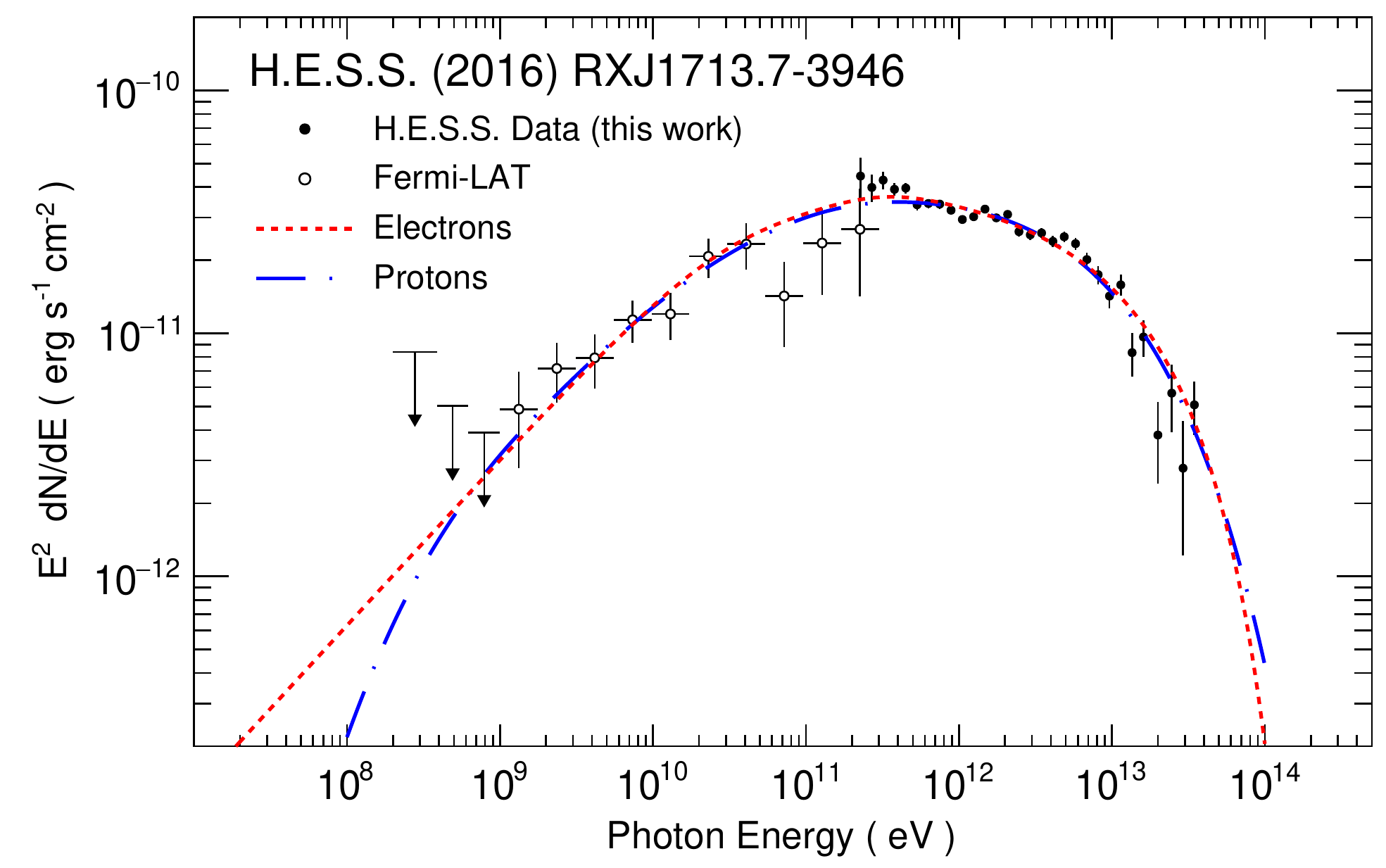}
  \end{subfigure}
  \begin{subfigure}[c]{0.48\textwidth}
    \centering
    \includegraphics[width=\textwidth]{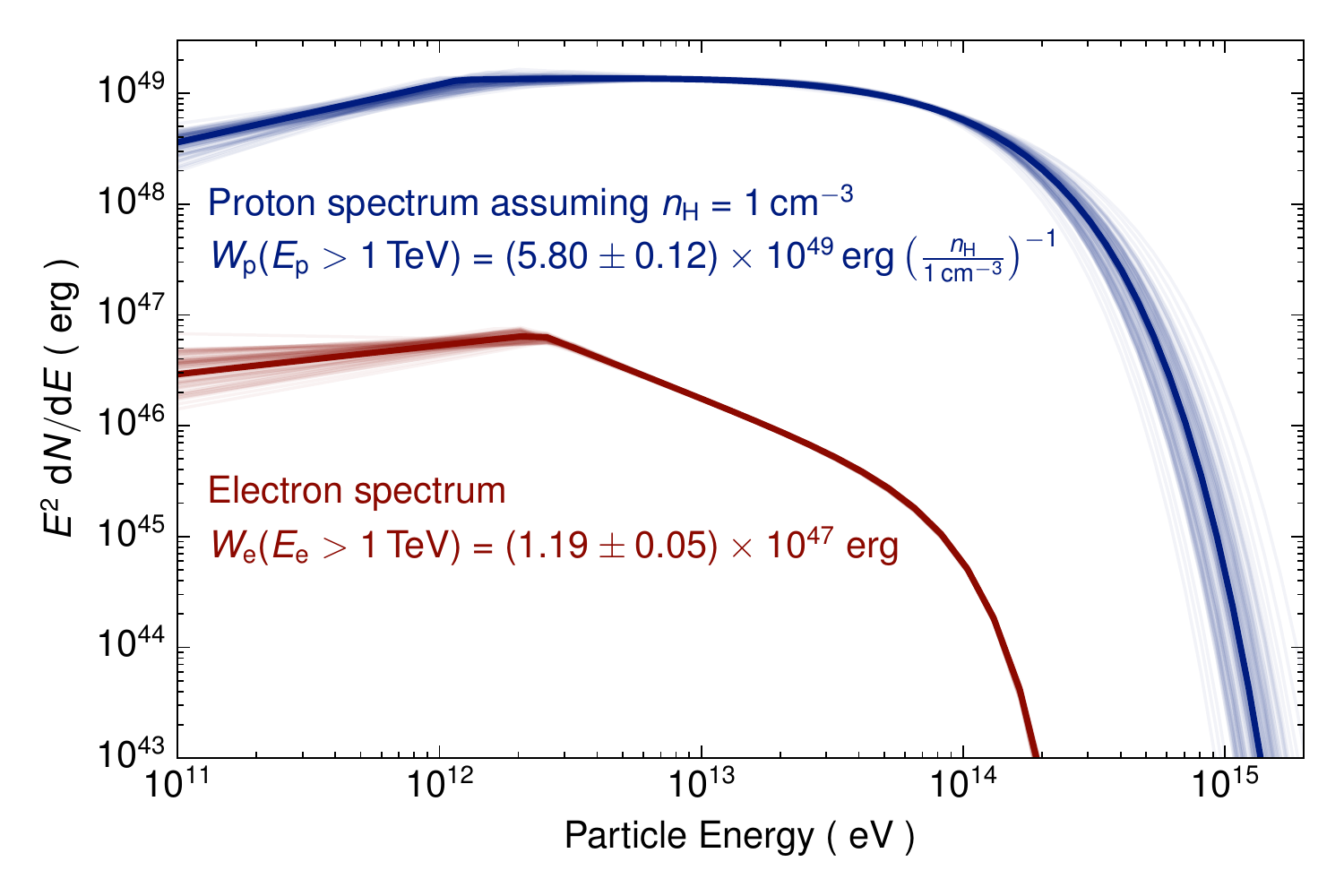}
  \end{subfigure}
  \caption{Gamma-ray model curves and parent particle energy
    spectra. On the left, the best-fit electron and proton gamma-ray
    models (broken power laws with exponential cut-offs) are compared
    to the \fermi\ and \hess\ data. The data points and model curves
    are the same as in Fig.~\ref{fig:sed:full}. On the right, the
    corresponding best-fit parent particle energy spectra are
    shown. The electron model is derived from a combined fit to both
    the X-ray and gamma-ray data.}
  \label{fig:parent:spectra}
\end{figure*}

The full-remnant SED at gamma-ray energies shown in
Fig.~\ref{fig:sed:full} exhibits a hard spectrum in the GeV regime, a
flattening between $\sim$100 GeV and a few TeV, and an exponential
cut-off above $\sim$10 TeV. In a hadronic scenario, this spectrum is
best fit with a particle distribution consisting of a broken power law
and an exponential cut-off at high energies. The results of the fit can
be seen in Table~\ref{tab:sed} and Fig.~\ref{fig:sed:full}~(top left)
and Fig.~\ref{fig:parent:spectra}, where the resulting best-fit
gamma-ray models (left) and parent particle energy spectra (right) are
shown. In Table~\ref{tab:sed}, the difference between the two particle
indices of the broken power law is significant, indicating a break in
the proton energy distribution at an energy of $1.4\pm0.5$\,TeV. For a
target density of $n_\mathrm{H} / 1\,\mathrm{cm}^{-3}$, a total energy
in protons of
$W_p = (5.80\pm0.12) \times 10^{49} (n_\mathrm{H} /
1\,\mathrm{cm}^{-3})^{-1}$\,erg
above 1 TeV is required to explain the measured gamma-ray flux.

As in the hadronic scenario, the observed gamma-ray spectrum cannot be
explained with an electron population described by a single power
law. This is clearly seen on the right-hand side of
Fig.~\ref{fig:sed:full}, where the best-fit power-law electron model
is shown to be incompatible with the gamma-ray data even when taking
all uncertainties into account. Fitting a broken power-law electron
distribution to the X-ray and gamma-ray emission from the full remnant
results in a break at $E_b = 2.4\pm0.3$\,TeV and a difference between
the particle indices of $\Delta\Gamma_e = 1.16\pm0.14$ (see
Table~\ref{tab:sed}). The magnetic field strength required to reproduce
the X-ray and gamma-ray spectra is $B=14.2\pm0.2\,\mu$G.

To illustrate the need for a low-energy break in the particle energy
spectrum, the Akaike information
criterion~\citep[AIC;][]{akaike1974new} is also given in
Table~\ref{tab:sed} as measure for the relative quality of both
spectral models, the simple power law with exponential cut-off and the
broken power law with exponential cut-off. A lower AIC value
corresponds to the more likely model, the relative likelihood also
given in the table is defined as
$\exp{((AIC_\mathrm{min}-AIC_\mathrm{max})/2)}$. In all cases, the
broken power law is clearly preferred over the simple power law. We
also tested fitting a broken power law with a smooth instead of a hard
transition,
\begin{equation*}
E^2\,\times\,\frac{dN}{dE}\, = E^2\,\times\,F_o \, E^{-\Gamma^1}\, \left(1 +
  \left(\frac{E}{E^{\mathrm{break}}}\right)^{-\frac{\Gamma^1-\Gamma^2}{\beta}}
\right)^{-\beta} \, \mathrm{,} 
\end{equation*}
plus a high-energy exponential cut-off, but find that the addition of
one more parameter to our results is not justified. The hard
transition, $\beta \rightarrow 0$, is mildly favoured at the
$1-2\sigma$ level over a smoother transition,
$\beta_\mathrm{fit} \approx 0.3$, for the SED of the entire SNR in
both the hadronic and leptonic models. The data cannot thus 
discriminate between these two versions of a broken power law. We
therefore use the simpler version with a hard break, which has one
parameter less.

To test the impact of the X-ray data and see which fit parameters are
affected more by these than the gamma-ray data, we have also performed
the broadband leptonic fits only to the gamma-ray data (losing any
handle on the magnetic field). The resulting parameters are shown in
Table~\ref{tab:sed}. Also in this case, a broken power law instead of a
single power law is needed to fit the gamma-ray data, the resulting particle
indices and break energy are compatible with the full broadband
fit. The exponential cut-off of the parent particle spectrum, on the
other hand, is significantly lower: $65\pm7$\,TeV compared to
$88.4\pm1.2$\,TeV when including the X-ray data.

From the particle spectra shown in Fig.~\ref{fig:parent:spectra}, one
can see that electrons via IC emission are much more efficient in
producing VHE gamma rays than protons via $\pi^0$
decay~\citep{2016EPJWC.12104001G}. A proton spectrum about 100 times
higher is needed to produce nearly identical
gamma-ray curves as shown in Fig.~\ref{fig:sed:full}.

\subsubsection{Half remnant}
Splitting the remnant ad hoc into the dim eastern and bright western
halves, we can test for spatial differences in the broadband parent
particle spectra within the remnant region while including the \fermi\
data. Using similar models to those described above, we find that for
a hadronic origin of the gamma-ray emission a broken power law is
statistically required to explain the GeV and TeV spectra for both
halves of the remnant. The corresponding plots are shown in the
appendix (Fig.~\ref{fig:sed:halves}). As can be seen in
Table~\ref{tab:sed}, the particle indices for the power laws from the
remnant halves are compatible with the high-energy particle index of
the full-remnant broken power-law spectrum, confirming that, like for
the gamma-ray spectra, there is no spectral variation seen in the
derived proton spectra either.

Assuming a leptonic scenario, the western half of the remnant shows a
slightly stronger magnetic field strength with
$B_\mathrm{W}=16.7\pm0.2\,\mu$G, compared to a strength of
$B_\mathrm{E}=12.0\pm0.2\,\mu$G in the eastern
half~(Table~\ref{tab:sed}). In addition, the electron high-energy
cut-off measured is significantly lower in the western half,
$E\mathrm{_{c,W}^e}=88.4\pm1.2$\,TeV, compared to
$E\mathrm{_{c,E}^e}=120\pm3$\,TeV in the eastern half. The inverse
dependency between the magnetic field strength and cut-off energy is
consistent with electron acceleration limited by synchrotron losses at
the highest energies.  Given that the X-ray emission is produced by
electrons of higher energies than the TeV emission, the energy of the
exponential cut-off is constrained strongly by the X-ray spectrum. To
demonstrate the impact of this, we also fit the electron spectrum only
to the gamma-ray data, see Table~\ref{tab:sed}. From this fit the
cut-off energy increases and has much larger uncertainties. This can
be explained by synchrotron losses constrained by the X-ray data. If
some small regions have a magnetic field strength that is
significantly higher than the average field strength, these regions
can dominate the X-ray data and cause differences in the cut-off
energies.

\subsubsection{Spatially resolved particle distribution}
The deep \hess\ observations allow us to fit the broadband X-ray and
VHE gamma-ray spectra from the 29 smaller subregions defined in
Sect.~\ref{subsec:Spatially-Resolved-Spectrum} to probe the particle
distribution and environment properties by averaging over much smaller
physical regions of 1.4~pc (for a distance to the SNR of
1~kpc). However, in VHE gamma rays the resolvable scale is still much
larger than some of the features observed in
X-rays~\citep{2007Natur.449..576U}. It is therefore unlikely that the
regions probed here encompass a completely homogeneous environment,
and information is lost due to the averaging. In addition, the
projection of the near and far section of the remnant, and in fact the
interior, along the line of sight into the same two-dimensional region
adds an uncertainty when assessing the physical origin of the observed
spectrum. This degeneracy is only broken for the rim of the remnant
where the projection effects are minimal, and we know that the
observed spectrum is emitted close to the shock.  As before, we
consider both the leptonic and hadronic scenarios for the origin of
VHE gamma-ray emission.

\begin{figure*}
  \centering
  \begin{subfigure}[c]{0.48\textwidth}
    \centering
    \includegraphics[width=\textwidth]{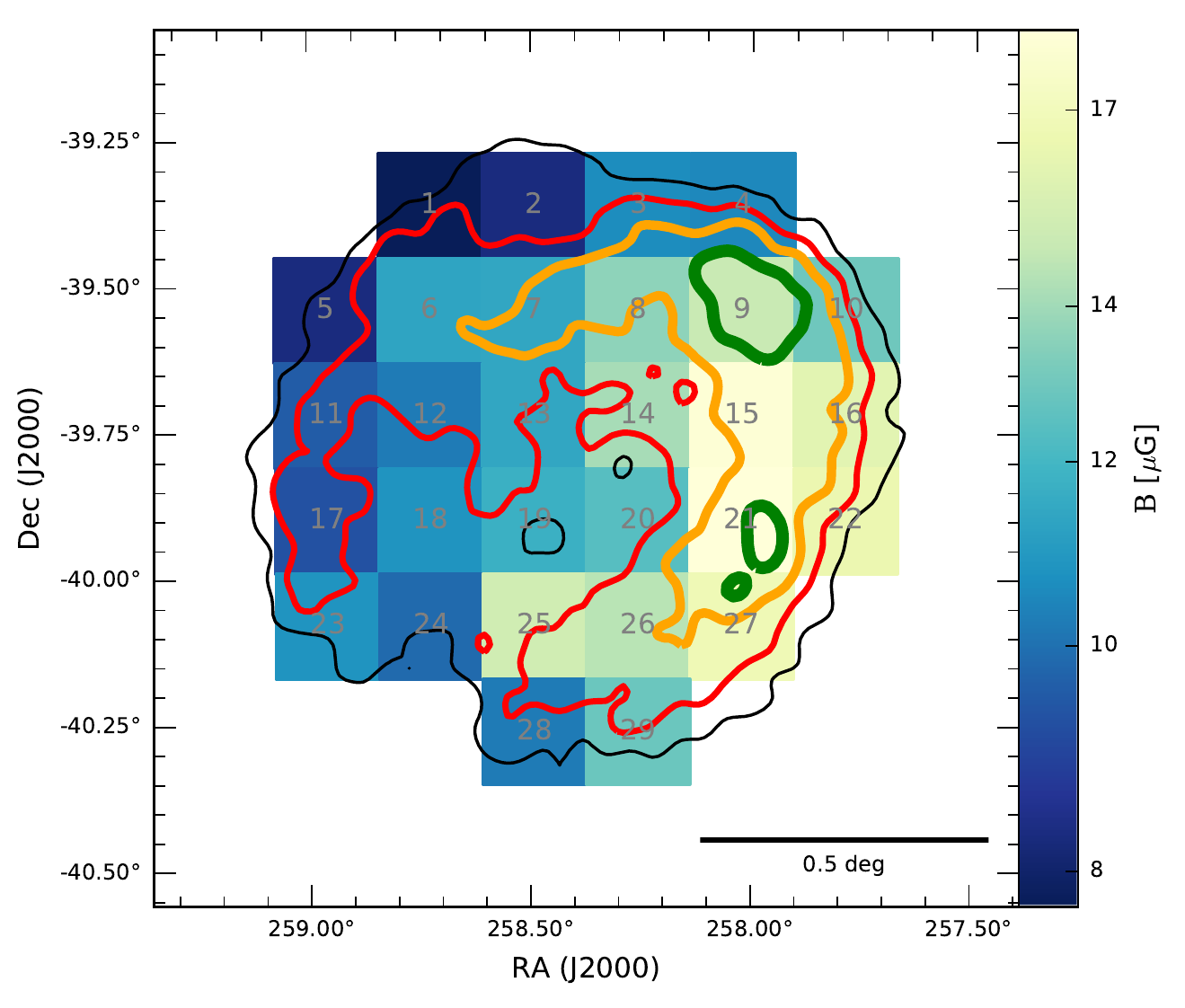}
  \end{subfigure}
  \begin{subfigure}[c]{0.48\textwidth}
    \centering
    \includegraphics[width=\textwidth]{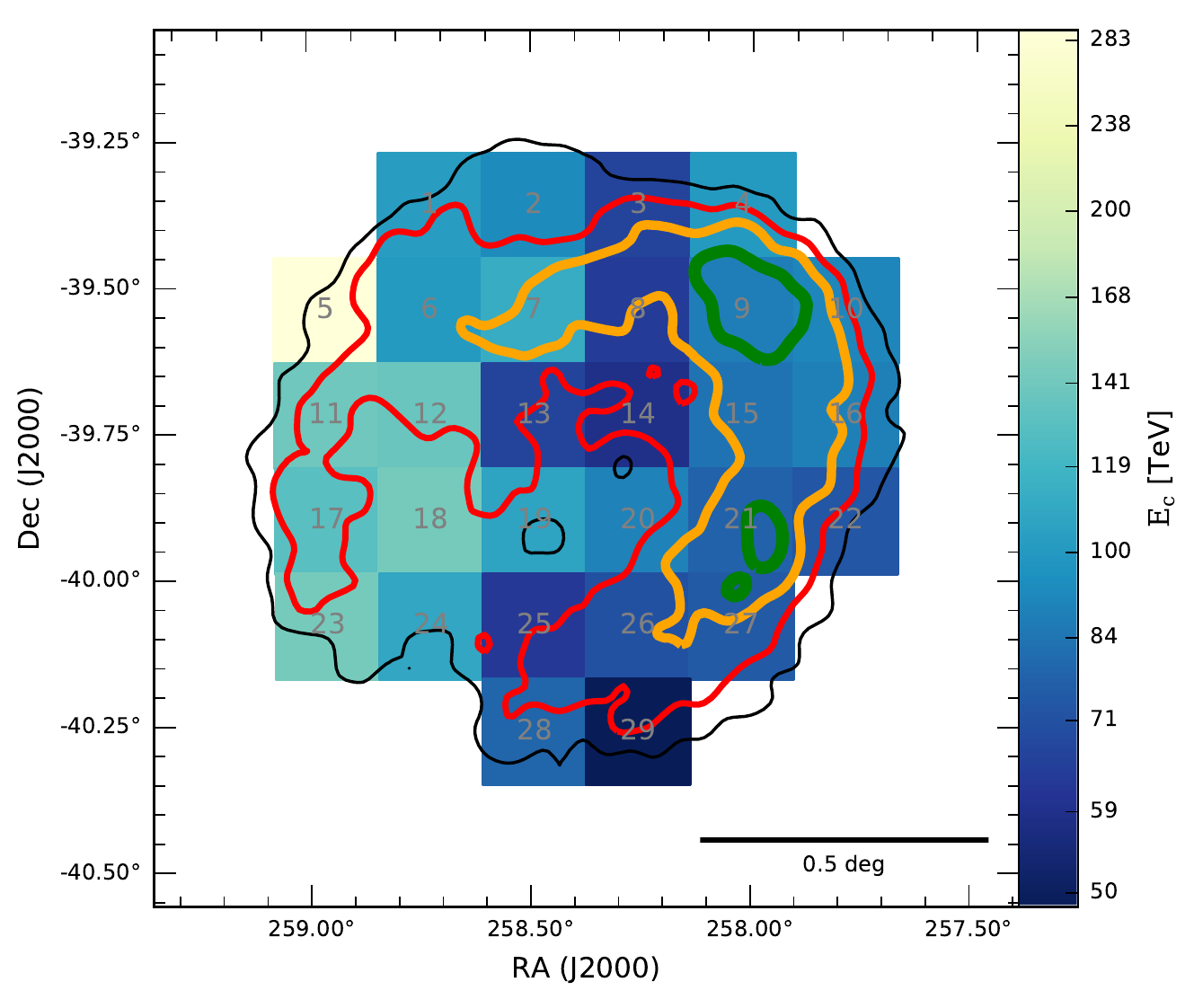}
  \end{subfigure}
  \begin{subfigure}[c]{0.48\textwidth}
    \centering
    \includegraphics[width=\textwidth]{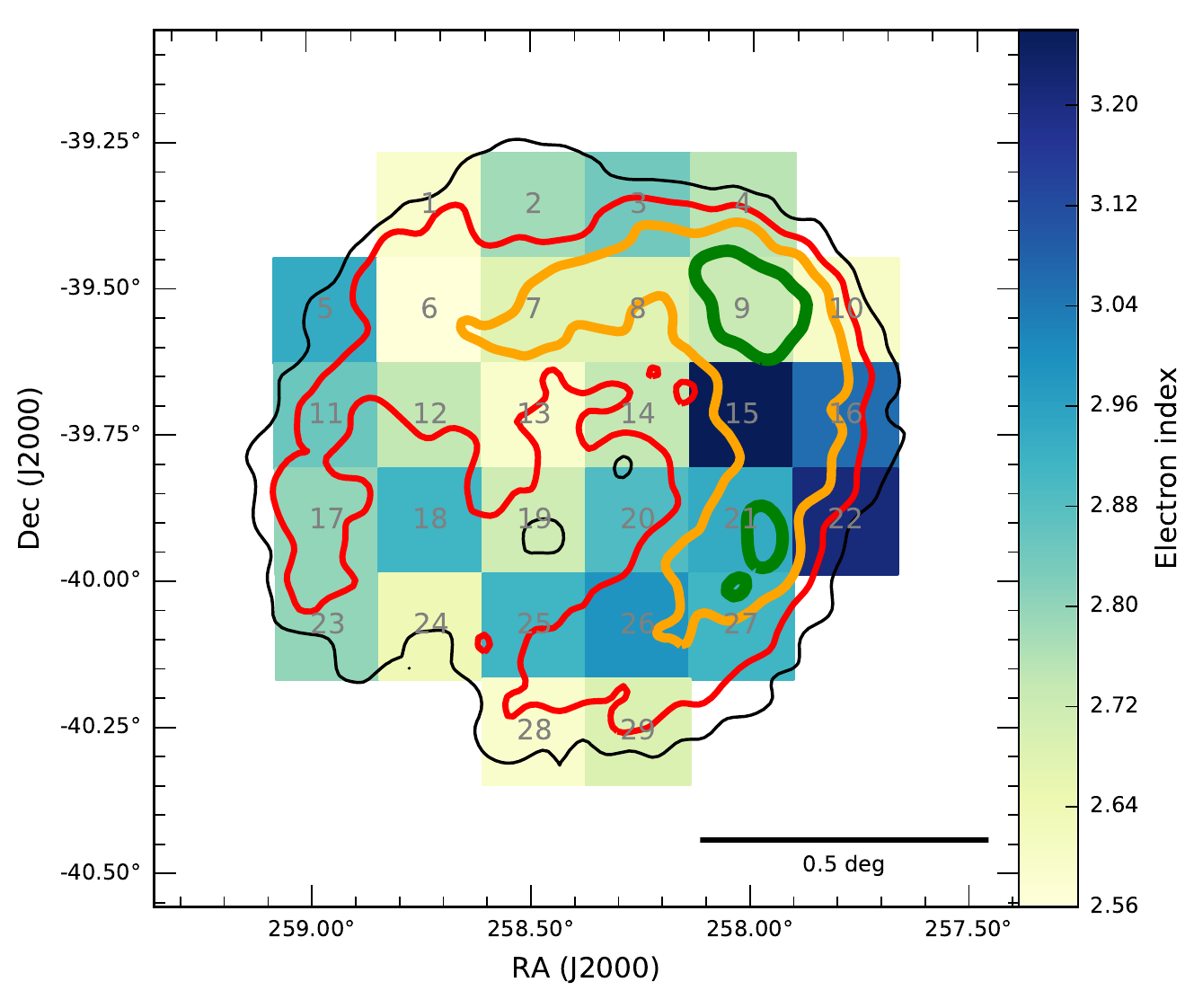}
  \end{subfigure}
  \begin{subfigure}[c]{0.48\textwidth}
    \centering
    \includegraphics[width=\textwidth]{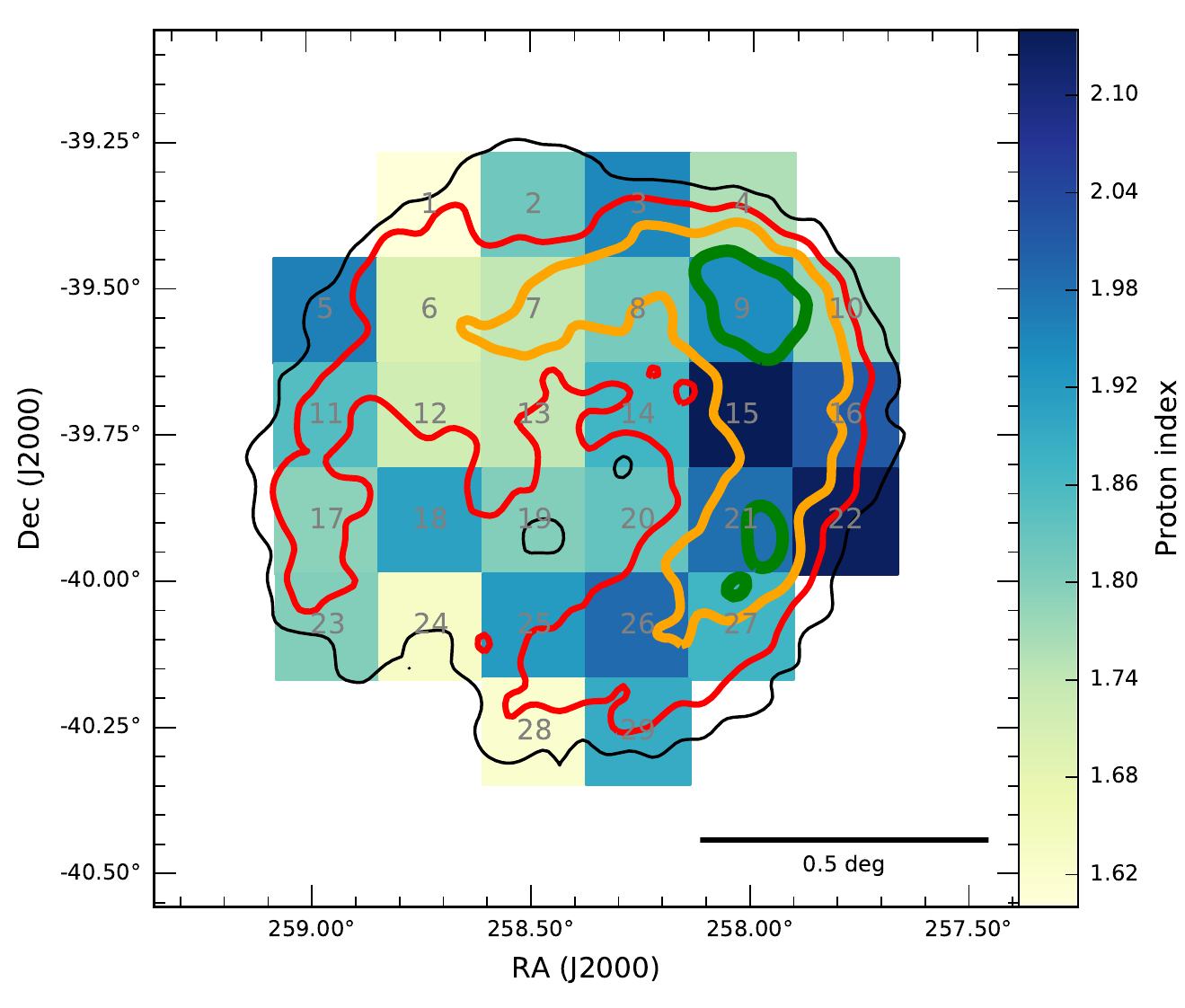}
  \end{subfigure}
  \caption{Spatial distribution of physical best-fit parameters across
    the SNR, overlaid on the \hess\ gamma-ray significance contours at
    3, 5, 7, and $9\,\sigma$ in black, red, orange, and green. For the
    leptonic model, colour codes are shown for the magnetic field
    strength (top left), exponential cut-off energies (top right), and
    particle indices (bottom left). For the hadronic models, only the
    particle indices (bottom right) are relevant and shown here. The
    29 subregions labelled with grey numbers are boxes of side lengths
    $0.18^\circ$ or 10.8 arcminutes. To judge whether the differences
    region to region are significant, the statistical uncertainties
    listed in
    Table~\ref{tab:elec-regions}~and~\ref{tab:proton-regions} have to
    be taken into account, and ultimately the \hess\ systematic
    measurement uncertainties discussed above as well. When doing
    this, the spectral indices show no variation across the SNR in
    either scenario.}
  \label{fig:regions:fitresults}
\end{figure*}

\begin{table}
  \begin{center}
    \begin{tabular}{rllll} 
      \hline\hline\noalign{\smallskip} 
      Reg. & Part.~Index& $E_\mathrm{c}$& $W_\mathrm{e}$~($>$1\,TeV)& B\\
           & & (TeV)& ($10^{45}$ erg)& ($\mu$G)\\
      \noalign{\smallskip}\hline\noalign{\smallskip} 
      1 & $2.59 \pm 0.18$ & $102 \pm 15$ & $1.7^{+0.5}_{-0.4}$ & $7.7 \pm 0.8$ \\ 
      2 & $2.78 \pm 0.14$ & $93 \pm 9$ & $3.4 \pm 0.7$ & $8.3 \pm 0.7$ \\ 
      3 & $2.84 \pm 0.13$ & $66^{+6}_{-5}$ & $3.7 \pm 0.7$ & $10.6 \pm 0.7$ \\ 
      4 & $2.75 \pm 0.14$ & $100 \pm 10$ & $2.8 \pm 0.6$ & $10.5^{+0.9}_{-0.7}$ \\ 
      5 & $2.94 \pm 0.11$ & $290^{+180}_{-70}$ & $3.3 \pm 0.6$ & $8.3 \pm 0.6$ \\ 
      6 & $2.56 \pm 0.11$ & $100 \pm 7$ & $4.1 \pm 0.6$ & $11.3 \pm 0.6$ \\ 
      7 & $2.68 \pm 0.09$ & $112 \pm 8$ & $4.9 \pm 0.7$ & $11.4 \pm 0.6$ \\ 
      8 & $2.68 \pm 0.08$ & $64 \pm 3$ & $6.5 \pm 0.7$ & $13.6 \pm 0.5$ \\ 
      9 & $2.73 \pm 0.06$ & $87 \pm 3$ & $7.8 \pm 0.8$ & $14.9 \pm 0.5$ \\ 
      10 & $2.60 \pm 0.09$ & $90 \pm 4$ & $4.2^{+0.7}_{-0.5}$ & $12.9^{+0.7}_{-0.5}$ \\ 
      11 & $2.85 \pm 0.09$ & $140^{+17}_{-12}$ & $4.6 \pm 0.6$ & $9.5 \pm 0.5$ \\ 
      12 & $2.74 \pm 0.10$ & $137^{+18}_{-14}$ & $4.1 \pm 0.6$ & $10.2 \pm 0.6$ \\ 
      13 & $2.59 \pm 0.09$ & $66 \pm 3$ & $4.0 \pm 0.5$ & $11.4 \pm 0.5$ \\ 
      14 & $2.74 \pm 0.09$ & $58 \pm 3$ & $4.9^{+0.7}_{-0.5}$ & $14.1 \pm 0.6$ \\ 
      15 & $3.26 \pm 0.10$ & $83^{+9}_{-7}$ & $12.4 \pm 1.8$ & $18.2 \pm 1.0$ \\ 
      16 & $3.06 \pm 0.10$ & $88 \pm 6$ & $8.4 \pm 1.2$ & $16.0 \pm 1.0$ \\ 
      17 & $2.80 \pm 0.08$ & $129 \pm 13$ & $4.6 \pm 0.6$ & $9.3^{+0.5}_{-0.4}$ \\ 
      18 & $2.91 \pm 0.13$ & $144^{+30}_{-20}$ & $3.9 \pm 0.8$ & $10.8 \pm 0.9$ \\ 
      19 & $2.72 \pm 0.10$ & $106 \pm 8$ & $3.7 \pm 0.5$ & $11.7 \pm 0.7$ \\ 
      20 & $2.89 \pm 0.14$ & $89 \pm 8$ & $4.8 \pm 0.9$ & $12.4 \pm 1.0$ \\ 
      21 & $2.94^{+0.06}_{-0.08}$ & $77 \pm 4$ & $8.6 \pm 0.9$ & $18.4 \pm 0.6$ \\ 
      22 & $3.21 \pm 0.15$ & $73^{+7}_{-5}$ & $8.5 \pm 1.7$ & $16.4 \pm 1.5$ \\ 
      23 & $2.80 \pm 0.09$ & $144 \pm 16$ & $4.0 \pm 0.5$ & $10.8 \pm 0.6$ \\ 
      24 & $2.64 \pm 0.13$ & $108 \pm 11$ & $2.4 \pm 0.5$ & $9.8 \pm 0.7$ \\ 
      25 & $2.91 \pm 0.10$ & $62^{+4}_{-3}$ & $4.8 \pm 0.6$ & $15.2 \pm 0.8$ \\ 
      26 & $2.99 \pm 0.13$ & $71 \pm 4$ & $8.6^{+1.7}_{-1.3}$ & $14.5 \pm 1.0$ \\ 
      27 & $2.91 \pm 0.09$ & $75 \pm 4$ & $8.6 \pm 1.1$ & $16.7 \pm 0.8$ \\ 
      28 & $2.6 \pm 0.2$ & $78 \pm 9$ & $2.8^{+1.1}_{-0.8}$ & $10.2^{+1.7}_{-1.1}$ \\ 
      29 & $2.69 \pm 0.10$ & $49 \pm 3$ & $4.5 \pm 0.6$ & $12.8 \pm 0.7$ \\ 
      \noalign{\smallskip}\hline\noalign{\smallskip}
    \end{tabular} 
  \end{center}
  \caption{Electron distribution and magnetic field parameters derived from
    the VHE and X-ray spectra of the 29 regions under the
    assumption of an IC origin of the gamma-ray emission.
    The total energy in electrons $W_e$ is 
    computed for electron energies above 1\,TeV.} 
  \label{tab:elec-regions}
\end{table}

In the leptonic scenario, the \emph{Suzaku} X-ray spectra are used
together with the \hess\ gamma-ray data in the fits. This allows us to
derive the magnetic field per subregion in addition to the parameters
of the electron energy distribution. Given that the \fermi\ GeV
spectra cannot be obtained in such small regions, only electrons above
$\sim$5\,TeV are probed by the VHE gamma-ray and X-ray spectra, and we
can only infer the properties of the high-energy part of the particle
spectra, i.e.\ the power-law slope and its cut-off. No information
about the break energy or the low-energy power law can be extracted in
the subregions. In the leptonic scenario, the VHE gamma-ray emission
probes the electron spatial distribution, whereas the X-ray emission
probes the electron distribution times $B^2$, causing regions with
enhanced magnetic field to be over-represented in the X-ray spectrum.

We find that in all regions the emission from an electron distribution
with a power law and an exponential cut-off reproduces the spectral
shape in both X-ray and VHE gamma-ray
energies. Table~\ref{tab:elec-regions} and
Fig.~\ref{fig:regions:fitresults} show the results of these fits. The
electron particle index for all the regions is in the range 2.56 to
3.26 and is compatible with the average full-remnant particle index of
2.93. Such steep particle indices, which are significantly larger than the
canonical acceleration index of about 2, indicate that the accelerated
electron population at these energies ($E_e\gtrsim5$\,TeV) has
undergone modifications, i.e.\ cooling through synchrotron
losses. However, neither the age of the remnant of O(1000 years) nor
the derived average magnetic field are high enough for the electrons
to have cooled down to such energies. Explaining this spectral shape
is thus a challenge for the leptonic scenario, which is discussed
further in
Sect.~\ref{subsec:interpretation:1}. Figure~\ref{fig:regions:fitresults}
(bottom left) shows that the spatial distribution of the electron
index is not entirely uniform, even when taking the statistical
uncertainties given in Table~\ref{tab:elec-regions} into account the
indices in the brighter western part of the shell tend to be
larger. Such a trend is also seen in the distribution of the
high-energy exponential cut-off energy (in the range 50--200\,TeV) and
the average magnetic field strength (in the range 8--20\,$\mu$G) shown
in the same figure. The western half of the remnant shows higher
values of the magnetic field strength and lower values of the cut-off
with the opposite behaviour seen in the eastern half (see top left and
right of Fig.~\ref{fig:regions:fitresults}). In a
synchrotron-loss-limited acceleration scenario, the maximum energy
achievable at a given shock is proportional to $B^{-1/2}$, so that the
anti-correlation between cut-off energy and magnetic field strength is
to be expected.
\begin{table}
  \begin{center}
    \begin{tabular}{rll} 
      \hline\hline\noalign{\smallskip} 
      Reg. & Part.~Index& $W_\mathrm{p}$~($>$1\,TeV)\\
           & & ($10^{48}$ erg\,cm$^{-3}$)\\
      \noalign{\smallskip}\hline\noalign{\smallskip} 
      1 & $1.60^{+0.14}_{-0.19}$ & $0.91 \pm 0.09$ \\ 
      2 & $1.82 \pm 0.11$ & $1.51 \pm 0.11$ \\ 
      3 & $1.95 \pm 0.10$ & $1.52 \pm 0.12$ \\ 
      4 & $1.76 \pm 0.11$ & $1.24 \pm 0.10$ \\ 
      5 & $1.96 \pm 0.11$ & $1.27 \pm 0.10$ \\ 
      6 & $1.70 \pm 0.09$ & $2.37 \pm 0.17$ \\ 
      7 & $1.74 \pm 0.08$ & $2.48 \pm 0.15$ \\ 
      8 & $1.81 \pm 0.05$ & $3.07 \pm 0.13$ \\ 
      9 & $1.94 \pm 0.06$ & $3.9 \pm 0.2$ \\ 
      10 & $1.78 \pm 0.10$ & $2.29 \pm 0.15$ \\ 
      11 & $1.85 \pm 0.08$ & $1.91 \pm 0.11$ \\ 
      12 & $1.72 \pm 0.09$ & $1.91 \pm 0.13$ \\ 
      13 & $1.74 \pm 0.07$ & $2.07 \pm 0.11$ \\ 
      14 & $1.87 \pm 0.06$ & $2.19 \pm 0.11$ \\ 
      15 & $2.14 \pm 0.07$ & $3.3 \pm 0.2$ \\ 
      16 & $2.01 \pm 0.08$ & $2.71 \pm 0.19$ \\ 
      17 & $1.79 \pm 0.07$ & $1.96 \pm 0.11$ \\ 
      18 & $1.91 \pm 0.10$ & $1.56 \pm 0.14$ \\ 
      19 & $1.80 \pm 0.09$ & $1.72 \pm 0.10$ \\ 
      20 & $1.83 \pm 0.11$ & $1.79 \pm 0.15$ \\ 
      21 & $1.98 \pm 0.05$ & $3.19 \pm 0.15$ \\ 
      22 & $2.13 \pm 0.10$ & $2.4 \pm 0.2$ \\ 
      23 & $1.80 \pm 0.08$ & $1.74 \pm 0.10$ \\ 
      24 & $1.63 \pm 0.14$ & $1.18 \pm 0.09$ \\ 
      25 & $1.92 \pm 0.07$ & $1.72 \pm 0.11$ \\ 
      26 & $1.99 \pm 0.08$ & $3.0 \pm 0.2$ \\ 
      27 & $1.87 \pm 0.07$ & $3.21 \pm 0.19$ \\ 
      28 & $1.62 \pm 0.16$ & $1.44 \pm 0.19$ \\ 
      29 & $1.89 \pm 0.07$ & $2.09 \pm 0.12$ \\ 
      \noalign{\smallskip}\hline\noalign{\smallskip}
    \end{tabular} 
  \end{center}
  \caption{Proton distribution parameters derived from the VHE
    gamma-ray spectra of the 29 subregions, assuming a neutral pion
    decay origin of the gamma-ray emission. The total energy in protons is
    computed for proton energies above 1~TeV. The cut-off energy is
    fixed to 93\,TeV, which is the value obtained for the full-remnant
    spectrum, to overcome fit convergence problems due to limited
    statistics in dim regions of the SNR.}
  \label{tab:proton-regions}
\end{table}

In a hadronic scenario we only consider radiation from primary protons
without considering secondary X-ray emission from charged pions
produced in interactions of protons with ambient
matter~\citep{2013SAAS...40....1A}. Using only the \hess\ spectra, we
find that the proton cut-off energy is not constrained for many of the
regions. We therefore fix the cut-off energy when fitting the
subregions spectrum to the value found for the full SNR spectrum:
$E_\mathrm{c}=93$\,TeV. Under this assumption, all the regions are
well fit by a neutral pion decay spectrum with the parameters shown in
Table~\ref{tab:proton-regions}. The proton particle indices for all
the regions cover a range between 1.60 and 2.14 as shown in
Fig.~\ref{fig:regions:fitresults}~(bottom right) and listed in
Table~\ref{tab:proton-regions}. As already found above for the
gamma-ray spectral
fits~(Sect.~\ref{subsec:Spatially-Resolved-Spectrum}), the maximum
difference between the particle spectral indices of different regions
is at the level of 3\,$\sigma$ -- the indices do not vary across the
SNR when taking the \hess\ spectral index systematic uncertainty in
addition to the statistical uncertainties given in
Table~\ref{tab:proton-regions} into account. In the hadronic scenario
proposed by~\citet{2014MNRAS.445L..70G} for \rxj, these particle index
values are from energy spectra of protons that have completely
diffused within dense small clumps of interstellar matter and are
therefore altered by diffusion effects resulting in a hardening of the
spectral shapes compared to the standard index of 2 expected from DSA.

\section{Interpretation}
\label{sec:interpretation}
\subsection{Leptonic versus hadronic origin}
\label{subsec:interpretation:1}
 The observational data presented in this work provide the deepest
data set available to date to evaluate the physical origin of the VHE
gamma-ray emission from the shell-type SNR \rxj\ to find out whether
the emission has a leptonic (IC upscatter of external radiation fields
by relativistic electrons) or hadronic (pion decay emission from
interaction of relativistic protons with ambient gas) origin. The
discussion of the origin is intimately tied to the question of whether
the present-age electron or proton distribution required by the
observed gamma-ray spectra can be achieved considering the physical
properties of the supernova remnant. Therefore, the derivation of the
relativistic parent particle energy distributions presented in
Sect.~\ref{sec:pdist-derivation} is a crucial tool to evaluate
theoretical scenarios for the VHE radiation of \rxj.

A notable feature in both the electron and proton energy distributions
inferred in Sect.~\ref{sec:pdist-derivation} is the spectral break at
a few TeV. \citet{2014MNRAS.445L..70G}, following
\citet{2010ApJ...708..965Z} and \citet{2012ApJ...744...71I}, put
forward an explanation for such a break in the hadronic scenario as
the result of diffusion of protons in high-density cold clumps within
the SNR: higher energy protons diffuse faster and interact with the
highest density regions within the clumps, whereas lower energy
protons only probe the outer, lower density regions and therefore have
a lower emissivity.  This scenario would arise from a massive star
exploding in a molecular cloud that itself has been swept away by a
wind of the progenitor star, resulting in a rarefied cavity with dense
clumps. On the passage of the SNR shock, the high density ratio
between the clumps and cavity effectively stalls the shock at the
surface of the clumps avoiding their
disruption~\citep{2012ApJ...744...71I}. The break energy depends on
the age of the SNR and the density profile of the
clouds~\citep{2014MNRAS.445L..70G}. Considering a density within the
clumps of $10^3\,\mathrm{cm}^{-3}$, the break energy is of the order
of 1--5\,TeV for the age of \rxj, which is consistent with the
best-fit value of $E\mathrm{_p^{break}}=1.4\pm0.5$\,TeV as shown in
Sect.~\ref{sec:full-remnant-pdist}. This scenario would explain the
very low level of thermal X-ray emission recently
reported~\citep{2015ApJ...814...29K}, and it is also consistent with
the average density of $130\,\mathrm{cm}^{-3}$ integrated over the SNR
and the cavity walls reported in \citet{2012ApJ...746...82F}.  In this
cold clump scenario, the proton energetics
given in Table~\ref{tab:sed} for $n = 1\,\mathrm{cm}^{-3}$ would only
correspond to the protons interacting within these clumps of
$n = 10^3\,\mathrm{cm}^{-3}$. To obtain the total energy in protons, a
filling factor correction based on the combined clumps and SNR shell
volume ratio ($V_\mathrm{clumps} / V_\mathrm{shell}$) is needed.

In the leptonic scenario, a break in the electron spectrum could arise due to
synchrotron losses: electrons at higher energies suffer faster synchrotron
cooling and therefore a break is introduced at the energy for which the
synchrotron loss timescale and SNR age are equal,
\begin{equation}
  \label{eq:cooling}
    E_b \simeq 1.25 \left(\frac{B}{100\,\mu\mathrm{G}}\right)^{-2}
    \left( \frac{t_0}{10^3\,\mathrm{yr}} \right)^{-1}\ \mathrm{TeV.}
\end{equation}
Considering an SNR age of the order of 1000\,yr, a magnetic field of
$\sim$70\,$\mu$G would be required for a cooling break at 2.5\,TeV,
the best-fit value found in Sect.~\ref{sec:full-remnant-pdist}. While
there are indeed indications for such high magnetic fields from X-ray
variability measurements of small
filaments~\citep{2007Natur.449..576U}, or the X-ray width of the
filaments itself~\citep[see for example][and references
therein]{2012A&ARv..20...49V}, the residence time of electrons in
these small regions is much shorter than the age of the SNR and thus
too short for significant synchrotron cooling. The best probe we have
of the relevant magnetic field strength that may explain such a
cooling break, averaged over a volume large enough so that the
electron residence time is sufficiently long, is the simultaneous
fitting of X-ray and gamma-ray data of the whole SNR presented
here. The results in Sect.~\ref{sec:full-remnant-pdist} indicate a
present-age average magnetic field strength of $B=14.3\pm0.2\,\mu$G
for the SNR, which is much less than the 70\,$\mu$G required to
explain the energy break according to Eq.\,\ref{eq:cooling}. This
remains a challenge in the leptonic scenario. In particular,
experimental systematic uncertainties are also unable to explain this
energy break, as clearly shown in Fig.~\ref{fig:sed:full}~(right).

Two alternative explanations of the flattening between 10\,GeV and a
few TeV of the SED of \rxj\ are discussed in the literature. Firstly,
a second population of VHE electrons is suggested for example by
\citet{2012ApJ...751...65F}. With different electron populations the
relevant physical parameters may be tuned in a way that would exactly
reproduce the flat spectral shape of \rxj. Alternatively, a single
power-law electron population in the presence of an additional optical
seed photon field, as discussed in \citet{2008ApJ...685..988T}, could
produce the broad measured shape. We argue that this explanation is
unlikely since such a photon field would require an unrealistically
large energy density of $\sim$140\,eV\,cm$^{-3}$, which is more than
two orders of magnitude above the standard estimates, for example
implemented in GALPROP~\citep{2006ApJ...648L..29P} at the position of
the remnant. Beyond such simplified models, approaches taking the
temporal and spatial SNR evolution into account have also been shown
to reproduce the GeV to TeV gamma-ray data in a leptonic
scenario~\citep[see for example][]{2012ApJ...744...39E}.

\subsection{Particles beyond the shock}
\label{subsec:escape}
The X-ray synchrotron emission from \rxj\ is expected to be mostly
confined to the region within the shock front. Very high-energy
electrons must also be present beyond the shock, but the magnetic
field in the unshocked medium is a factor
$\mathcal{R}_\mathrm{B} \approx 3$~\citep[see for
example][]{2006A&A...453..387P} smaller than in the shocked
medium. $\mathcal{R}_\mathrm{B}$ is the magnetic field compression
ratio. It depends on the magnetic field orientation and is generally
comparable but smaller than the shock compression ratio $\mathcal{R}$,
which is $\mathcal{R} = 4$ for a strong shock. Since the synchrotron
emissivity scales with $B^\Gamma$ \citep{ginzburg65}, where
$\Gamma \approx 2$~\citep{AceroXMM2009} is typically the synchrotron
X-ray photon index here, the X-ray synchrotron emissivity is expected
to drop by a factor $\mathcal{R}_\mathrm{B}^{\Gamma} = 9$ at the
shock.  The boundary of the X-ray emission therefore traces the shock
front of \rxj. The evidence presented in Sect.~\ref{subsec:morph2} for
gamma-ray emission outside the X-ray boundary requires the presence of
accelerated particles beyond the shock front.

Since the electrons outside the shock experience the same radiation
energy density as those within the shock boundary, the emissivity does
not sharply drop at the shock in the case that the leptonic emission
scenario (IC scattering) applies to the VHE gamma-ray emission of
\rxj. For standard hadronic emission scenarios, the emissivity should
change at the shock boundary as the density of the ambient medium
increases across the shock and drops again beyond it. But this
standard hadronic scenario is not relevant for \rxj\ since it fails to
explain the hard gamma-ray emission in the GeV band. The alternative
scenario mentioned in Sect.~\ref{subsec:interpretation:1} requires the
presence of dense clumps, which are to first order not affected by the
passage of the shock. The ambient density within, at, and beyond the
shock boundary in this case is therefore constant. The gamma-ray
emissivity in both the leptonic and the considered hadronic scenario
is therefore constant across and beyond the shock. The gamma-ray
emission measured from the unshocked medium beyond the shock front
then solely traces the density of accelerated particles, be it
electrons or protons.

The existence of such accelerated particles in the unshocked medium
producing gamma-ray emission beyond the SNR shell is a long-standing
prediction and might be interpreted either as the detection of the
so-called CR precursor ahead of the shock, characteristic of
DSA~\citep{2005ApJ...624L..37M,2010ApJ...708..965Z}, or as the result
of the escape of particles from the
SNR~\citep{1996A&A...309..917A,2009MNRAS.396.1629G,2010PASJ...62.1127C,2013ApJ...768...73M}.

In fact, even though the mechanism of particle escape from shocks is
far from being understood, it is clear that it has to be intimately
connected to the process of acceleration itself, i.e.\ to the way in
which particles are confined upstream of the
shock~\citep{2011MNRAS.415.1807D,2013MNRAS.431..415B}. While the shock
wave is decreasing in velocity the particles upstream of the shock
have a smaller probability of re-entering the SNR shell. There is thus
a gradual transition from acceleration to escape, which is expected to
be energy dependent: in general, the highest energy particles have a
larger mean free path length and detach earlier from the shock. The
escape and CR precursor scenarios are therefore not completely
distinct. With our new results from deep \hess\ observations we can
now probe these highly unknown aspects of shock acceleration for the
first time.
 
The extraction of the three-dimensional spatial distribution of the
charged particles ahead of the shock from the measured two-dimensional
gamma-ray data would require an accurate and realistic modelling of
the physical shock and its direct environment, which is clearly beyond
our scope here. We therefore restrict the discussion to some general
considerations. We also emphasise here that the extent of gamma-ray
emission from around \rxj\ varies considerably, which likely reflects
different particle acceleration conditions around the shell.

The observations reveal the presence of gamma rays from parsec scale
regions of size $\Delta r$ upstream of the shock.  If the VHE gamma
rays are from IC scattering of electrons, the spatial distribution of
the gamma-ray emission simply traces the distribution of electrons
(the target photon field density is not likely to vary on such small
scales). If the emission is due to neutral pion decay, its morphology
results from the convolution of the spatial distributions of CR
protons and interstellar medium gas.  In both cases, a rough estimate
of the maximal extension of the TeV emission outside of the SNR can be
obtained by computing the diffusion length of multi-TeV particles
ahead of the shock.

To do this, we use $\delta r_{1/e}$ listed in
Tab.~\ref{table:extension} as the typical extent of the particle
distribution upstream of the shock. In theoretical assessments the
diffusion length scale is usually taken to be the distance from the
shock at which the particle density has dropped to $1/e$. Even though
the physical diffusion length scale is in addition subject to
projection effects, for our purpose of an order of magnitude
assessment we assume that the difference in $\delta r_{1/e}$ between
X-rays and gamma rays is equivalent to the diffusion length
scale. From Tab.~\ref{table:extension}, this angular difference in
regions 2, 3, and 4 is
$\Delta r \equiv (\delta r_{1/e}^{\mathrm{gamma\, rays}} - \delta
r_{1/e}^{\mathrm{X-rays}})$,
and we therefore obtain a maximum of $\Delta r = 0.05^\circ$ for
region 3, which corresponds to $0.87\, d_1$~pc, where $d_1$ is the
distance to the SNR in units of 1~kpc. In the precursor scenario, the
diffusion length scale is given by
\begin{equation}
\label{eq:pre:1}
\ell_\mathrm{p} \approx \frac{D(E)}{u_\mathrm{shock}}.
\end{equation}
For the escape scenario the typical length scale over which the
particles diffuse is given by the diffusion length scale
\begin{equation}
\label{eq:esc:1}
\ell_\mathrm{e} \approx \sqrt{2\, D(E)\, \Delta t}\, .
\end{equation}
Here, $u_\mathrm{shock}$ is the shock velocity, and $\Delta t$ is the
escape time.  $D(E)$ is the energy dependent diffusion coefficient,
which we parameterise as
\begin{equation}
D(E) = \eta(E) \frac{1}{3} \frac{cE}{eB}.
\end{equation}
$\eta$ is the ratio between the mean free path of the particles and
their gyroradius. In general, $\eta$ is an energy dependent parameter
that expresses the deviation from Bohm diffusion, which itself is thus
defined as $\eta=1$.  Its value in regions associated with the SNR
should in any case be close to $\eta=1$ for particle energies of
10-100 TeV in order to explain the fact that \rxj\ is a source of
X-synchrotron emission~\citep[see][]{aharonian99}.

Assuming that the diffusion length scale in both cases is equal
to the measured parameter $\Delta r$ we arrive at 
\begin{equation}
\label{eq:pre:2}
\frac{B}{\eta} \approx 0.36\, \left( \frac{E}{10\,\mathrm{TeV}}
\right) \left( \frac{u_\mathrm{shock}}{3000\,\mathrm{km\,s}^{-1}}
\right)^{-1}\,\left( \frac{\Delta r}{\mathrm{pc}} \right)^{-1} \mu
\mathrm{G}
\end{equation}
for the precursor scenario.  For the escape scenario we should take
into account that the shock itself will also have displaced during a
time $\Delta t$. So we have
$\Delta r= \ell_\mathrm{e}-u_\mathrm{shock}\Delta t$.  However, for
escape it holds that $\ell_\mathrm{e}> u_\mathrm{shock}\Delta t$,
since escape implies that diffusion is more important than advection,
and even more so since during the time $\Delta t$ the shock slows down
and hence $u_\mathrm{shock}$ decreases. Dropping terms with
$u_\mathrm{shock}^2\Delta t^2/\Delta r^2$ we find that
\begin{align}
\label{eq:esc:2}
  \frac{B}{\eta} \approx 1.1 \,\left( \frac{E}{10\,\mathrm{TeV}} \right)
  \left( \frac{\Delta t}{500\,\mathrm{yr}} \right) \left( \frac{\Delta
  r}{\mathrm{pc}} \right)^{-2} \left[1 + \frac{u_\mathrm{shock}\Delta t}{\Delta r}\right]^{-1}\mu \mathrm{G},
\end{align}
with $B$ the magnetic field upstream of the shock and $\eta$ again the
mean free path of the particles in units of the gyroradius. The factor in
square brackets is $\lesssim1.5$.  For the shock velocity of \rxj, an
upper limit of 4500~km\,s$^{-1}$ has been derived from \emph{Chandra}
data~\citep{2007Natur.449..576U} and from \emph{Suzaku} data the
velocity is estimated to be
$3300\eta^{1/2}$~km\,s$^{-1}$~\citep{2008ApJ...685..988T}. For
particles in the shock or shock precursor region, \rxj\ therefore
operates at or close to the Bohm regime since the synchrotron X-ray
data require $\eta=1-1.8$ for shock velocities of
$3300$--$4500$~km\,s$^{-1}$. Taking this into account, for $\eta=2$,
we obtain for region 3: $B=0.8\, \mu\mathrm{G}$ in the precursor
scenario. In the escape scenario where the particles have left the
shock region, $\eta$ is not constrained by the X-ray emission any more
and in particular it can be larger ($\eta > 1$). We therefore derive
in more general terms $B\lesssim \eta\,2.8\, \mu\mathrm{G}$ in the
escape scenario. In the standard DSA paradigm, and in the absence of
further magnetic field amplification through
turbulences~\citep[discussed for example in][]{2007ApJ...663L..41G},
the expected magnetic field compression at the shock would result in
downstream magnetic fields a factor of $\mathcal{R}_\mathrm{B}=3-4$
higher than those upstream, that is, up to $B=3.2\, \mu\mathrm{G}$ and
$B=\eta\,11.2\, \mu\mathrm{G}$ for region 3 in the precursor and
escape scenario, respectively.

Whilst the escape scenario is compatible with our broadband leptonic
fits, in the precursor scenario the downstream magnetic field value is
lower than the values obtained with these fits (see
Fig.~\ref{fig:regions:fitresults} and Tab.~\ref{tab:elec-regions}).
In particular, $B=3.2\, \mu\mathrm{G}$ downstream is somewhat lower
than expected in the DSA paradigm, unless we invoke a recent sudden
increase of $\eta$ to values well above 2 or a decrease of
$u_\mathrm{shock}$ to well below $3300$~km\,s$^{-1}$ to recover higher
downstream magnetic field values. Such sudden changes must occur on
timescales smaller than the synchrotron radiation loss time of the
downstream electrons, since $\eta \lesssim 5$ is needed to explain
X-ray synchrotron radiation from the shell in these regions
\citep{2008ApJ...685..988T}.  We therefore require that the timescale
for substantial changes in the upstream diffusion properties,
$\Delta t$, must satisfy
\begin{equation}
\tau_\mathrm{loss} = \frac{634}{B^2E}\ \mathrm{s} > \Delta t,
\end{equation}
with $\tau_\mathrm{loss} =|E/(dE/dt)|$~\citep{ginzburg65}. The typical
X-ray synchrotron photon energy is given by
$\epsilon = 7.4 E^2 B$\,keV~\citep{ginzburg65}, so that the condition
for the presence of X-ray emission from the shell at 1 keV for a given
timescale $\Delta t$ is
\begin{equation}
B\lesssim 23 \left(\frac{\Delta t}{500\,\mathrm{yr}}\right)^{-2/3}~\mu \mathrm{G}.
\end{equation}
This condition is fully consistent with the leptonic emission
scenario, but requires for the hadronic emission scenario timescales
shorter than $\Delta t=500$~yr.

To summarise, the significant extension of the gamma-ray emission
beyond the X-ray defined shock in some regions of \rxj\ requires
either low magnetic fields or diffusion length scales much larger than
for Bohm diffusion, irrespective of whether the gamma rays are from
particles originating in the shock precursor or escaping the remnant
diffusively. In both cases, the length scales are in fact governed by
diffusion. 

The relative length scale of the gamma-ray emission measured beyond
the shock is rather large, $\Delta r/r_\mathrm{SNR} \approx 13$\%, for
a precursor scenario. One can estimate the typical relative length
scale of a shock precursor by starting from Eq.\,3.39 of
\citet{0034-4885-46-8-002} for the particle acceleration time
$\tau_\mathrm{acc}$: 
\begin{equation}
\tau_\mathrm{acc} = \frac{3}{u_1-u_2}\, \left(\frac{D_1}{u_1} +
  \frac{D_2}{u_2}\right), 
\end{equation}
with the subscript 1 and 2 referring to the diffusion coefficients and
velocities of the upstream and downstream regions, respectively. We note
that $u_\mathrm{shock}=u_1$. With the compression ratio at the shock
$\mathcal{R}=u_1/u_2$, we obtain
\begin{equation}
  \tau_\mathrm{acc} =
  \frac{3}{u^2_1}\,\frac{\mathcal{R}}{\mathcal{R}-1}\,D_1\,\left(1+\frac{D_2}{D_1}\mathcal{R}\right).  
\end{equation}
Assuming Bohm diffusion for $D_1$ and $D_2$, their ratio is $D_2/D_1=1$
for a parallel shock and $D_2/D_1=1/\mathcal{R}$ for a perpendicular
shock. With this, and a compression ratio of $\mathcal{R}=4$, we get
\begin{equation}
  \tau_\mathrm{acc} = \kappa\,\frac{D_1}{u_1^2},
\end{equation}
with $\kappa=8$ for a perpendicular and $\kappa=20$ for a parallel
shock. The following relation connects the shock velocity of SNRs with
their radius over long stretches of
time~\citep{chevalier82,truelove99}:
\begin{align}
r \propto t_\mathrm{age}^m   \Rightarrow 
u_\mathrm{shock} = m \frac{r}{t_\mathrm{age}},
\end{align}
where $m=0.4$ for the Sedov-Taylor phase and $m=0.5-0.7$ for younger
remnants like \rxj. Since the age of the SNR $t_\mathrm{age}$
corresponds to the maximum possible acceleration time of particles,
and hence $\tau_\mathrm{acc}<t_\mathrm{age}$, the maximum precursor
length scale can now be calculated as
\begin{equation}
  \ell_\mathrm{p}= \frac{D_1(E)}{u_\mathrm{shock}}=
  \frac{\tau_\mathrm{acc} u_\mathrm{shock}}{\kappa} <
  \frac{t_\mathrm{age} u_\mathrm{shock}}{\kappa} = \frac{m}{\kappa}\,
  r = 0.0875\,r,
\end{equation}
with $m=0.7$ and $\kappa=8$ for a perpendicular shock. This estimate
of the maximum precursor size of about 10\% of the SNR radius is
conservatively large as most particles have not been accelerated from
the date of the explosion, but considerably later, and thus
$\tau_\mathrm{acc}<t_\mathrm{age}$. We therefore conclude that the
measured length scale of 13\% is of the order of the maximum possible
scale expected for a shock precursor. More precise measurements and
modelling of the precursor or diffusion region, including line of
sight effects, are needed to assess whether the extended emission we
measure is from the shock precursor or from particles escaping the
shock region.

This discussion only pertains to certain regions of \rxj; there are
other regions where the gamma-ray size does not exceed the X-ray
size. Keeping in mind that \rxj\ is argued to be a supernova remnant
evolving in a cavity~\citep{2010ApJ...708..965Z}, the shock wave could
be starting to interact with a positive density gradient associated
with the edges of the cavity in those regions where the gamma-ray
emission extends farther out. As a result of the density gradient, the
shock wave velocity and/or the magnetic field turbulence are
decreasing and the VHE particles start diffusing out farther ahead of
the shock, close to, or already beyond the escape limit.

The above analysis is somewhat simplified, and we are left with one
surprising observational fact: within the current uncertainties, the
gamma-ray emission beyond the shell is energy
independent~(Sect.~\ref{subsec:morph2}), whereas one would expect that
the diffusion length scale is larger for more energetic particles.
This is true for both the precursor and the escape scenario. The
energy dependence is therefore either too small to be measurable with
\hess; for instance, only for pure Bohm diffusion would one expect
that $D\propto E$. More generically, one expects $D = E^{\delta}$, so
perhaps $\delta<1$ in the regions with extended emission. Or else the
energy dependence of the diffusion coefficient could be suppressed as
recently argued in \citet{2013ApJ...768...73M}, where a model is
developed for older SNRs interacting with molecular clouds. Elements
of this model may also be relevant for the interaction of \rxj\ with
the cavity wall.  Given the potential evidence for escape and the
surprising lack of any energy dependence of the gamma-ray emission and
therefore the diffusion coefficient, \rxj\ will remain a key priority
target for the future Cherenkov Telescope Array (CTA)
observatory~\citep{2013APh....43....3A,2015arXiv150806052N}.

\section{Summary}
\label{sec:summary}
The new \hess\ measurement of \rxj\ reaches unprecedented precision
and sensitivity for this source. With an angular resolution of
$0.048^\circ$ (2.9 arcminutes) above gamma-ray energies of 250 GeV,
and $0.036^\circ$ (2.2 arcminutes) above energies of 2 TeV, the new
\hess\ map is the most precise image of any cosmic gamma-ray source at
these energies. The energy spectrum of the entire SNR confirms our
previous measurements at better statistical precision and is most
compatible with a power law with an exponential cut-off, both a linear
power-law model at gamma-ray energies of 12.9\,TeV and a quadratic
model at 16.5\,TeV.

A spatially resolved spectral analysis is performed in a regular grid
of 29 small rectangular boxes of $0.18^\circ$ (10.8 arcminutes) side
lengths, confirming our previous finding of the lack of spectral shape
variation across the SNR. 

The broadband emission spectra of \rxj\ from various regions are fit
with present age parent particle spectra in both a hadronic and
leptonic scenario, using \emph{Suzaku} X-ray and \hess\ gamma-ray
data. From the resolved spectra in the 29 small boxes in the leptonic
scenario, we derive magnetic field, energy cut-off, and particle index
maps of the SNR. For the latter parameter, we do the same for the
hadronic scenario. The leptonic and hadronic parent particle spectra
of the entire remnant are also derived without further detailed
assumptions about the acceleration process. These particle spectra
reveal that the \fermi\ and \hess\ gamma-ray data require a
two-component power-law with a break at 1-3\,TeV, challenging our
standard ideas about diffusive particle acceleration in shocks. In
either leptonic or hadronic scenarios, approaches more involved than
one or two zone models are needed to explain such a spectral
shape. Neither of the two scenarios (leptonic or hadronic), or a mix
of both, can currently be concluded to explain the data
unambiguously. Either better gamma-ray measurements with the future
CTA, with much improved angular resolution and much higher energy
coverage, or high sensitivity VHE neutrino measurements will
eventually settle this case for \rxj.
 
Comparing the gamma-ray to the \xmm\ X-ray image of \rxj, we find
significant differences between these two energy regimes. As
concluded before by~\citet{2008ApJ...685..988T}, the bright X-ray
hotspots in the western part of the shell appear relatively brighter
than the \hess\ gamma-ray data. The most exciting new finding of our
analysis is that in some regions of \rxj\ the SNR is larger in gamma
rays than it is in X-rays -- the gamma-ray shell emission extends
radially farther out than the X-ray shell emission in these
regions. We interpret this as VHE particles leaking out of the actual
shock acceleration region -- we either see the shock precursor or
particles escaping the shock region. Such signs of escaping particles
are a longstanding prediction of DSA, and we find the first such
observational evidence with our current measurement.

\section*{Acknowledgments}
The support of the Namibian authorities and of the University of
Namibia in facilitating the construction and operation of H.E.S.S. is
gratefully acknowledged, as is the support by the German Ministry for
Education and Research (BMBF), the Max Planck Society, the German
Research Foundation (DFG), the French Ministry for Research, the
CNRS-IN2P3 and the Astroparticle Interdisciplinary Programme of the
CNRS, the U.K. Science and Technology Facilities Council (STFC), the
IPNP of the Charles University, the Czech Science Foundation, the
Polish Ministry of Science and Higher Education, the South African
Department of Science and Technology and National Research Foundation,
the University of Namibia, the Innsbruck University, the Austrian
Science Fund (FWF), and the Austrian Federal Ministry for Science,
Research and Economy, and by the University of Adelaide and the
Australian Research Council. We appreciate the excellent work of the
technical support staff in Berlin, Durham, Hamburg, Heidelberg,
Palaiseau, Paris, Saclay, and in Namibia in the construction and
operation of the equipment. This work benefited from services provided
by the H.E.S.S. Virtual Organisation, supported by the national
resource providers of the EGI Federation.
 
The \fermi\ Collaboration acknowledges generous ongoing support from a
number of agencies and institutes that have supported both the
development and the operation of the LAT as well as scientific data
analysis.  These include the National Aeronautics and Space
Administration and the Department of Energy in the United States, the
Commissariat \`a l'Energie Atomique and the Centre National de la
Recherche Scientifique / Institut National de Physique Nucl\'eaire et
de Physique des Particules in France, the Agenzia Spaziale Italiana
and the Istituto Nazionale di Fisica Nucleare in Italy, the Ministry
of Education, Culture, Sports, Science and Technology (MEXT), High
Energy Accelerator Research Organization (KEK) and Japan Aerospace
Exploration Agency (JAXA) in Japan, and the K.~A.~Wallenberg
Foundation, the Swedish Research Council, and the Swedish National
Space Board in Sweden.

Additional support for science analysis during the operations phase is
gratefully acknowledged from the Istituto Nazionale di Astrofisica in
Italy and the Centre National d'\'Etudes Spatiales in France.

\bibliographystyle{aa}
\bibliography{paper}

\begin{thebibliography}{81}
\expandafter\ifx\csname natexlab\endcsname\relax\def\natexlab#1{#1}\fi

\bibitem[{{Abdo} {et~al.}(2011){Abdo}, {Ackermann}, {Ajello}, {Allafort},
  {Baldini}, {Ballet}, {Barbiellini}, {Baring}, {Bastieri}, {Bellazzini},
  {Berenji}, {Blandford}, {Bloom}, {Bonamente}, {Borgland}, {Bouvier},
  {Brandt}, {Bregeon}, {Brigida}, {Bruel}, {Buehler}, {Buson}, {Caliandro},
  {Cameron}, {Caraveo}, {Casandjian}, {Cecchi}, {Chaty}, {Chekhtman}, {Cheung},
  {Chiang}, {Cillis}, {Ciprini}, {Claus}, {Cohen-Tanugi}, {Conrad}, {Corbel},
  {Cutini}, {de Angelis}, {de Palma}, {Dermer}, {Digel}, {Silva}, {Drell},
  {Drlica-Wagner}, {Dubois}, {Dumora}, {Favuzzi}, {Ferrara}, {Fortin},
  {Frailis}, {Fukazawa}, {Fukui}, {Funk}, {Fusco}, {Gargano}, {Gasparrini},
  {Gehrels}, {Germani}, {Giglietto}, {Giordano}, {Giroletti}, {Glanzman},
  {Godfrey}, {Grenier}, {Grondin}, {Guiriec}, {Hadasch}, {Hanabata}, {Harding},
  {Hayashida}, {Hayashi}, {Hays}, {Horan}, {Jackson}, {J{\'o}hannesson},
  {Johnson}, {Kamae}, {Katagiri}, {Kataoka}, {Kerr}, {Kn{\"o}dlseder}, {Kuss},
  {Lande}, {Latronico}, {Lee}, {Lemoine-Goumard}, {Longo}, {Loparco},
  {Lovellette}, {Lubrano}, {Madejski}, {Makeev}, {Mazziotta}, {McEnery},
  {Michelson}, {Mignani}, {Mitthumsiri}, {Mizuno}, {Moiseev}, {Monte},
  {Monzani}, {Morselli}, {Moskalenko}, {Murgia}, {Naumann-Godo}, {Nolan},
  {Norris}, {Nuss}, {Ohsugi}, {Okumura}, {Orlando}, {Ormes}, {Paneque},
  {Parent}, {Pelassa}, {Pesce-Rollins}, {Pierbattista}, {Piron}, {Pohl},
  {Porter}, {Rain{\`o}}, {Rando}, {Razzano}, {Reimer}, {Reposeur}, {Ritz},
  {Romani}, {Roth}, {Sadrozinski}, {Saz Parkinson}, {Sgr{\`o}}, {Smith},
  {Smith}, {Spandre}, {Spinelli}, {Strickman}, {Tajima}, {Takahashi},
  {Takahashi}, {Tanaka}, {Thayer}, {Thayer}, {Thompson}, {Tibaldo}, {Tibolla},
  {Torres}, {Tosti}, {Tramacere}, {Troja}, {Uchiyama}, {Vandenbroucke},
  {Vasileiou}, {Vianello}, {Vilchez}, {Vitale}, {Waite}, {Wang}, {Winer},
  {Wood}, {Yamamoto}, {Yamazaki}, {Yang}, \& {Ziegler}}]{FermiRXJ}
{Abdo}, A.~A., {Ackermann}, M., {Ajello}, M., {et~al.} 2011, \apj, 734, 28

\bibitem[{{Abdo} {et~al.}(2009){Abdo}, {Ackermann}, {Ajello}, {Atwood},
  {Baldini}, {Ballet}, {Barbiellini}, {Bastieri}, {Baughman}, {Bechtol},
  {Bellazzini}, {Berenji}, {Bloom}, {Bonamente}, {Borgland}, {Bouvier},
  {Bregeon}, {Brez}, {Brigida}, {Bruel}, {Buehler}, {Burnett}, {Buson},
  {Caliandro}, {Cameron}, {Caraveo}, {Casandjian}, {Cecchi}, {{\c C}elik},
  {Charles}, {Chekhtman}, {Chiang}, {Ciprini}, {Claus}, {Cohen-Tanugi},
  {Conrad}, {de Palma}, {Digel}, {Do Couto E Silva}, {Drell}, {Dubois},
  {Dumora}, {Farnier}, {Favuzzi}, {Fegan}, {Focke}, {Fortin}, {Frailis},
  {Fukazawa}, {Funk}, {Fusco}, {Gargano}, {Gehrels}, {Germani}, {Giebels},
  {Giglietto}, {Giordano}, {Glanzman}, {Godfrey}, {Grenier}, {Grondin},
  {Grove}, {Guillemot}, {Guiriec}, {Hays}, {Horan}, {Hughes},
  {J{\'o}hannesson}, {Johnson}, {Johnson}, {Johnson}, {Kamae}, {Katagiri},
  {Kataoka}, {Kawai}, {Kerr}, {Kn{\"o}dlseder}, {Kuss}, {Lande}, {Latronico},
  {Lemoine-Goumard}, {Longo}, {Loparco}, {Lott}, {Lovellette}, {Lubrano},
  {Makeev}, {Mazziotta}, {McEnery}, {Meurer}, {Michelson}, {Mitthumsiri},
  {Mizuno}, {Monte}, {Monzani}, {Morselli}, {Moskalenko}, {Murgia}, {Nolan},
  {Norris}, {Nuss}, {Ohsugi}, {Okumura}, {Omodei}, {Orlando}, {Ormes},
  {Paneque}, {Panetta}, {Parent}, {Pelassa}, {Pepe}, {Pesce-Rollins}, {Piron},
  {Porter}, {Rain{\`o}}, {Rando}, {Razzano}, {Reimer}, {Reimer}, {Reposeur},
  {Rochester}, {Rodriguez}, {Roth}, {Sadrozinski}, {Sander}, {Saz Parkinson},
  {Sgr{\`o}}, {Share}, {Siskind}, {Smith}, {Smith}, {Spandre}, {Spinelli},
  {Strickman}, {Suson}, {Takahashi}, {Tanaka}, {Thayer}, {Thayer}, {Thompson},
  {Tibaldo}, {Torres}, {Tosti}, {Tramacere}, {Uchiyama}, {Usher}, {Vasileiou},
  {Vilchez}, {Vitale}, {Waite}, {Wang}, {Winer}, {Wood}, {Ylinen}, \&
  {Ziegler}}]{FermiEarthLimb}
{Abdo}, A.~A., {Ackermann}, M., {Ajello}, M., {et~al.} 2009, \prd, 80, 122004

\bibitem[{{Abramowski} {et~al.}(2011){Abramowski}, {Acero}, {Aharonian},
  {Akhperjanian}, {Anton}, {Barnacka}, {Barres de Almeida}, {Bazer-Bachi},
  {Becherini}, {Becker}, {Behera}, {Bernl{\"o}hr}, {Bochow}, {Boisson},
  {Bolmont}, {Bordas}, {Borrel}, {Brucker}, {Brun}, {Brun}, {Bulik},
  {B{\"u}sching}, {Carrigan}, {Casanova}, {Cerruti}, {Chadwick}, {Charbonnier},
  {Chaves}, {Cheesebrough}, {Chounet}, {Clapson}, {Coignet}, {Conrad},
  {Dalton}, {Daniel}, {Davids}, {Degrange}, {Deil}, {Dickinson},
  {Djannati-Ata{\"i}}, {Domainko}, {Drury}, {Dubois}, {Dubus}, {Dyks}, {Dyrda},
  {Egberts}, {Eger}, {Espigat}, {Fallon}, {Farnier}, {Fegan}, {Feinstein},
  {Fernandes}, {Fiasson}, {Fontaine}, {F{\"o}rster}, {F{\"u}{\ss}ling},
  {Gallant}, {Gast}, {G{\'e}rard}, {Gerbig}, {Giebels}, {Glicenstein},
  {Gl{\"u}ck}, {Goret}, {G{\"o}ring}, {Hague}, {Hampf}, {Hauser}, {Heinz},
  {Heinzelmann}, {Henri}, {Hermann}, {Hinton}, {Hoffmann}, {Hofmann},
  {Hofverberg}, {Horns}, {Jacholkowska}, {de Jager}, {Jahn}, {Jamrozy}, {Jung},
  {Kastendieck}, {Katarzy{\'n}ski}, {Katz}, {Kaufmann}, {Keogh}, {Kerschhaggl},
  {Khangulyan}, {Kh{\'e}lifi}, {Klochkov}, {Klu{\'z}niak}, {Kneiske}, {Komin},
  {Kosack}, {Kossakowski}, {Laffon}, {Lamanna}, {Lennarz}, {Lohse}, {Lopatin},
  {Lu}, {Marandon}, {Marcowith}, {Masbou}, {Maurin}, {Maxted}, {McComb},
  {Medina}, {M{\'e}hault}, {Moderski}, {Moulin}, {Naumann}, {Naumann-Godo}, {de
  Naurois}, {Nedbal}, {Nekrassov}, {Nguyen}, {Nicholas}, {Niemiec}, {Nolan},
  {Ohm}, {Olive}, {de O{\~n}a Wilhelmi}, {Opitz}, {Ostrowski}, {Panter}, {Paz
  Arribas}, {Pedaletti}, {Pelletier}, {Petrucci}, {Pita}, {P{\"u}hlhofer},
  {Punch}, {Quirrenbach}, {Raue}, {Rayner}, {Reimer}, {Reimer}, {Renaud}, {de
  Los Reyes}, {Rieger}, {Ripken}, {Rob}, {Rosier-Lees}, {Rowell}, {Rudak},
  {Rulten}, {Ruppel}, {Ryde}, {Sahakian}, {Santangelo}, {Schlickeiser},
  {Sch{\"o}ck}, {Sch{\"o}nwald}, {Schwanke}, {Schwarzburg}, {Schwemmer},
  {Shalchi}, {Sikora}, {Skilton}, {Sol}, {Spengler}, {Stawarz}, {Steenkamp},
  {Stegmann}, {Stinzing}, {Sushch}, {Szostek}, {Tavernet}, {Terrier},
  {Tibolla}, {Tluczykont}, {Valerius}, {van Eldik}, {Vasileiadis}, {Venter},
  {Vialle}, {Viana}, {Vincent}, {Vivier}, {V{\"o}lk}, {Volpe}, {Vorobiov},
  {Vorster}, {Wagner}, {Ward}, {Wierzcholska}, {Zajczyk}, {Zdziarski}, {Zech},
  \& {Zechlin}}]{ReflectedPixelBG}
{Abramowski}, A., {Acero}, F., {Aharonian}, F.~A., {et~al.} 2011, Physical
  Review Letters, 106, 161301

\bibitem[{{Abramowski} {et~al.}(2010){Abramowski}, {Acero}, {Aharonian},
  {Akhperjanian}, {Anton}, {Barres de Almeida}, {Bazer-Bachi}, {Becherini},
  {Benbow}, {Bernl{\"o}hr}, {Bochow}, {Boisson}, {Bolmont}, {Borrel},
  {Brucker}, {Brun}, {Brun}, {B{\"u}hler}, {Bulik}, {B{\"u}sching},
  {Boutelier}, {Chadwick}, {Charbonnier}, {Chaves}, {Cheesebrough}, {Chounet},
  {Clapson}, {Coignet}, {Conrad}, {Costamante}, {Dalton}, {Daniel}, {Davids},
  {Degrange}, {Deil}, {Dickinson}, {Djannati-Ata{\"i}}, {Domainko},
  {O'C.~Drury}, {Dubois}, {Dubus}, {Dyks}, {Dyrda}, {Egberts}, {Eger},
  {Espigat}, {Fallon}, {Farnier}, {Fegan}, {Feinstein}, {Fernandes}, {Fiasson},
  {F{\"o}rster}, {Fontaine}, {F{\"u}{\ss}ling}, {Gabici}, {Gallant},
  {G{\'e}rard}, {Gerbig}, {Giebels}, {Glicenstein}, {Gl{\"u}ck}, {Goret},
  {G{\"o}ring}, {Hampf}, {Hauser}, {Heinz}, {Heinzelmann}, {Henri}, {Hermann},
  {Hinton}, {Hoffmann}, {Hofmann}, {Hofverberg}, {Holleran}, {Hoppe}, {Horns},
  {Jacholkowska}, {de Jager}, {Jahn}, {Jung}, {Katarzy{\'n}ski}, {Katz},
  {Kaufmann}, {Kerschhaggl}, {Khangulyan}, {Kh{\'e}lifi}, {Keogh}, {Klochkov},
  {Klu{\'z}niak}, {Kneiske}, {Komin}, {Kosack}, {Kossakowski}, {Lamanna},
  {Lenain}, {Lohse}, {Lu}, {Marandon}, {Marcowith}, {Masbou}, {Maurin},
  {McComb}, {Medina}, {M{\'e}hault}, {Moderski}, {Moulin}, {Naumann-Godo}, {de
  Naurois}, {Nedbal}, {Nekrassov}, {Nguyen}, {Nicholas}, {Niemiec}, {Nolan},
  {Ohm}, {Olive}, {de O{\~n}a Wilhelmi}, {Opitz}, {Orford}, {Ostrowski},
  {Panter}, {Paz Arribas}, {Pedaletti}, {Pelletier}, {Petrucci}, {Pita},
  {P{\"u}hlhofer}, {Punch}, {Quirrenbach}, {Raubenheimer}, {Raue}, {Rayner},
  {Reimer}, {Renaud}, {de los Reyes}, {Rieger}, {Ripken}, {Rob}, {Rosier-Lees},
  {Rowell}, {Rudak}, {Rulten}, {Ruppel}, {Ryde}, {Sahakian}, {Santangelo},
  {Schlickeiser}, {Sch{\"o}ck}, {Sch{\"o}nwald}, {Schwanke}, {Schwarzburg},
  {Schwemmer}, {Shalchi}, {Sushch}, {Sikora}, {Skilton}, {Sol}, {Stawarz},
  {Steenkamp}, {Stegmann}, {Stinzing}, {Superina}, {Szostek}, {Tam},
  {Tavernet}, {Terrier}, {Tibolla}, {Tluczykont}, {Valerius}, {van Eldik},
  {Vasileiadis}, {Venter}, {Venter}, {Vialle}, {Viana}, {Vincent}, {Vivier},
  {V{\"o}lk}, {Volpe}, {Vorobiov}, {Wagner}, {Ward}, {Zdziarski}, {Zech}, \&
  {Zechlin}}]{2010A&A...520A..83H}
{Abramowski}, A., {Acero}, F., {Aharonian}, F.~A., {et~al.} 2010, \aap, 520,
  A83

\bibitem[{{Acero} {et~al.}(2015){Acero}, {Ackermann}, {Ajello}, {Albert},
  {Atwood}, {Axelsson}, {Baldini}, {Ballet}, {Barbiellini}, {Bastieri},
  {Belfiore}, {Bellazzini}, {Bissaldi}, {Blandford}, {Bloom}, {Bogart},
  {Bonino}, {Bottacini}, {Bregeon}, {Britto}, {Bruel}, {Buehler}, {Burnett},
  {Buson}, {Caliandro}, {Cameron}, {Caputo}, {Caragiulo}, {Caraveo},
  {Casandjian}, {Cavazzuti}, {Charles}, {Chaves}, {Chekhtman}, {Cheung},
  {Chiang}, {Chiaro}, {Ciprini}, {Claus}, {Cohen-Tanugi}, {Cominsky}, {Conrad},
  {Cutini}, {D'Ammando}, {de Angelis}, {DeKlotz}, {de Palma}, {Desiante},
  {Digel}, {Di Venere}, {Drell}, {Dubois}, {Dumora}, {Favuzzi}, {Fegan},
  {Ferrara}, {Finke}, {Franckowiak}, {Fukazawa}, {Funk}, {Fusco}, {Gargano},
  {Gasparrini}, {Giebels}, {Giglietto}, {Giommi}, {Giordano}, {Giroletti},
  {Glanzman}, {Godfrey}, {Grenier}, {Grondin}, {Grove}, {Guillemot}, {Guiriec},
  {Hadasch}, {Harding}, {Hays}, {Hewitt}, {Hill}, {Horan}, {Iafrate}, {Jogler},
  {J{\'o}hannesson}, {Johnson}, {Johnson}, {Johnson}, {Johnson}, {Kamae},
  {Kataoka}, {Katsuta}, {Kuss}, {La Mura}, {Landriu}, {Larsson}, {Latronico},
  {Lemoine-Goumard}, {Li}, {Li}, {Longo}, {Loparco}, {Lott}, {Lovellette},
  {Lubrano}, {Madejski}, {Massaro}, {Mayer}, {Mazziotta}, {McEnery},
  {Michelson}, {Mirabal}, {Mizuno}, {Moiseev}, {Mongelli}, {Monzani},
  {Morselli}, {Moskalenko}, {Murgia}, {Nuss}, {Ohno}, {Ohsugi}, {Omodei},
  {Orienti}, {Orlando}, {Ormes}, {Paneque}, {Panetta}, {Perkins},
  {Pesce-Rollins}, {Piron}, {Pivato}, {Porter}, {Racusin}, {Rando}, {Razzano},
  {Razzaque}, {Reimer}, {Reimer}, {Reposeur}, {Rochester}, {Romani},
  {Salvetti}, {S{\'a}nchez-Conde}, {Saz Parkinson}, {Schulz}, {Siskind},
  {Smith}, {Spada}, {Spandre}, {Spinelli}, {Stephens}, {Strong}, {Suson},
  {Takahashi}, {Takahashi}, {Tanaka}, {Thayer}, {Thayer}, {Thompson},
  {Tibaldo}, {Tibolla}, {Torres}, {Torresi}, {Tosti}, {Troja}, {Van Klaveren},
  {Vianello}, {Winer}, {Wood}, {Wood}, {Zimmer}, \& {Fermi-LAT
  Collaboration}}]{2015ApJS..218...23A}
{Acero}, F., {Ackermann}, M., {Ajello}, M., {et~al.} 2015, \apjs, 218, 23

\bibitem[{{Acero} {et~al.}(2016{\natexlab{a}}){Acero}, {Ackermann}, {Ajello},
  {Albert}, {Baldini}, {Ballet}, {Barbiellini}, {Bastieri}, {Bellazzini},
  {Bissaldi}, {Bloom}, {Bonino}, {Bottacini}, \&
  {Brandt}}]{2016ApJS..223...26A}
{Acero}, F., {Ackermann}, M., {Ajello}, M., {et~al.} 2016{\natexlab{a}}, \apjs,
  223, 26

\bibitem[{{Acero} {et~al.}(2016{\natexlab{b}}){Acero}, {Ackermann}, {Ajello},
  {Baldini}, {Ballet}, {Barbiellini}, {Bastieri}, {Bellazzini}, {Bissaldi},
  {Blandford}, {Bloom}, {Bonino}, {Bottacini}, \&
  {Brandt}}]{2016ApJS..224....8A}
{Acero}, F., {Ackermann}, M., {Ajello}, M., {et~al.} 2016{\natexlab{b}}, \apjs,
  224, 8

\bibitem[{{Acero} {et~al.}(2009){Acero}, {Ballet}, {Decourchelle},
  {Lemoine-Goumard}, {Ortega}, {Giacani}, {Dubner}, \&
  {Cassam-Chena{\"i}}}]{AceroXMM2009}
{Acero}, F., {Ballet}, J., {Decourchelle}, A., {et~al.} 2009, \aap, 505, 157

\bibitem[{{Acharya} {et~al.}(2013){Acharya}, {Actis}, {Aghajani}, {Agnetta},
  {Aguilar}, {Aharonian}, {Ajello}, {Akhperjanian}, {Alcubierre},
  {Aleksi{\'c}}, \& et~al.}]{2013APh....43....3A}
{Acharya}, B.~S., {Actis}, M., {Aghajani}, T., {et~al.} 2013, Astroparticle
  Physics, 43, 3

\bibitem[{{Ackermann} {et~al.}(2013){Ackermann}, {Ajello}, {Allafort},
  {Baldini}, {Ballet}, {Barbiellini}, {Baring}, {Bastieri}, {Bechtol},
  {Bellazzini}, {Blandford}, {Bloom}, {Bonamente}, {Borgland}, {Bottacini},
  {Brandt}, {Bregeon}, {Brigida}, {Bruel}, {Buehler}, {Busetto}, {Buson},
  {Caliandro}, {Cameron}, {Caraveo}, {Casandjian}, {Cecchi}, {{\c C}elik},
  {Charles}, {Chaty}, {Chaves}, {Chekhtman}, {Cheung}, {Chiang}, {Chiaro},
  {Cillis}, {Ciprini}, {Claus}, {Cohen-Tanugi}, {Cominsky}, {Conrad}, {Corbel},
  {Cutini}, {D'Ammando}, {de Angelis}, {de Palma}, {Dermer}, {do Couto e
  Silva}, {Drell}, {Drlica-Wagner}, {Falletti}, {Favuzzi}, {Ferrara},
  {Franckowiak}, {Fukazawa}, {Funk}, {Fusco}, {Gargano}, {Germani},
  {Giglietto}, {Giommi}, {Giordano}, {Giroletti}, {Glanzman}, {Godfrey},
  {Grenier}, {Grondin}, {Grove}, {Guiriec}, {Hadasch}, {Hanabata}, {Harding},
  {Hayashida}, {Hayashi}, {Hays}, {Hewitt}, {Hill}, {Hughes}, {Jackson},
  {Jogler}, {J{\'o}hannesson}, {Johnson}, {Kamae}, {Kataoka}, {Katsuta},
  {Kn{\"o}dlseder}, {Kuss}, {Lande}, {Larsson}, {Latronico}, {Lemoine-Goumard},
  {Longo}, {Loparco}, {Lovellette}, {Lubrano}, {Madejski}, {Massaro}, {Mayer},
  {Mazziotta}, {McEnery}, {Mehault}, {Michelson}, {Mignani}, {Mitthumsiri},
  {Mizuno}, {Moiseev}, {Monzani}, {Morselli}, {Moskalenko}, {Murgia},
  {Nakamori}, {Nemmen}, {Nuss}, {Ohno}, {Ohsugi}, {Omodei}, {Orienti},
  {Orlando}, {Ormes}, {Paneque}, {Perkins}, {Pesce-Rollins}, {Piron}, {Pivato},
  {Rain{\`o}}, {Rando}, {Razzano}, {Razzaque}, {Reimer}, {Reimer}, {Ritz},
  {Romoli}, {S{\'a}nchez-Conde}, {Schulz}, {Sgr{\`o}}, {Simeon}, {Siskind},
  {Smith}, {Spandre}, {Spinelli}, {Stecker}, {Strong}, {Suson}, {Tajima},
  {Takahashi}, {Takahashi}, {Tanaka}, {Thayer}, {Thayer}, {Thompson},
  {Thorsett}, {Tibaldo}, {Tibolla}, {Tinivella}, {Troja}, {Uchiyama}, {Usher},
  {Vandenbroucke}, {Vasileiou}, {Vianello}, {Vitale}, {Waite}, {Werner},
  {Winer}, {Wood}, {Wood}, {Yamazaki}, {Yang}, \& {Zimmer}}]{FermiPion}
{Ackermann}, M., {Ajello}, M., {Allafort}, A., {et~al.} 2013, Science, 339, 807

\bibitem[{{Aharonian}(2013{\natexlab{a}})}]{2013SAAS...40....1A}
{Aharonian}, F. 2013{\natexlab{a}}, Astrophysics at Very High Energies,
  Saas-Fee Advanced Course, Volume 40.~ISBN 978-3-642-36133-3.~Springer-Verlag
  Berlin Heidelberg, 2013, p.~1, 40, 1

\bibitem[{{Aharonian}(2013{\natexlab{b}})}]{2013APh....43...71A}
{Aharonian}, F.~A. 2013{\natexlab{b}}, Astroparticle Physics, 43, 71

\bibitem[{{Aharonian} {et~al.}(2004){Aharonian}, {Akhperjanian}, {Aye},
  {Bazer-Bachi}, {Beilicke}, {Benbow}, {Berge}, {Berghaus}, {Bernl{\"o}hr},
  {Bolz}, {Boisson}, {Borgmeier}, {Breitling}, {Brown}, {Bussons Gordo},
  {Chadwick}, {Chitnis}, {Chounet}, {Cornils}, {Costamante}, {Degrange},
  {Djannati-Ata{\"i}}, {Drury}, {Ergin}, {Espigat}, {Feinstein}, {Fleury},
  {Fontaine}, {Funk}, {Gallant}, {Giebels}, {Gillessen}, {Goret}, {Guy},
  {Hadjichristidis}, {Hauser}, {Heinzelmann}, {Henri}, {Hermann}, {Hinton},
  {Hofmann}, {Holleran}, {Horns}, {de Jager}, {Jung}, {Kh{\'e}lifi}, {Komin},
  {Konopelko}, {Latham}, {Le Gallou}, {Lemoine}, {Lemi{\`e}re}, {Leroy},
  {Lohse}, {Marcowith}, {Masterson}, {McComb}, {de Naurois}, {Nolan},
  {Noutsos}, {Orford}, {Osborne}, {Ouchrif}, {Panter}, {Pelletier}, {Pita},
  {Pohl}, {P{\"u}hlhofer}, {Punch}, {Raubenheimer}, {Raue}, {Raux}, {Rayner},
  {Redondo}, {Reimer}, {Reimer}, {Ripken}, {Rivoal}, {Rob}, {Rolland},
  {Rowell}, {Sahakian}, {Saug{\'e}}, {Schlenker}, {Schlickeiser}, {Schuster},
  {Schwanke}, {Siewert}, {Sol}, {Steenkamp}, {Stegmann}, {Tavernet},
  {Th{\'e}oret}, {Tluczykont}, {van der Walt}, {Vasileiadis}, {Vincent},
  {Visser}, {V{\"o}lk}, \& {Wagner}}]{Hess1713a}
{Aharonian}, F.~A., {Akhperjanian}, A.~G., {Aye}, K.-M., {et~al.} 2004, \nat,
  432, 75

\bibitem[{{Aharonian} {et~al.}(2008){Aharonian}, {Akhperjanian}, {Bazer-Bachi},
  {Behera}, {Beilicke}, {Benbow}, {Berge}, {Bernl{\"o}hr}, {Boisson}, {Bolz},
  {Borrel}, {Braun}, {Brion}, {Brown}, {B{\"u}hler}, {Bulik}, {B{\"u}sching},
  {Boutelier}, {Carrigan}, {Chadwick}, {Chounet}, {Clapson}, {Coignet},
  {Cornils}, {Costamante}, {Degrange}, {Dickinson}, {Djannati-Ata{\"i}},
  {Domainko}, {O'C.~Drury}, {Dubus}, {Dyks}, {Egberts}, {Emmanoulopoulos},
  {Espigat}, {Farnier}, {Feinstein}, {Fiasson}, {F{\"o}rster}, {Fontaine},
  {Fukui}, {Funk}, {Funk}, {F{\"u}{\ss}ling}, {Gallant}, {Giebels},
  {Glicenstein}, {Gl{\"u}ck}, {Goret}, {Hadjichristidis}, {Hauser}, {Hauser},
  {Heinzelmann}, {Henri}, {Hermann}, {Hinton}, {Hoffmann}, {Hofmann},
  {Holleran}, {Hoppe}, {Horns}, {Jacholkowska}, {de Jager}, {Kendziorra},
  {Kerschhaggl}, {Kh{\'e}lifi}, {Komin}, {Kosack}, {Lamanna}, {Latham}, {Le
  Gallou}, {Lemi{\`e}re}, {Lemoine-Goumard}, {Lenain}, {Lohse}, {Martin},
  {Martineau-Huynh}, {Marcowith}, {Masterson}, {Maurin}, {McComb}, {Moderski},
  {Moriguchi}, {Moulin}, {de Naurois}, {Nedbal}, {Nolan}, {Olive}, {Orford},
  {Osborne}, {Ostrowski}, {Panter}, {Pedaletti}, {Pelletier}, {Petrucci},
  {Pita}, {P{\"u}hlhofer}, {Punch}, {Ranchon}, {Raubenheimer}, {Raue},
  {Rayner}, {Reimer}, {Renaud}, {Ripken}, {Rob}, {Rolland}, {Rosier-Lees},
  {Rowell}, {Rudak}, {Ruppel}, {Sahakian}, {Santangelo}, {Saug{\'e}},
  {Schlenker}, {Schlickeiser}, {Schr{\"o}der}, {Schwanke}, {Schwarzburg},
  {Schwemmer}, {Shalchi}, {Sol}, {Spangler}, {Stawarz}, {Steenkamp},
  {Stegmann}, {Superina}, {Takeuchi}, {Tam}, {Tavernet}, {Terrier}, {van
  Eldik}, {Vasileiadis}, {Venter}, {Vialle}, {Vincent}, {Vivier}, {V{\"o}lk},
  {Volpe}, {Wagner}, \& {Ward}}]{2008A&A...481..401A}
{Aharonian}, F.~A., {Akhperjanian}, A.~G., {Bazer-Bachi}, A.~R., {et~al.} 2008,
  \aap, 481, 401

\bibitem[{{Aharonian} {et~al.}(2006{\natexlab{a}}){Aharonian}, {Akhperjanian},
  {Bazer-Bachi}, {Beilicke}, {Benbow}, {Berge}, {Bernl{\"o}hr}, {Boisson},
  {Bolz}, {Borrel}, {Braun}, {Breitling}, {Brown}, {B{\"u}hler},
  {B{\"u}sching}, {Carrigan}, {Chadwick}, {Chounet}, {Cornils}, {Costamante},
  {Degrange}, {Dickinson}, {Djannati-Ata{\"i}}, {O'C.~Drury}, {Dubus},
  {Egberts}, {Emmanoulopoulos}, {Espigat}, {Feinstein}, {Ferrero}, {Fiasson},
  {Fontaine}, {Funk}, {Funk}, {Gallant}, {Giebels}, {Glicenstein}, {Goret},
  {Hadjichristidis}, {Hauser}, {Hauser}, {Heinzelmann}, {Henri}, {Hermann},
  {Hinton}, {Hofmann}, {Holleran}, {Horns}, {Jacholkowska}, {de Jager},
  {Kh{\'e}lifi}, {Komin}, {Konopelko}, {Kosack}, {Latham}, {Le Gallou},
  {Lemi{\`e}re}, {Lemoine-Goumard}, {Lohse}, {Martin}, {Martineau-Huynh},
  {Marcowith}, {Masterson}, {McComb}, {de Naurois}, {Nedbal}, {Nolan},
  {Noutsos}, {Orford}, {Osborne}, {Ouchrif}, {Panter}, {Pelletier}, {Pita},
  {P{\"u}hlhofer}, {Punch}, {Raubenheimer}, {Raue}, {Rayner}, {Reimer},
  {Reimer}, {Ripken}, {Rob}, {Rolland}, {Rowell}, {Sahakian}, {Saug{\'e}},
  {Schlenker}, {Schlickeiser}, {Schwanke}, {Sol}, {Spangler}, {Spanier},
  {Steenkamp}, {Stegmann}, {Superina}, {Tavernet}, {Terrier}, {Th{\'e}oret},
  {Tluczykont}, {van Eldik}, {Vasileiadis}, {Venter}, {Vincent}, {V{\"o}lk},
  {Wagner}, \& {Ward}}]{HessCrab}
{Aharonian}, F.~A., {Akhperjanian}, A.~G., {Bazer-Bachi}, A.~R., {et~al.}
  2006{\natexlab{a}}, \aap, 457, 899

\bibitem[{{Aharonian} {et~al.}(2006{\natexlab{b}}){Aharonian}, {Akhperjanian},
  {Bazer-Bachi}, {Beilicke}, {Benbow}, {Berge}, {Bernl{\"o}hr}, {Boisson},
  {Bolz}, {Borrel}, {Braun}, {Breitling}, {Brown}, {Chadwick}, {Chounet},
  {Cornils}, {Costamante}, {Degrange}, {Dickinson}, {Djannati-Ata{\"i}},
  {O'C.~Drury}, {Dubus}, {Emmanoulopoulos}, {Espigat}, {Feinstein}, {Fontaine},
  {Fuchs}, {Funk}, {Gallant}, {Giebels}, {Glicenstein}, {Goret},
  {Hadjichristidis}, {Hauser}, {Hauser}, {Heinzelmann}, {Henri}, {Hermann},
  {Hinton}, {Hofmann}, {Holleran}, {Horns}, {Jacholkowska}, {de Jager},
  {Kh{\'e}lifi}, {Klages}, {Komin}, {Konopelko}, {Latham}, {Le Gallou},
  {Lemi{\`e}re}, {Lemoine-Goumard}, {Lohse}, {Martin}, {Martineau-Huynh},
  {Marcowith}, {Masterson}, {McComb}, {de Naurois}, {Nedbal}, {Nolan},
  {Noutsos}, {Orford}, {Osborne}, {Ouchrif}, {Panter}, {Pelletier}, {Pita},
  {P{\"u}hlhofer}, {Punch}, {Raubenheimer}, {Raue}, {Rayner}, {Reimer},
  {Reimer}, {Ripken}, {Rob}, {Rolland}, {Rowell}, {Sahakian}, {Saug{\'e}},
  {Schlenker}, {Schlickeiser}, {Schuster}, {Schwanke}, {Siewert}, {Sol},
  {Spangler}, {Steenkamp}, {Stegmann}, {Superina}, {Tavernet}, {Terrier},
  {Th{\'e}oret}, {Tluczykont}, {van Eldik}, {Vasileiadis}, {Venter}, {Vincent},
  {V{\"o}lk}, \& {Wagner}}]{Hess1713b}
{Aharonian}, F.~A., {Akhperjanian}, A.~G., {Bazer-Bachi}, A.~R., {et~al.}
  2006{\natexlab{b}}, \aap, 449, 223

\bibitem[{{Aharonian} {et~al.}(2007){Aharonian}, {Akhperjanian}, {Bazer-Bachi},
  {Beilicke}, {Benbow}, {Berge}, {Bernl{\"o}hr}, {Boisson}, {Bolz}, {Borrel},
  {Braun}, {Brion}, {Brown}, {B{\"u}hler}, {B{\"u}sching}, {Carrigan},
  {Chadwick}, {Chounet}, {Coignet}, {Cornils}, {Costamante}, {Degrange},
  {Dickinson}, {Djannati-Ata{\"i}}, {O'C.~Drury}, {Dubus}, {Egberts},
  {Emmanoulopoulos}, {Espigat}, {Feinstein}, {Ferrero}, {Fiasson}, {Fontaine},
  {Funk}, {Funk}, {F{\"u}{\ss}ling}, {Gallant}, {Giebels}, {Glicenstein},
  {Gl{\"u}ck}, {Goret}, {Hadjichristidis}, {Hauser}, {Hauser}, {Heinzelmann},
  {Henri}, {Hermann}, {Hinton}, {Hoffmann}, {Hofmann}, {Holleran}, {Hoppe},
  {Horns}, {Jacholkowska}, {de Jager}, {Kendziorra}, {Kerschhaggl},
  {Kh{\'e}lifi}, {Komin}, {Konopelko}, {Kosack}, {Lamanna}, {Latham}, {Le
  Gallou}, {Lemi{\`e}re}, {Lemoine-Goumard}, {Lohse}, {Martin},
  {Martineau-Huynh}, {Marcowith}, {Masterson}, {Maurin}, {McComb}, {Moulin},
  {de Naurois}, {Nedbal}, {Nolan}, {Noutsos}, {Olive}, {Orford}, {Osborne},
  {Panter}, {Pelletier}, {Pita}, {P{\"u}hlhofer}, {Punch}, {Ranchon},
  {Raubenheimer}, {Raue}, {Rayner}, {Reimer}, {Reimer}, {Ripken}, {Rob},
  {Rolland}, {Rosier-Lees}, {Rowell}, {Sahakian}, {Santangelo}, {Saug{\'e}},
  {Schlenker}, {Schlickeiser}, {Schr{\"o}der}, {Schwanke}, {Schwarzburg},
  {Schwemmer}, {Shalchi}, {Sol}, {Spangler}, {Spanier}, {Steenkamp},
  {Stegmann}, {Superina}, {Tam}, {Tavernet}, {Terrier}, {Tluczykont}, {van
  Eldik}, {Vasileiadis}, {Venter}, {Vialle}, {Vincent}, {V{\"o}lk}, {Wagner},
  \& {Ward}}]{Hess1713c}
{Aharonian}, F.~A., {Akhperjanian}, A.~G., {Bazer-Bachi}, A.~R., {et~al.} 2007,
  \aap, 464, 235

\bibitem[{{Aharonian} \& {Atoyan}(1996)}]{1996A&A...309..917A}
{Aharonian}, F.~A. \& {Atoyan}, A.~M. 1996, \aap, 309, 917

\bibitem[{{Aharonian} \& {Atoyan}(1999)}]{aharonian99}
{Aharonian}, F.~A. \& {Atoyan}, A.~M. 1999, \aap, 351, 330

\bibitem[{Akaike(1974)}]{akaike1974new}
Akaike, H. 1974, IEEE transactions on automatic control, 19, 716

\bibitem[{{Axford} {et~al.}(1977){Axford}, {Leer}, \&
  {Skadron}}]{1977ICRC...11..132A}
{Axford}, W.~I., {Leer}, E., \& {Skadron}, G. 1977, International Cosmic Ray
  Conference, 11, 132

\bibitem[{{Bell}(1978)}]{1978MNRAS.182..147B}
{Bell}, A.~R. 1978, \mnras, 182, 147

\bibitem[{{Bell} {et~al.}(2013){Bell}, {Schure}, {Reville}, \&
  {Giacinti}}]{2013MNRAS.431..415B}
{Bell}, A.~R., {Schure}, K.~M., {Reville}, B., \& {Giacinti}, G. 2013, \mnras,
  431, 415

\bibitem[{{Berezhko} \& {V{\"o}lk}(2006)}]{2006A&A...451..981B}
{Berezhko}, E.~G. \& {V{\"o}lk}, H.~J. 2006, \aap, 451, 981

\bibitem[{{Berge} {et~al.}(2007){Berge}, {Funk}, \& {Hinton}}]{BackgroundPaper}
{Berge}, D., {Funk}, S., \& {Hinton}, J. 2007, \aap, 466, 1219

\bibitem[{{Blandford} \& {Ostriker}(1978)}]{1978ApJ...221L..29B}
{Blandford}, R.~D. \& {Ostriker}, J.~P. 1978, \apjl, 221, L29

\bibitem[{{Blasi}(2013)}]{2013A&ARv..21...70B}
{Blasi}, P. 2013, \aapr, 21, 70

\bibitem[{{Casanova} {et~al.}(2010){Casanova}, {Jones}, {Aharonian}, {Fukui},
  {Gabici}, {Kawamura}, {Onishi}, {Rowell}, {Sano}, {Torii}, \&
  {Yamamoto}}]{2010PASJ...62.1127C}
{Casanova}, S., {Jones}, D.~I., {Aharonian}, F.~A., {et~al.} 2010, \pasj, 62,
  1127

\bibitem[{{Cassam-Chena{\"i}} {et~al.}(2004){Cassam-Chena{\"i}},
  {Decourchelle}, {Ballet}, {Sauvageot}, {Dubner}, \& {Giacani}}]{CassamXMM}
{Cassam-Chena{\"i}}, G., {Decourchelle}, A., {Ballet}, J., {et~al.} 2004, A\&A,
  427, 199

\bibitem[{{Chan} \& {Vese}(2001)}]{ChanVese}
{Chan}, T.~F. \& {Vese}, L.~A. 2001, IEEE Transactions on Image Processing, 10,
  266

\bibitem[{{Chevalier}(1982)}]{chevalier82}
{Chevalier}, R.~A. 1982, \apj, 258, 790

\bibitem[{{de Naurois} \& {Rolland}(2009)}]{modelAna}
{de Naurois}, M. \& {Rolland}, L. 2009, Astroparticle Physics, 32, 231

\bibitem[{Drury(1983)}]{0034-4885-46-8-002}
Drury, L.~O. 1983, Reports on Progress in Physics, 46, 973

\bibitem[{{Drury}(2011)}]{2011MNRAS.415.1807D}
{Drury}, L.~O. 2011, \mnras, 415, 1807

\bibitem[{{Ellison} {et~al.}(2010){Ellison}, {Patnaude}, {Slane}, \&
  {Raymond}}]{2010ApJ...712..287E}
{Ellison}, D.~C., {Patnaude}, D.~J., {Slane}, P., \& {Raymond}, J. 2010, \apj,
  712, 287

\bibitem[{{Ellison} {et~al.}(2012){Ellison}, {Slane}, {Patnaude}, \&
  {Bykov}}]{2012ApJ...744...39E}
{Ellison}, D.~C., {Slane}, P., {Patnaude}, D.~J., \& {Bykov}, A.~M. 2012, \apj,
  744, 39

\bibitem[{{Federici} {et~al.}(2015){Federici}, {Pohl}, {Telezhinsky},
  {Wilhelm}, \& {Dwarkadas}}]{2015A&A...577A..12F}
{Federici}, S., {Pohl}, M., {Telezhinsky}, I., {Wilhelm}, A., \& {Dwarkadas},
  V.~V. 2015, \aap, 577, A12

\bibitem[{{Finke} \& {Dermer}(2012)}]{2012ApJ...751...65F}
{Finke}, J.~D. \& {Dermer}, C.~D. 2012, \apj, 751, 65

\bibitem[{{Fukui} {et~al.}(2003){Fukui}, {Moriguchi}, {Tamura}, {Yamamoto},
  {Tawara}, {Mizuno}, {Onishi}, {Mizuno}, {Uchiyama}, {Hiraga}, {Takahashi},
  {Yamashita}, \& {Ikeuchi}}]{2003PASJ...55L..61F}
{Fukui}, Y., {Moriguchi}, Y., {Tamura}, K., {et~al.} 2003, \pasj, 55, L61

\bibitem[{{Fukui} {et~al.}(2012){Fukui}, {Sano}, {Sato}, {Torii}, {Horachi},
  {Hayakawa}, {McClure-Griffiths}, {Rowell}, {Inoue}, {Inutsuka}, {Kawamura},
  {Yamamoto}, {Okuda}, {Mizuno}, {Onishi}, {Mizuno}, \&
  {Ogawa}}]{2012ApJ...746...82F}
{Fukui}, Y., {Sano}, H., {Sato}, J., {et~al.} 2012, \apj, 746, 82

\bibitem[{{Funk} {et~al.}(2004){Funk}, {Hermann}, {Hinton}, {Berge},
  {Bernl{\"o}hr}, {Hofmann}, {Nayman}, {Toussenel}, \&
  {Vincent}}]{FunkTriggerPaper}
{Funk}, S., {Hermann}, G., {Hinton}, J., {et~al.} 2004, Astroparticle Physics,
  22, 285

\bibitem[{{Gabici} \& {Aharonian}(2016)}]{2016EPJWC.12104001G}
{Gabici}, S. \& {Aharonian}, F. 2016, in European Physical Journal Web of
  Conferences, Vol. 121, European Physical Journal Web of Conferences, 04001

\bibitem[{{Gabici} \& {Aharonian}(2014)}]{2014MNRAS.445L..70G}
{Gabici}, S. \& {Aharonian}, F.~A. 2014, \mnras, 445, L70

\bibitem[{{Gabici} {et~al.}(2009){Gabici}, {Aharonian}, \&
  {Casanova}}]{2009MNRAS.396.1629G}
{Gabici}, S., {Aharonian}, F.~A., \& {Casanova}, S. 2009, \mnras, 396, 1629

\bibitem[{{Giacalone} \& {Jokipii}(2007)}]{2007ApJ...663L..41G}
{Giacalone}, J. \& {Jokipii}, J.~R. 2007, \apjl, 663, L41

\bibitem[{{Ginzburg} \& {Syrovatskii}(1965)}]{ginzburg65}
{Ginzburg}, V.~L. \& {Syrovatskii}, S.~I. 1965, \araa, 3, 297

\bibitem[{{Grenier} {et~al.}(2015){Grenier}, {Black}, \&
  {Strong}}]{2015ARA&A..53..199G}
{Grenier}, I.~A., {Black}, J.~H., \& {Strong}, A.~W. 2015, \araa, 53, 199

\bibitem[{{Hahn} {et~al.}(2014){Hahn}, {de los Reyes}, {Bernl{\"o}hr},
  {Kr{\"u}ger}, {Lo}, {Chadwick}, {Daniel}, {Deil}, {Gast}, {Kosack}, \&
  {Marandon}}]{2014APh....54...25H}
{Hahn}, J., {de los Reyes}, R., {Bernl{\"o}hr}, K., {et~al.} 2014,
  Astroparticle Physics, 54, 25

\bibitem[{{H.E.S.S.~Collaboration}
  {et~al.}(2016{\natexlab{a}}){H.E.S.S.~Collaboration}, {Abdalla},
  {Abramowski}, {Aharonian}, {Ait Benkhali}, \& {Akhperjanian}}]{VelaJrForth}
{H.E.S.S.~Collaboration}, {Abdalla}, H., {Abramowski}, A., {et~al.}
  2016{\natexlab{a}}, Deeper H.E.S.S. Observations of Vela Junior
  (RX\,J0852.0$−$4622): Morphology Studies and Resolved Spectroscopy, \aap\
  forthcoming

\bibitem[{{H.E.S.S.~Collaboration}
  {et~al.}(2016{\natexlab{b}}){H.E.S.S.~Collaboration}, {Abdalla},
  {Abramowski}, {Aharonian}, {Ait Benkhali}, \& {Akhperjanian}}]{HGPSForth}
{H.E.S.S.~Collaboration}, {Abdalla}, H., {Abramowski}, A., {et~al.}
  2016{\natexlab{b}}, {The H.E.S.S. Galactic plane survey}, \aap\ forthcoming

\bibitem[{{H.E.S.S.~Collaboration}
  {et~al.}(2016{\natexlab{c}}){H.E.S.S.~Collaboration}, {Abdalla},
  {Abramowski}, {Aharonian}, {Ait Benkhali}, \& {Akhperjanian}}]{W49BForth}
{H.E.S.S.~Collaboration}, {Abdalla}, H., {Abramowski}, A., {et~al.}
  2016{\natexlab{c}}, {The supernova remnant W49B as seen with H.E.S.S. and
  \emph{Fermi}-LAT}, \aap\ forthcoming, arXiv:1609.00600

\bibitem[{{Inoue} {et~al.}(2012){Inoue}, {Yamazaki}, {Inutsuka}, \&
  {Fukui}}]{2012ApJ...744...71I}
{Inoue}, T., {Yamazaki}, R., {Inutsuka}, S.-i., \& {Fukui}, Y. 2012, \apj, 744,
  71

\bibitem[{{Jogler} \& {Funk}(2016)}]{2016ApJ...816..100J}
{Jogler}, T. \& {Funk}, S. 2016, \apj, 816, 100

\bibitem[{{Kafexhiu} {et~al.}(2014){Kafexhiu}, {Aharonian}, {Taylor}, \&
  {Vila}}]{2014PhRvD..90l3014K}
{Kafexhiu}, E., {Aharonian}, F.~A., {Taylor}, A.~M., \& {Vila}, G.~S. 2014,
  \prd, 90, 123014

\bibitem[{{Katsuda} {et~al.}(2015){Katsuda}, {Acero}, {Tominaga}, {Fukui},
  {Hiraga}, {Koyama}, {Lee}, {Mori}, {Nagataki}, {Ohira}, {Petre}, {Sano},
  {Takeuchi}, {Tamagawa}, {Tsuji}, {Tsunemi}, \&
  {Uchiyama}}]{2015ApJ...814...29K}
{Katsuda}, S., {Acero}, F., {Tominaga}, N., {et~al.} 2015, \apj, 814, 29

\bibitem[{{Kelner} {et~al.}(2006){Kelner}, {Aharonian}, \&
  {Bugayov}}]{2006PhRvD..74c4018K}
{Kelner}, S.~R., {Aharonian}, F.~A., \& {Bugayov}, V.~V. 2006, \prd, 74, 034018

\bibitem[{{Khangulyan} {et~al.}(2014){Khangulyan}, {Aharonian}, \&
  {Kelner}}]{2014ApJ...783..100K}
{Khangulyan}, D., {Aharonian}, F.~A., \& {Kelner}, S.~R. 2014, \apj, 783, 100

\bibitem[{{Koyama} {et~al.}(1997){Koyama}, {Kinugasa}, {Matsuzaki},
  {Nishiuchi}, {Sugizaki}, {Torii}, {Yamauchi}, \& {Aschenbach}}]{Koyama}
{Koyama}, K., {Kinugasa}, K., {Matsuzaki}, K., {et~al.} 1997, \pasj, 49, L7

\bibitem[{{Krymskii}(1977)}]{1977DoSSR.234.1306K}
{Krymskii}, G.~F. 1977, Akademiia Nauk SSSR Doklady, 234, 1306

\bibitem[{{Lazendic} {et~al.}(2004){Lazendic}, {Slane}, {Gaensler}, {Reynolds},
  {Plucinsky}, \& {Hughes}}]{Lazendic}
{Lazendic}, J.~S., {Slane}, P.~O., {Gaensler}, B.~M., {et~al.} 2004, \apj, 602,
  271

\bibitem[{{Malkov} {et~al.}(2005){Malkov}, {Diamond}, \&
  {Sagdeev}}]{2005ApJ...624L..37M}
{Malkov}, M.~A., {Diamond}, P.~H., \& {Sagdeev}, R.~Z. 2005, \apjl, 624, L37

\bibitem[{{Malkov} {et~al.}(2013){Malkov}, {Diamond}, {Sagdeev}, {Aharonian},
  \& {Moskalenko}}]{2013ApJ...768...73M}
{Malkov}, M.~A., {Diamond}, P.~H., {Sagdeev}, R.~Z., {Aharonian}, F.~A., \&
  {Moskalenko}, I.~V. 2013, \apj, 768, 73

\bibitem[{{Maxted} {et~al.}(2013){Maxted}, {Rowell}, {Dawson}, {Burton},
  {Fukui}, {Lazendic}, {Kawamura}, {Horachi}, {Sano}, {Walsh}, {Yoshiike}, \&
  {Fukuda}}]{2013PASA...30...55M}
{Maxted}, N.~I., {Rowell}, G.~P., {Dawson}, B.~R., {et~al.} 2013, \pasa, 30, 55

\bibitem[{{Morlino} {et~al.}(2009){Morlino}, {Amato}, \&
  {Blasi}}]{2009MNRAS.392..240M}
{Morlino}, G., {Amato}, E., \& {Blasi}, P. 2009, \mnras, 392, 240

\bibitem[{{Nakamori} {et~al.}(2015){Nakamori}, {Katagiri}, {Sano}, {Yamazaki},
  {Ohira}, {Bamba}, {Fukui}, {Mori}, {Lee}, {Fujita}, {Tajima}, {Inoue},
  {Gunji}, {Hanabata}, {Hayashida}, {Kubo}, {Kushida}, {Inoue}, {Ioka},
  {Kohri}, {Murase}, {Nagataki}, {Naito}, {Okumura}, {Saito}, {Sawada},
  {Tanaka}, {Terada}, {Uchiyama}, {Yanagita}, {Yoshida}, {Yoshikoshi}, \& {CTA
  Consortium}}]{2015arXiv150806052N}
{Nakamori}, T., {Katagiri}, H., {Sano}, H., {et~al.} 2015, ArXiv e-prints

\bibitem[{{Ohm} {et~al.}(2009){Ohm}, {van Eldik}, \& {Egberts}}]{TMVA}
{Ohm}, S., {van Eldik}, C., \& {Egberts}, K. 2009, Astroparticle Physics, 31,
  383

\bibitem[{Olive {et~al.}(2014)}]{Agashe:2014kda}
Olive, K.~A. {et~al.} 2014, Chin. Phys., C38, 090001

\bibitem[{{Parizot} {et~al.}(2006){Parizot}, {Marcowith}, {Ballet}, \&
  {Gallant}}]{2006A&A...453..387P}
{Parizot}, E., {Marcowith}, A., {Ballet}, J., \& {Gallant}, Y.~A. 2006, \aap,
  453, 387

\bibitem[{{Parsons} \& {Hinton}(2014)}]{ImPACT}
{Parsons}, R.~D. \& {Hinton}, J.~A. 2014, Astroparticle Physics, 56, 26

\bibitem[{{Pfeffermann} \& {Aschenbach}(1996)}]{Pfeffermann}
{Pfeffermann}, E. \& {Aschenbach}, B. 1996, in Roentgenstrahlung from the
  Universe, 267--268

\bibitem[{{Porter} {et~al.}(2006){Porter}, {Moskalenko}, \&
  {Strong}}]{2006ApJ...648L..29P}
{Porter}, T.~A., {Moskalenko}, I.~V., \& {Strong}, A.~W. 2006, \apjl, 648, L29

\bibitem[{{Shibata} {et~al.}(2011){Shibata}, {Ishikawa}, \&
  {Sekiguchi}}]{2011ApJ...727...38S}
{Shibata}, T., {Ishikawa}, T., \& {Sekiguchi}, S. 2011, \apj, 727, 38

\bibitem[{{Slane} {et~al.}(1999){Slane}, {Gaensler}, {Dame}, {Hughes},
  {Plucinsky}, \& {Green}}]{Slane}
{Slane}, P., {Gaensler}, B.~M., {Dame}, T.~M., {et~al.} 1999, \apj, 525, 357

\bibitem[{{Snowden} {et~al.}(1997){Snowden}, {Egger}, {Freyberg}, {McCammon},
  {Plucinsky}, {Sanders}, {Schmitt}, {Tr{\"u}mper}, \& {Voges}}]{RASS}
{Snowden}, S.~L., {Egger}, R., {Freyberg}, M.~J., {et~al.} 1997, \apj, 485, 125

\bibitem[{{Tanaka} {et~al.}(2008){Tanaka}, {Uchiyama}, {Aharonian},
  {Takahashi}, {Bamba}, {Hiraga}, {Kataoka}, {Kishishita}, {Kokubun}, {Mori},
  {Nakazawa}, {Petre}, {Tajima}, \& {Watanabe}}]{2008ApJ...685..988T}
{Tanaka}, T., {Uchiyama}, Y., {Aharonian}, F.~A., {et~al.} 2008, \apj, 685, 988

\bibitem[{{Truelove} \& {McKee}(1999)}]{truelove99}
{Truelove}, J.~K. \& {McKee}, C.~F. 1999, \apjs, 120, 299

\bibitem[{{Uchiyama} {et~al.}(2003){Uchiyama}, {Aharonian}, \&
  {Takahashi}}]{UchiyamaChandra}
{Uchiyama}, Y., {Aharonian}, F.~A., \& {Takahashi}, T. 2003, \aap, 400, 567

\bibitem[{{Uchiyama} {et~al.}(2007){Uchiyama}, {Aharonian}, {Tanaka},
  {Takahashi}, \& {Maeda}}]{2007Natur.449..576U}
{Uchiyama}, Y., {Aharonian}, F.~A., {Tanaka}, T., {Takahashi}, T., \& {Maeda},
  Y. 2007, \nat, 449, 576

\bibitem[{{Vink}(2012)}]{2012A&ARv..20...49V}
{Vink}, J. 2012, \aapr, 20, 49

\bibitem[{{Zabalza}(2015)}]{2015arXiv150903319Z}
{Zabalza}, V. 2015, in Proceedings of the 34th International Cosmic Ray
  Conference (ICRC 2015), in press, arXiv:1509.03319

\bibitem[{{Zirakashvili} \& {Aharonian}(2010)}]{2010ApJ...708..965Z}
{Zirakashvili}, V.~N. \& {Aharonian}, F.~A. 2010, \apj, 708, 965

\end{thebibliography}

\clearpage
\begin{appendix}
\label{sec:appendix}
\section{Fits to radial profiles}
\label{subsec:appendix-radial-fits}
\begin{figure*}[hbp!]
  \resizebox{0.98\hsize}{!}{\includegraphics[clip=]{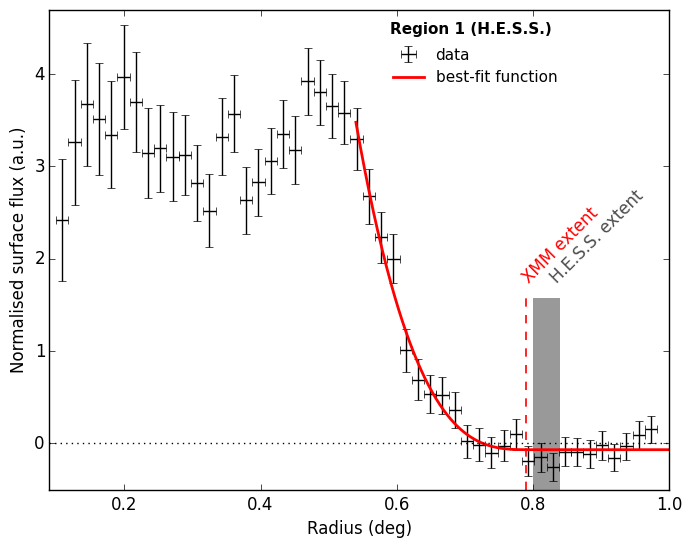}
    \includegraphics[clip=]{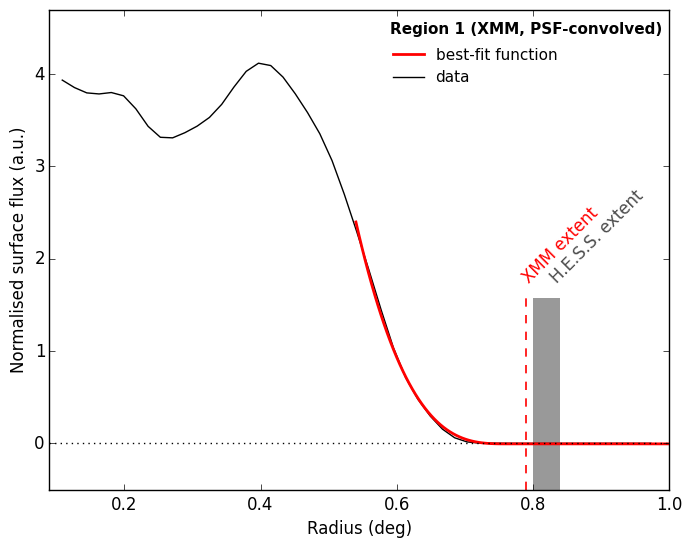}}
  \phantomcaption
\end{figure*}
\begin{figure*}
  \ContinuedFloat
  \resizebox{0.98\hsize}{!}{\includegraphics[clip=]{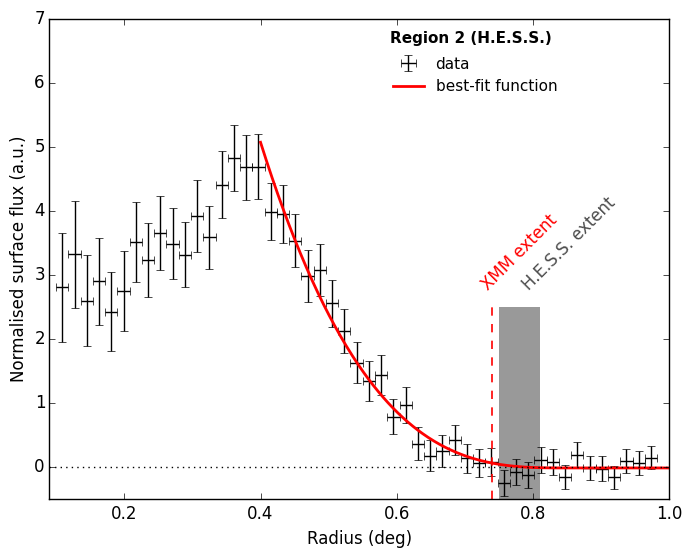}
    \includegraphics[clip=]{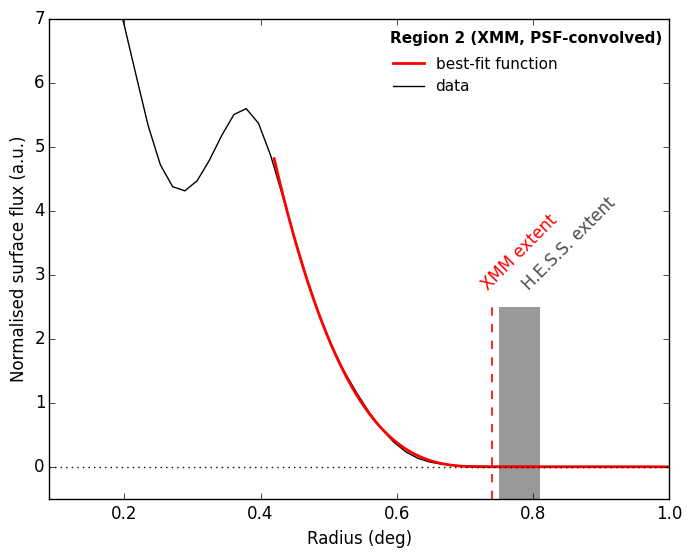}}
  \caption{\hess\ radial profiles from \rxj\ compared to the \xmm\
    data. For wedges 1 and 2, the profiles are shown on the left in
    both panels. On the right, the PSF-convolved \xmm\ profile is
    shown. For both profiles, the empirical best-fit function is
    overlaid as red solid line, see Sect.~\ref{sec:morphology} for
    more details.}
  \label{fig:radial-profile-fits:1}
\end{figure*}
Figure~\ref{fig:radial-profile-fits:1} shows the radial profiles from
the H.E.S.S and \xmm\ maps separately together with the best-fit
model functions discussed in Sect.~\ref{sec:morphology}.
\clearpage
\begin{figure*}
  \ContinuedFloat
  \resizebox{0.98\hsize}{!}{\includegraphics[clip=]{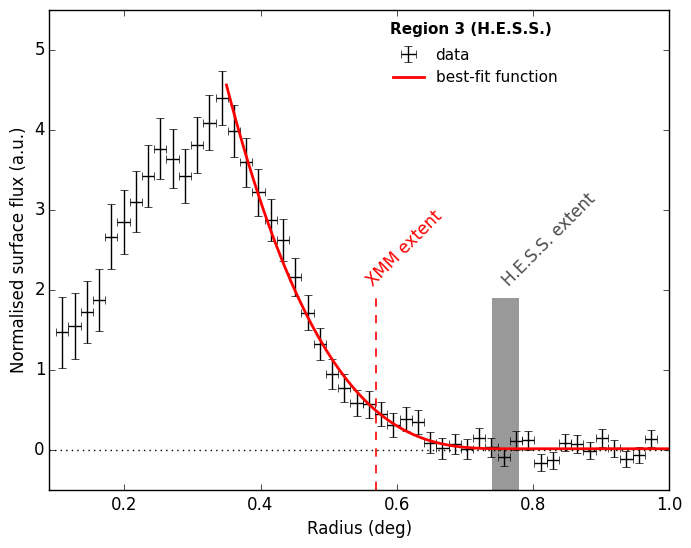}
    \includegraphics[clip=]{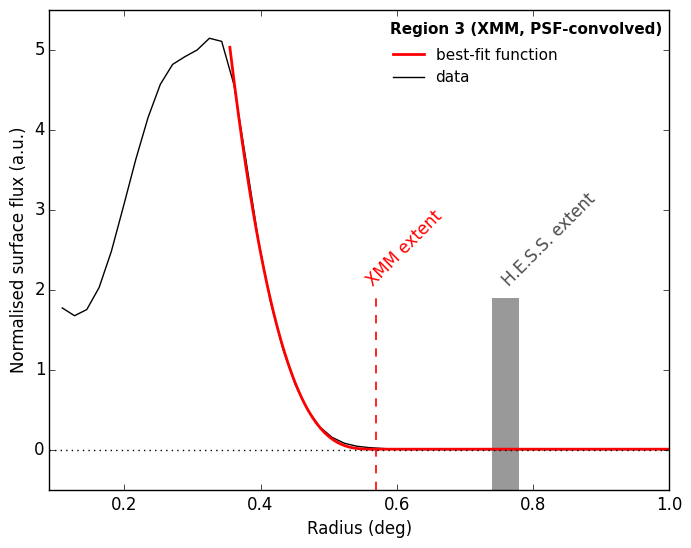}}
  \phantomcaption
\end{figure*}
\begin{figure*}
  \ContinuedFloat
  \resizebox{0.98\hsize}{!}{\includegraphics[clip=]{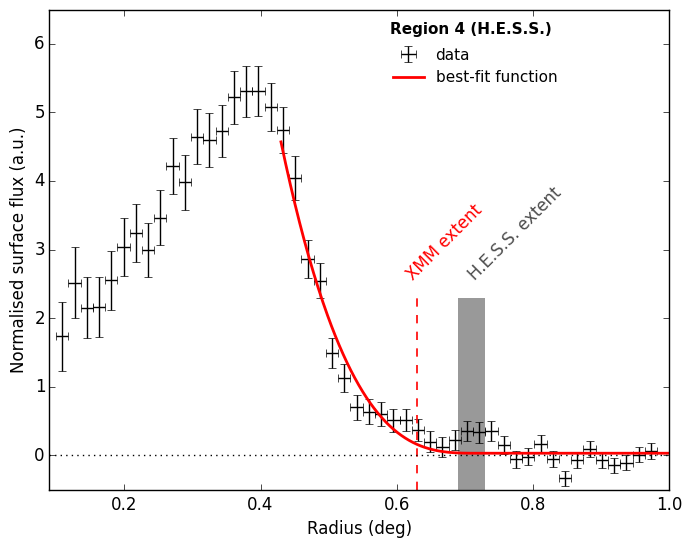}
    \includegraphics[clip=]{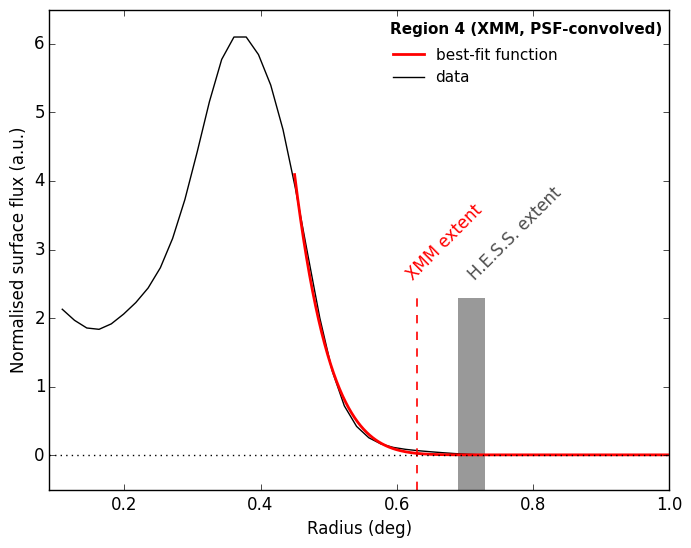}}
  \phantomcaption
\end{figure*}
\begin{figure*}
  \ContinuedFloat
  \resizebox{0.98\hsize}{!}{\includegraphics[clip=]{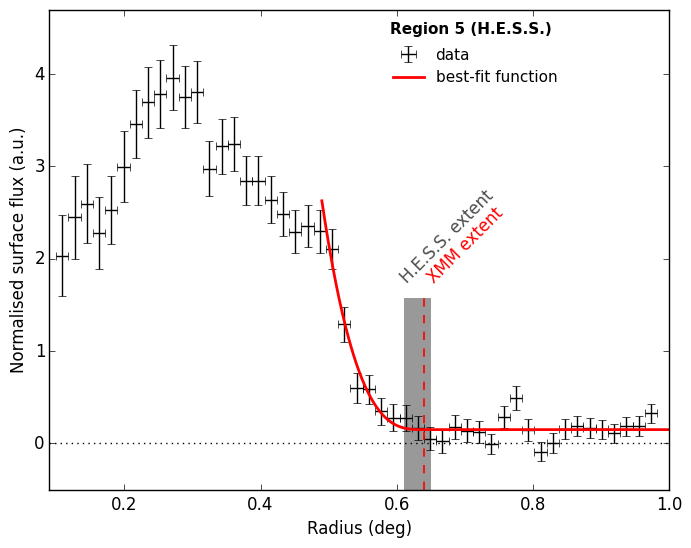}
    \includegraphics[clip=]{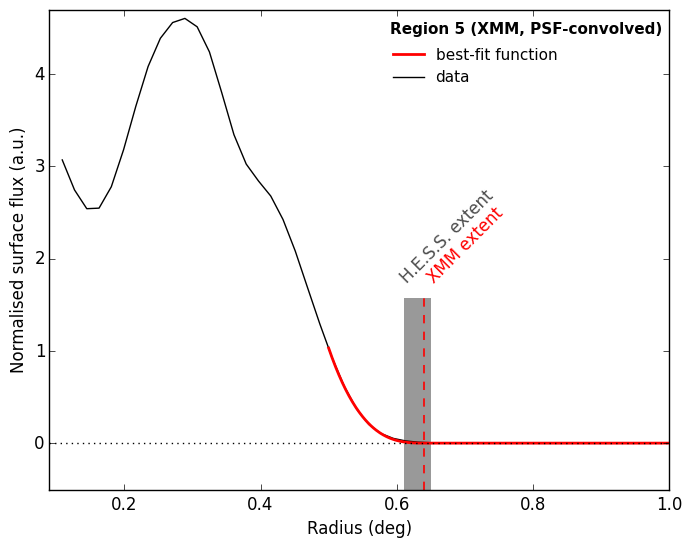}}
  \caption{continued: For wedges 3-5, the \hess\ radial profiles from
    \rxj\ are shown on the left in all three panels. On the right, the
    PSF-convolved \xmm\ profile is shown. For both profiles, the
    empirical best-fit function is overlaid as red solid line; see
    Sect.~\ref{sec:morphology} for more details.}
  \label{fig:radial-profile-fits:2}
\end{figure*}
\clearpage

\section{\hess\ image with overlaid XMM contours}
\label{subsec:appendix-hess-xmm}
\begin{figure*}[htbp]
\centering
  \begin{subfigure}{0.6\textwidth}
    \includegraphics[width=0.95\textwidth]{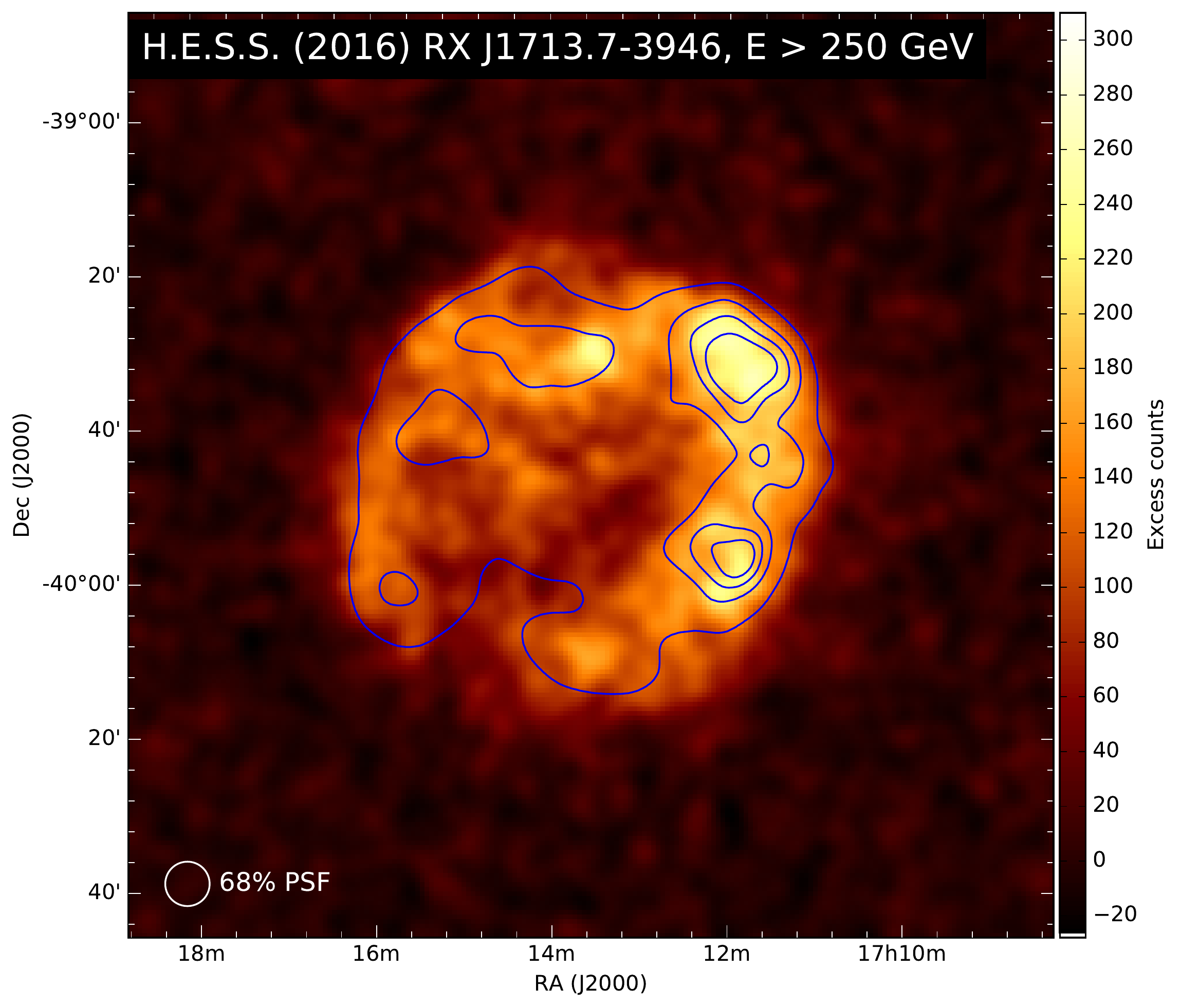}
  \end{subfigure}
  \caption{\hess\ gamma-ray excess image of \rxj\ with
    overlaid \xmm\ contours (1--10\,keV).}
  \label{fig:hess-xmm}
\end{figure*}
\clearpage

\section{Results from a border-detection algorithm}
\label{subsec:appendix-border-finder}
\begin{figure*}[htbp]
\centering
  \begin{subfigure}{0.48\textwidth}
    \includegraphics[width=\textwidth]{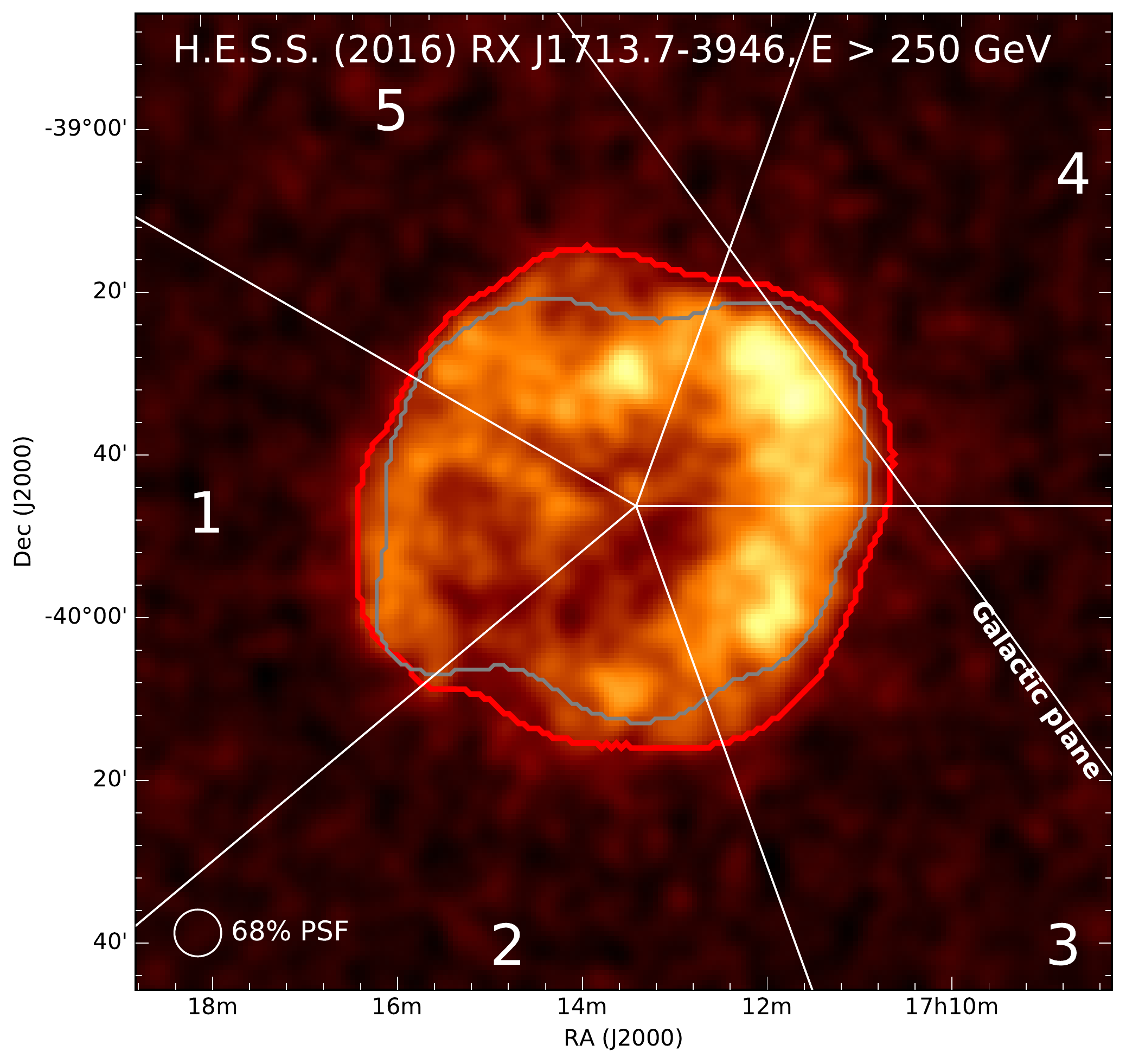}
  \end{subfigure}
  \begin{subfigure}{0.48\textwidth}
    \includegraphics[width=\textwidth]{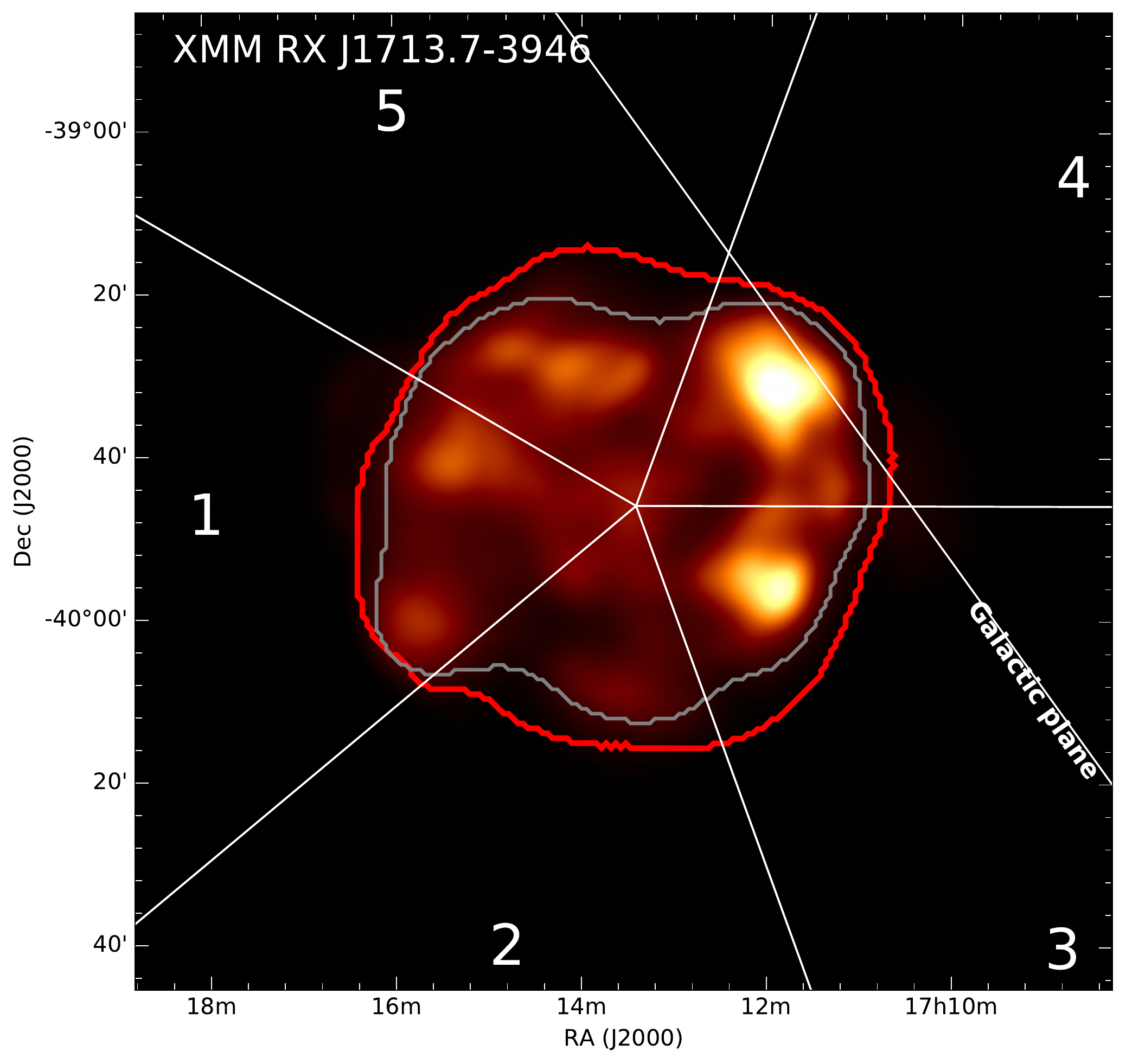}
  \end{subfigure}
  \caption{Results of the border-finder algorithm. On the left, the
    \hess\ gamma-ray excess image of \rxj\ is shown with overlaid
    borders of the gamma-ray (red) and X-ray (grey) data detected by
    the border-finder algorithm described in
    \protect\citet{ChanVese}. The wedges in which the radial profiles
    in Sect.~\ref{subsec:morph1} are studied are also shown along with
    the Galactic plane. On the right, the same two borders are
    overlaid on the \xmm\ X-ray image for comparison.}
  \label{fig:border-finder}
\end{figure*}
As an alternative method to determine the extent of the SNR shell the
border-detection algorithm described by \citet{ChanVese} was used on
the \xmm\ and \hess\ maps. This method is widely used in image
analysis to separate complex features from
backgrounds. Figure~\ref{fig:border-finder} shows the \hess\ image
together with the contours of the detected borders. The largest
differences between the radial sizes appear towards the south-west and
towards the north.  In the south-west  the radial fitting method
(Region 3, see Sect.~\ref{sec:morphology}) shows the largest
differences between X-rays and VHE gamma rays. However, towards the
north (Region 5), the radial sizes are consistent in the fitting
method. In this area, the radial profiles are the most complex, and a
diffuse emission component along the Galactic plane may play an
important role. While the radial fitting approach tries to find the
absolute outer edge of the shell, the border-finder algorithm
interprets fainter outer structures in the X-ray map as background not
belonging to the SNR shell.

Tests with the \hess\ map showed that the results from the
border-detection algorithm are very stable against a large range of
different signal-to-noise levels as well as systematic changes of the
normalisation of the \hess\ background by up to 2\%.
\clearpage

\section{Comparison of the radial profile between \xmm\ and \rosat}
\label{subsec:appendix-xmm-vs-rosat}
\begin{figure}[h]
  \begin{center}
\resizebox{0.98\hsize}{!}{\includegraphics[clip=]{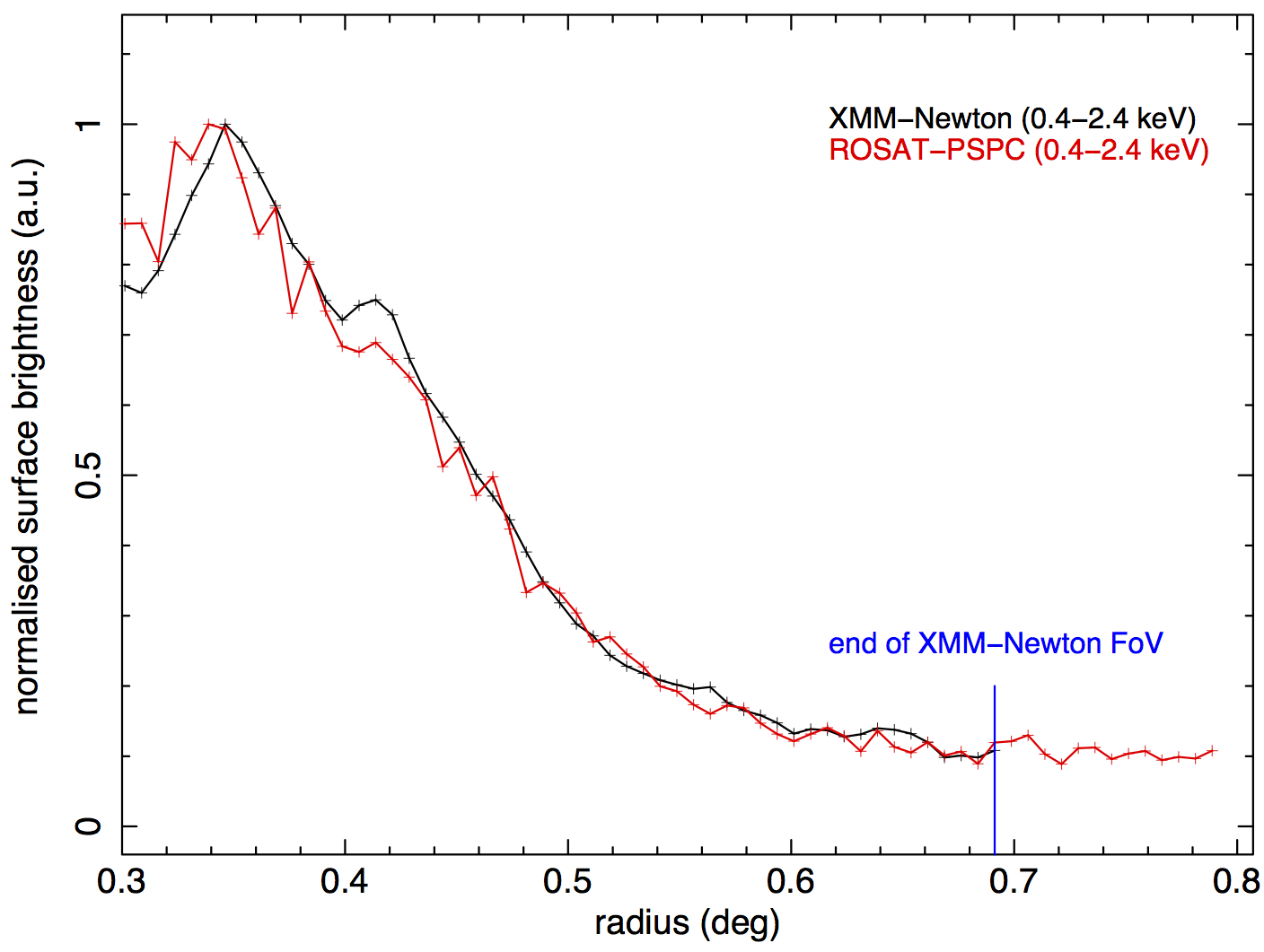}}
\caption{Radial profiles (full azimuth range) extracted from the \xmm\
  and \rosat\ maps and normalised to the peak flux.}
  \label{fig:xmm-vs-rosat}
  \end{center}
\end{figure}
 Figure~\ref{fig:xmm-vs-rosat} shows a comparison between radial
profiles from \xmm\ and \rosat~\citep{RASS}. The wider coverage of
\rosat\ is used here to confirm that the baseline Galactic diffuse surface
brightness level is already reached within the \xmm\ FoV.
\clearpage

\section{Spectral energy distributions of the western and eastern half}
\label{subsec:appendix:SEDs}
\begin{figure*}
\centering
  \begin{subfigure}{0.49\textwidth}
    \includegraphics[width=\textwidth]{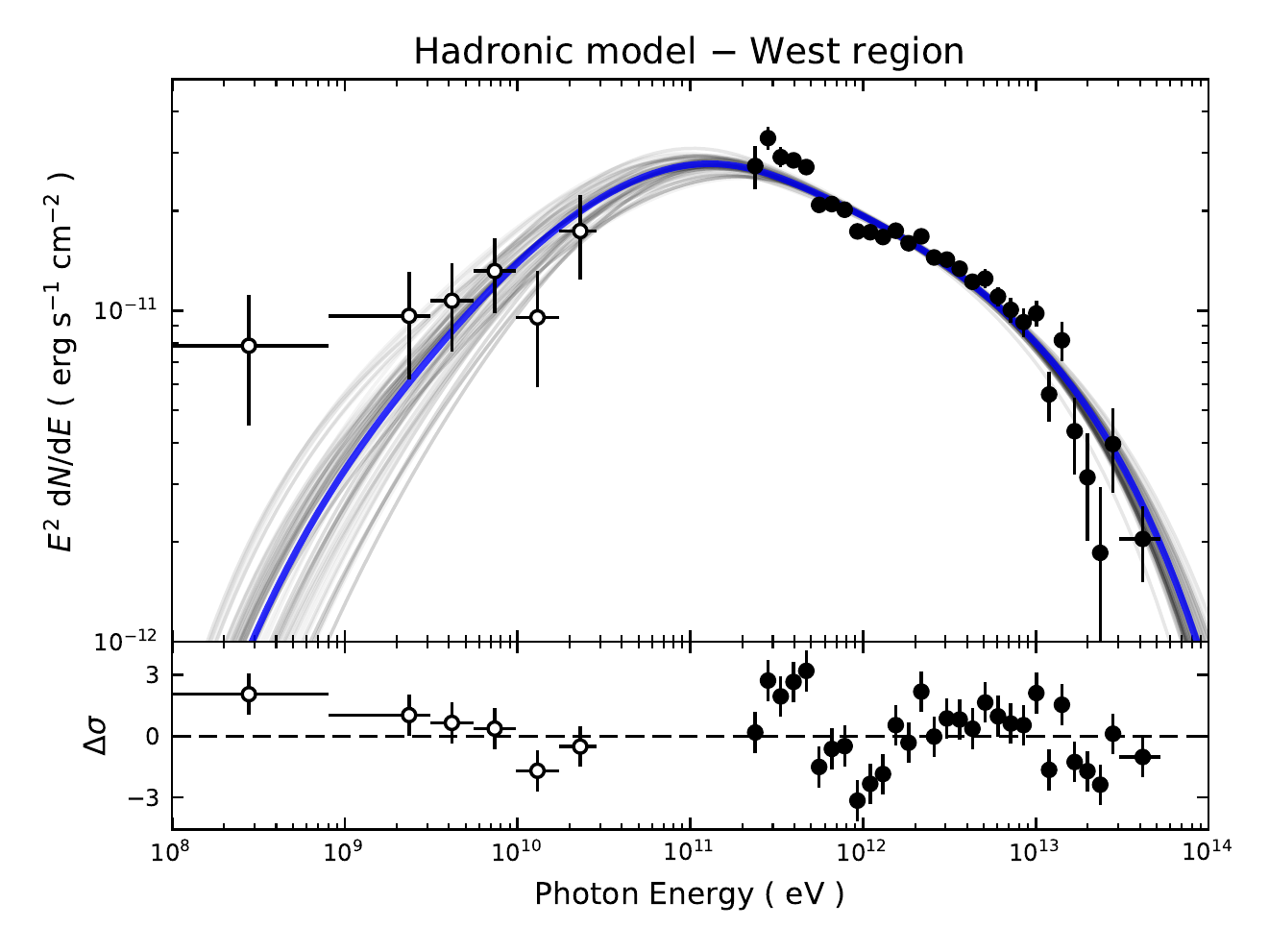}
  \end{subfigure}
  \begin{subfigure}{0.49\textwidth}
    \includegraphics[width=\textwidth]{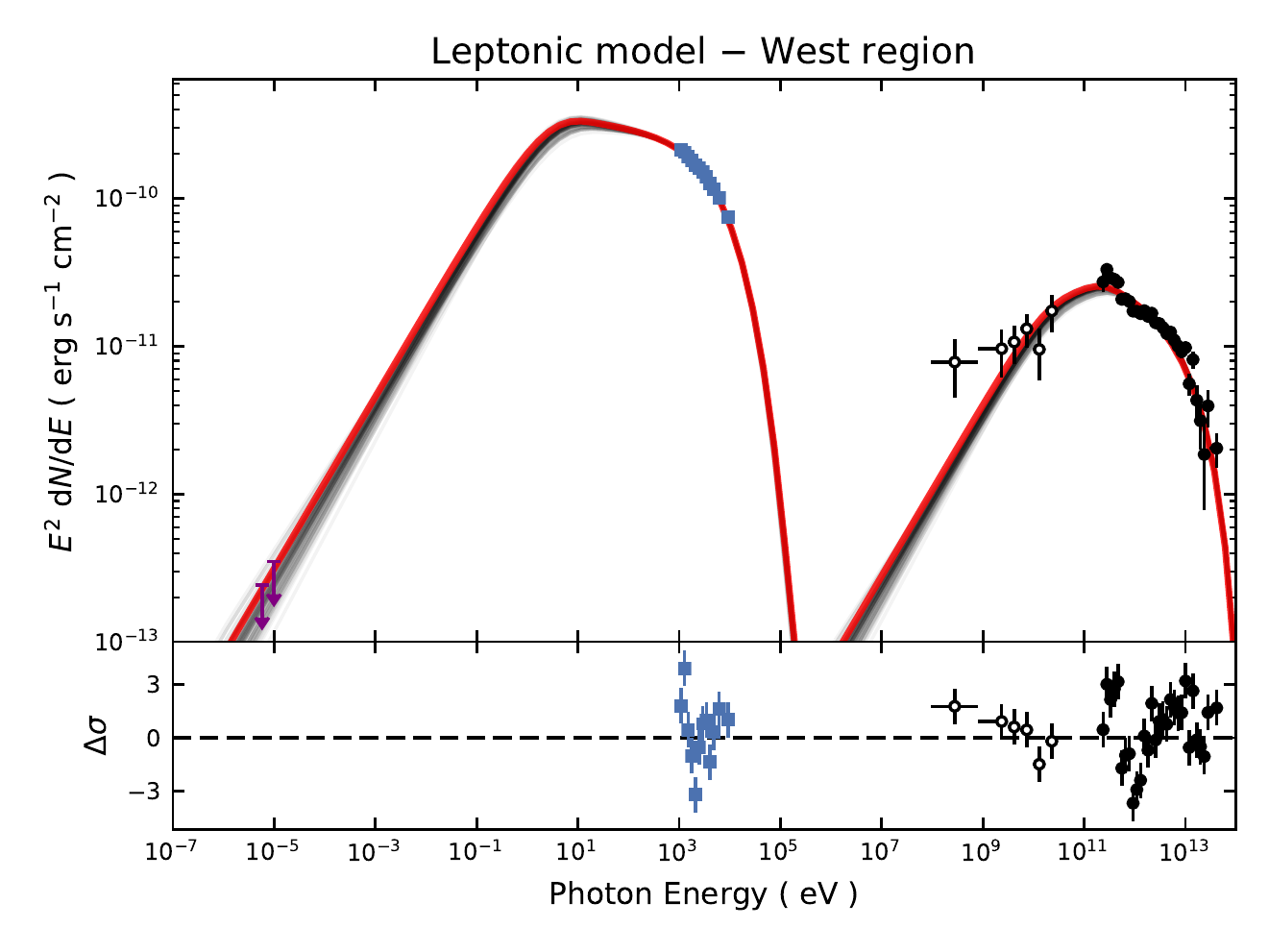}
  \end{subfigure}
  \begin{subfigure}{0.49\textwidth}
    \includegraphics[width=\textwidth]{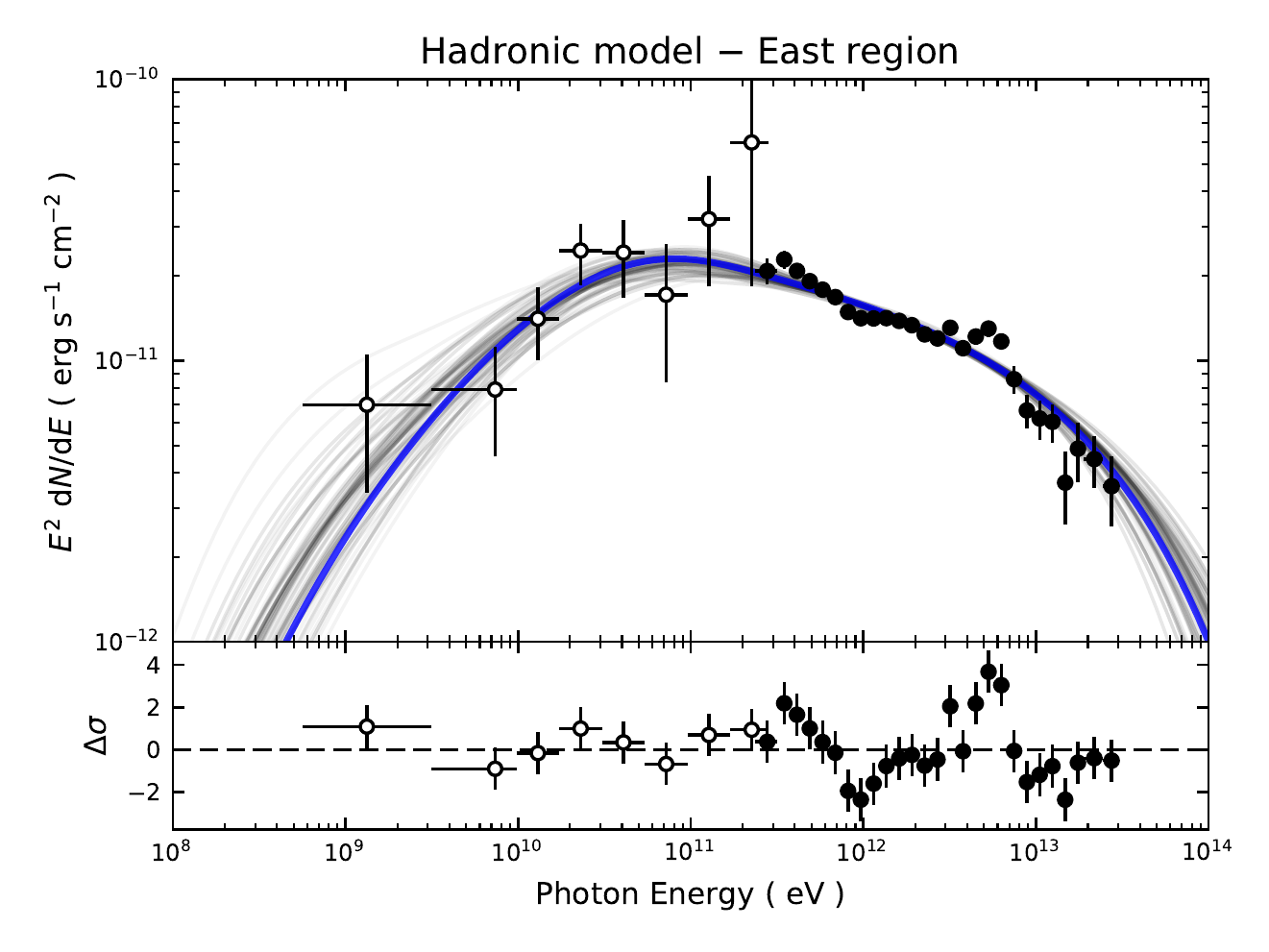}
  \end{subfigure}
  \begin{subfigure}{0.49\textwidth}
    \includegraphics[width=\textwidth]{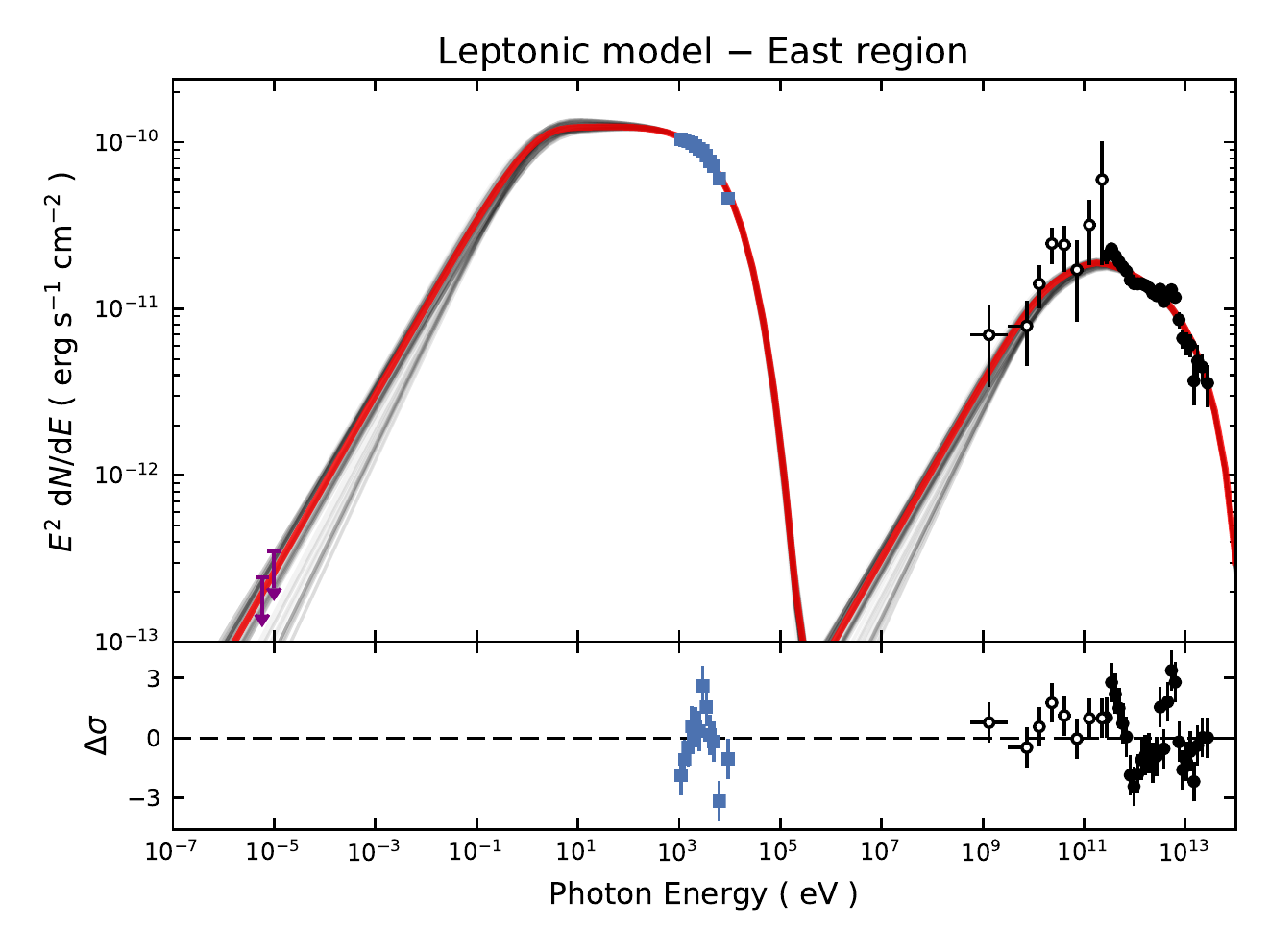}
  \end{subfigure}
  \caption{For the western~(top) and eastern~(bottom) halves of \rxj, the hadronic
    (left) and leptonic (right) gamma-ray models 
    obtained by our broadband fit are shown in these figures
    compared to the data. The thick blue and red lines indicate the
    maximum-likelihood models, and the grey lines surrounding them are
    the models for 100 samples of the MCMC chain and serve to illustrate
    the fit uncertainties. The ATCA radio data~\citep{Lazendic} of \rxj\
    plotted as magenta upper limits are determined for the
    north-west part of the SNR shell only and are scaled up by a factor
    of two here and included in the fit by constraining the model to
    stay below these values.
    \label{fig:sed:halves}}
\end{figure*}

\clearpage

\section{\hess\ energy flux points}
\label{appendix:subsection:spectra}
The energy flux points shown in Fig.~\ref{fig:full-remnant-spectrum}
are given below.

\clearpage

\begin{verbatim}
E (TeV)       E low (TeV)        E high (TeV)      Flux (erg cm-2 s-1)     Flux error (erg cm-2 s-1)
 0.23          0.21               0.25             4.44e-11                8.35e-12
 0.27          0.25               0.29             3.99e-11                5.17e-12
 0.32          0.29               0.35             4.28e-11                3.52e-12
 0.38          0.35               0.41             3.92e-11                2.41e-12
 0.45          0.41               0.49             3.96e-11                2.04e-12
 0.53          0.49               0.58             3.38e-11                1.73e-12
 0.63          0.58               0.69             3.43e-11                1.41e-12
 0.75          0.69               0.81             3.41e-11                1.34e-12
 0.89          0.81               0.96             3.21e-11                1.22e-12
 1.05          0.96               1.14             2.94e-11                1.12e-12
 1.25          1.14               1.36             3.02e-11                1.09e-12
 1.48          1.36               1.61             3.25e-11                1.07e-12
 1.76          1.61               1.91             2.99e-11                1.01e-12
 2.08          1.91               2.26             3.09e-11                1.19e-12
 2.47          2.26               2.68             2.62e-11                1.24e-12
 2.93          2.68               3.18             2.54e-11                1.17e-12
 3.47          3.18               3.77             2.59e-11                1.18e-12
 4.12          3.77               4.47             2.39e-11                1.28e-12
 4.88          4.47               5.29             2.50e-11                1.23e-12
 5.79          5.29               6.28             2.34e-11                1.31e-12
 6.86          6.28               7.44             2.01e-11                1.38e-12
 8.14          7.44               8.83             1.74e-11                1.51e-12
 9.65          8.83              10.47             1.42e-11                1.53e-12
11.44         10.47              12.41             1.58e-11                1.59e-12
13.56         12.41              14.71             8.34e-12                1.70e-12
16.08         14.71              17.45             9.67e-12                1.67e-12
19.99         17.45              22.53             3.82e-12                1.41e-12
24.62         22.53              26.71             5.68e-12                1.75e-12
29.19         26.71              31.67             2.79e-12                1.57e-12
34.61         31.67              37.55             5.08e-12                1.27e-12
\end{verbatim}
\clearpage
\end{appendix}
\end{document}